\documentclass[aps,onecolumn,showpacs,12pt,tightenlines,superscriptaddress,preprintnumbers,nofootinbib]{revtex4-1}
\usepackage{epsfig,amssymb,amsmath,psfrag,epstopdf,color}
\usepackage{graphicx}
\usepackage{subfigure}
\usepackage{caption}
\usepackage{graphics}
\usepackage{verbatim}

\allowdisplaybreaks
 
\def\beq{\begin{equation}}
\def\eeq{\end{equation}}
\def\bsp#1\esp{\begin{split}#1\end{split}}
\newcommand{\be}{\begin{equation}}
\newcommand{\ee}{\end{equation}}
\newcommand{\bea}{\begin{eqnarray}}
\newcommand{\eea}{\end{eqnarray}}

\def\Eqn#1{Eq.~(\ref{#1})}

\def\sect#1{section~{\ref{#1}}}

\def\Li{{\rm Li}}

\def\to{\rightarrow}

\def\ksl{\not{\hbox{\kern-2.3pt $k$}}}

\def\e{\epsilon}

\def\Ord{{\cal O}}

\def\Neqfour{{{\cal N}=4}}

\def\Li{\mathop{\rm Li}\nolimits}

\def\Eqn#1{Eq.~(\ref{#1})}

\def\spa#1.#2{\left\langle#1\,#2\right\rangle}
\def\spb#1.#2{\left[#1\,#2\right]}
\def\lor#1.#2{\left(#1\,#2\right)}
\def\sand#1.#2.#3{%
\left\langle\smash{#1}{\vphantom1}^{-}\right|{#2}%
\left|\smash{#3}{\vphantom1}^{-}\right\rangle}

\newcommand{\nn}{\nonumber}
\newcommand{\brk}{\nonumber\\&}
\newcommand{\nbrk}{\nonumber\\}

\newcommand{\Gcusp}{\Gamma_{\mathrm{cusp}}}

\newcommand{\gcusp}{\Gamma^{\mathrm{cusp}}}
\newcommand{\gsoft}{\gamma^{s}}
\newcommand{\grap}{\gamma^{r}}
\newcommand{\mcdot}{\!\cdot}

\newcommand{\nbar}{\bar{n}}

\usepackage{color}
\definecolor{darkblue}{rgb}{0,0,0.5}

\def\im{\mathrm{i}}

\def\sss{\scriptscriptstyle}
\def\qt{Q_{T}}
\def\vecqt{\vec{Q}_{T}}

\def\nuw{\nu}
\def\gae{{\gamma_{\sss E}}}

\def\vecb{\vec{b}_{\perp}}
\def\vecbsq{\vec{b}_{\perp}^{\, 2}}

\begin{document}

\preprint{FERMILAB-PUB-16-090-PPD-T}
\preprint{MIT-CTP-4795}
\title{An Exponential Regulator for Rapidity Divergences}
\author{Ye~Li}
\affiliation{Fermilab, PO Box 500, Batavia, IL 60510, USA}
\author{Duff Neill}
\author{Hua~Xing~Zhu}
\affiliation{Center for Theoretical Physics, Massachusetts Institute of Technology,
Cambridge, MA 02139, USA}

\begin{abstract}
\noindent Finding an efficient and compelling regularization of soft and collinear degrees of freedom at the same invariant mass scale, but separated in rapidity is a persistent problem in high-energy factorization. In the course of a calculation, one encounters divergences unregulated by dimensional regularization, often called rapidity divergences. Once regulated, a general framework exists for their renormalization, the rapidity renormalization group (RRG), leading to fully resummed calculations of transverse momentum (to the jet axis) sensitive quantities. We examine how this regularization can be implemented via a multi-differential factorization of the soft-collinear phase-space, leading to an (in principle) alternative non-perturbative regularization of rapidity divergences. As an example, we examine the fully-differential factorization of a color singlet's momentum spectrum in a hadron-hadron collision at threshold. We show how this factorization acts as a mother theory to both traditional threshold and transverse momentum resummation, recovering the classical results for both resummations. Examining the refactorization of the transverse momentum beam functions in the threshold region, we show that one can directly calculate the rapidity renormalized function, while shedding light on the structure of joint resummation. Finally, we show how using modern bootstrap techniques, the transverse momentum spectrum is determined by an expansion about the threshold factorization, leading to a viable higher loop scheme for calculating the relevant anomalous dimensions for the transverse momentum spectrum.
\end{abstract}

\maketitle

\section{Introduction}
Many phenomonologically important observables of Quantum Chromodynamics~(QCD) are transverse momentum sensitive. That is, they are defined as a measurement that directly puts a constraint on the momentum flowing perpendicular to some fiducial jet axis, without a corresponging cut on the rapidity. Examples include the transverse momentum~($\vecqt$) distribution of generic high-mass color-neutral systems~(Drell-Yan, Higgs, vector boson pair,\dots) in hadron-hadron collisions, semi-inclusive fragmentation of hadrons, the scalar sum of transverse momentum magnitudes as found in jet or beam broadening \cite{Chiu:2011qc,Chiu:2012ir,Becher:2011pf,Becher:2012qc,Tackmann:2012bt,Banfi:2012jm,Banfi:2015pju}, and vetoes on the transverse momenta of clustered jets \cite{Becher:2012qa,Becher:2013xia,Stewart:2013faa}\footnote{Not in this class are indirect restrictions on transverse-momenta, like beam thrust\cite{Stewart:2009yx,Stewart:2010pd,Berger:2010xi}.}. An universal feature of all such transverse momentum sensitive factorizations is the presence of rapidity divergences. In a naive soft or collinear sector, one encounters integrals over the light-cone components of the participating partons, which dimensional regularization fails to regulate. This is due to the fact that dimensional regularization breaks any possible dilatation invariance of a theory, but not the Poincare invariance. Thus classes of momenta with differing invariant masses can be distingushed via their relative scaling with respect to the dimensional regularization mass scale $\mu$. However, classes of momenta differing only by their boost are not distinguished by the boost invariant dimensional regularization. These rapidity divergences are then a necessity when soft and collinear modes exist at the same invariant mass scale, as found in a transverse momentum sensitive observable. From a more practical point of view, these rapidity divergences (an artefact of using a factorized form of the cross-section) are directly tied to a large logarithm of the fixed order QCD calculation (QCD itself is rapidity divergence free). Much literature has been devoted to the issue of a convenient scheme to regulate the light-cone integrals. Once regulated, the divergences are isolated, cancelled at the level of the physical cross-section, and the residual logarithms left from the isolation are exponentiated either by hand or by evolution-equations, thereby controlling the large logarithm found in fixed-order perturbation theory.

The transverse momentum spectrum for color-singlets in particular has long been a critical quantity to understand the factorization properties of QCD. One simply wishes to know the differential cross-section for the relative momentum of the color singlet object with respect to the beam axis, while being inclusive over all other radiation in the event. Taking the transverse momentum to be small relative to the hard scale ($Q$) involved in the production of the observed particles (the invariant mass scale of the Drell-Yan pair or the Higgs boson, for example) implicitly constrains how the recoiling QCD radiation moves with respect to the beam axis. In this limit, the cross-section is dominated by either soft radiation, or emissions collinear to the beam. Such soft and collinear emissions constitute the infra-red structure of perturbative QCD, which presents an underlying universality due to the fact that the emissions in these different kinematic regimes (hard, collinear, or soft) factorize from each other, and do not quantum mechanically interfere\footnote{Off-shell Glauber or Coloumb potential exchanges can violate this factorization when colored partons exist in the initial state, and cannot be absorbed into the soft exchanges \cite{Forshaw:2006fk,Rogers:2010dm, Catani:2011st, Forshaw:2012bi,Rothstein:2016bsq}. For a comprehensive discussion in the context of SCET, see \cite{Rothstein:2016bsq}. For the transverse momentum spectrum of color neutral states, Glauber exchanges have been argued to be irrelevant \cite{Collins:1984kg,Gaunt:2014ska}. }. Due to this infra-red sensitivity, the fixed order expansion for the cross-section becomes dominated by large logarithms of the hard production scale $Q$ to the infra-red recoil scale $\qt$. The factorization allows one to resum these large logarithmic contributions to all orders. When this resummation is combined with the fixed order distribution that is not singularly enhanced, one achieves a remarkably precise description of the QCD spectrum, giving a benchmark for theory versus experimental predictions. If we denote the potentially large logarithm as $L_{\qt}=\ln\frac{\qt}{Q}$, and assign the scaling $\alpha_s\sim L^{-1}_{\qt} $ when the logarithm is large, then the current state of the art for these resumations is the N$^2$LL+NLO accuracy, where one in the fixed order result includes contributions for up to two final state recoiling partons, and has resummed all logarithms up to contributions that scale as $\alpha_s^2 L_{\qt}$ \cite{Bozzi:2010xn,deFlorian:2011xf,Banfi:2012du,Becher:2012yn,Echevarria:2015uaa,Neill:2015roa,Bagnaschi:2015bop}\footnote{For example calculations necessary to fix the anomalous dimensions at two-loop accuracy with a variety of regularization procedures, see Refs. \cite{deFlorian:2000pr,deFlorian:2001zd,Gehrmann:2014yya,Echevarria:2015byo,Luebbert:2016itl}}.

In this paper, we introduce a new methodology for calculating the control quantity for rapiditiy divergences, the rapidity anomalous dimension, see Refs. \cite{Chiu:2011qc,Chiu:2012ir}\footnote{This is directly related to the Collins-Soper kernel of Refs. \cite{Collins:2011zzd, Aybat:2011zv} in TMD-PDF's.}. We exploit the fact that cross-sections often have multiple singular regions with distinct scalings of the low-scale modes, leading to distinct factorization formula, see Ref. \cite{Larkoski:2014tva}. These factorizations, however, while resumming different logarithms, must be consistent with each other at any fixed order in perturbation theory, since they describe the same cross-section. This allows us to calculate in the threshold region the transverse momentum spectrum, and through consistency with the more standard transverse momentum dependent parton distribution function (TMD-PDF's) factorization, extract the rapidity anomalous dimension. We can then combine the technology developed for threshold calculations (see Refs. \cite{Anastasiou:2013srw,Li:2013lsa,Duhr:2013msa,Li:2014bfa,Anastasiou:2014vaa,Zhu:2014fma,Anastasiou:2015yha}) with modern bootstrap techniques from amplitudes to push the calculation to three-loop order. In a companion paper, two of us will present the full three-loop rapidity anomalous dimension phenomologically relevant for collider experiments. Though we deal mainly with the transverse momentum spectrum in hadron-hadron collisions, however, we believe that our approach is widely adaptable to many transverse momentum sensitive observables, at least where one can understand the analytic structure of the fixed order calculation. We refer to this methodology as the ``exponential regulator,'' since it implements an exponential cutoff in the total energy of the final state. Alternatively, one may think of it as a threshold regulator, where one imposes in addition to the transverse momentum observable a constraint on the total energy of the soft radiation crossing the cut. It effectively acts as a gauge invariant cut-off on the rapidity integrals.

The factorization approach to resummation we will adopt is that of Soft Collinear Effective Field Theory, \cite{Bauer:2000yr, Bauer:2000ew, Bauer:2001ct}, which gives a precise set of rules for determining the all-orders form of the factorized formulae. In general one seeks to write a cross-section sensitive to a singular region of phase-space as a product of functions of the form:
\begin{align}
d\sigma \sim \sigma_0 H \otimes B_n\, \otimes B_{\bar{n}}\otimes_{i=1}^N J_i \otimes S+...\,.
\end{align} 
The $H$ denotes calculation of the hard process, the $B$ a beam function for the initial state radiation off of the colliding hadrons\footnote{It has become accepted in the QCD literature to combine enough of the soft function with the collinear matrix elements of the proton to cancel the rapidity divergences, and label the resulting entity the TMDPDF. Following the SCET literature, we will therefore call the strictly collinear part of the matrix element (defined with zero-bin subtractions) the transverse momentum dependent \emph{beam} function, in distinction to the TMDPDF.}, $J_i$ functions for any possible final state jets, and the $S$ the contribution from soft wide angle radiation. Each function has a field theoretic operator definition, and the $\otimes$ denotes a convolution over the contribution from each sector to the relevant observable and any possible momentum recoil. The factorized functions summarize the contribution from ``on-shell'' modes of QCD with a specified scaling, and can be calculated independently. This convolution structure is to be expanded according to the scaling of the modes to produce a formula that is homogenous in the power counting, a procedure known as the multipole expansion \cite{Grinstein:1997gv,Beneke:2002ph}. Since the multipole expansion enforces a homogeneous power counting in each convolution, one is prevented from developing a large logarithm in the effective theory matrix element. Instead, one is often rewarded with potentially multiple-divergences of the naive function. That is, one is trading the large logarithm of the perturbative expansion for an explict divergence in the calculation of effective theory matrix element. Several variations of the SCET formalism has been applied to transverse momentum distributions before, see Refs. \cite{Becher:2010tm, GarciaEchevarria:2011rb, GarciaEchevarria:2011rb, Echevarria:2012js, Chiu:2012ir}. In the operator based factorization literature, three approaches have appeared to accomplish this task, the collinear anomaly \cite{Becher:2010tm}, the Collins-Soper equation \cite{Collins:1981uk,Collins:1984kg,Collins:2011zzd,Aybat:2011zv}, and finally the framework of the rapidity renormalization group (RRG) \cite{Chiu:2011qc,Chiu:2012ir}. 

The outline of the paper is as follows: first we review the topic of transverse momentum resummation in the SCET formalism. For a general regularization scheme, we show that as long as the regulator is implemented symmetrically with appropriate subtractions in the different sectors, the rapidity resummation's scheme dependence is fixed by the hard function's scheme dependence. The subtractions themselves are regulator dependent. Since the hard function is free of rapidity divergences, this necessarily implies a universality to the rapidity anomalous dimension regardless of regulator+subtractions. Having established the factorization framework, we introduce the exponential regulator with the example of the one-loop calculation, which is defined by taking the limit of the fully differential soft functions. We then discuss the relation of transverse momentum and threshold factorizations from their connection with the fully differential functions, showing that the exponential regulator necessarily calculates a rapidity renormalized transverse momentum soft function. Lastly we show that the utility of the regulator at higher loops lies in reducing the calculational problem of the integrals to that of the threshold soft function. Using the extensive work done on this subject, we make an ansatz of the fully differential soft function in terms of harmonic polylogarithms (HPL's), and demonstrate how to reproduce existing results at one and two loops by bootstrapping. We also present partial results at three loops using this technique, and the full result is deferred to a companion paper. Finally, we conclude with thoughts on future directions. Some technical details are collected in appendicies.

\section{Review of Transverse Momentum Factorization}\label{sec:review}
\begin{figure}[ht]\centering
\includegraphics[scale=.5]{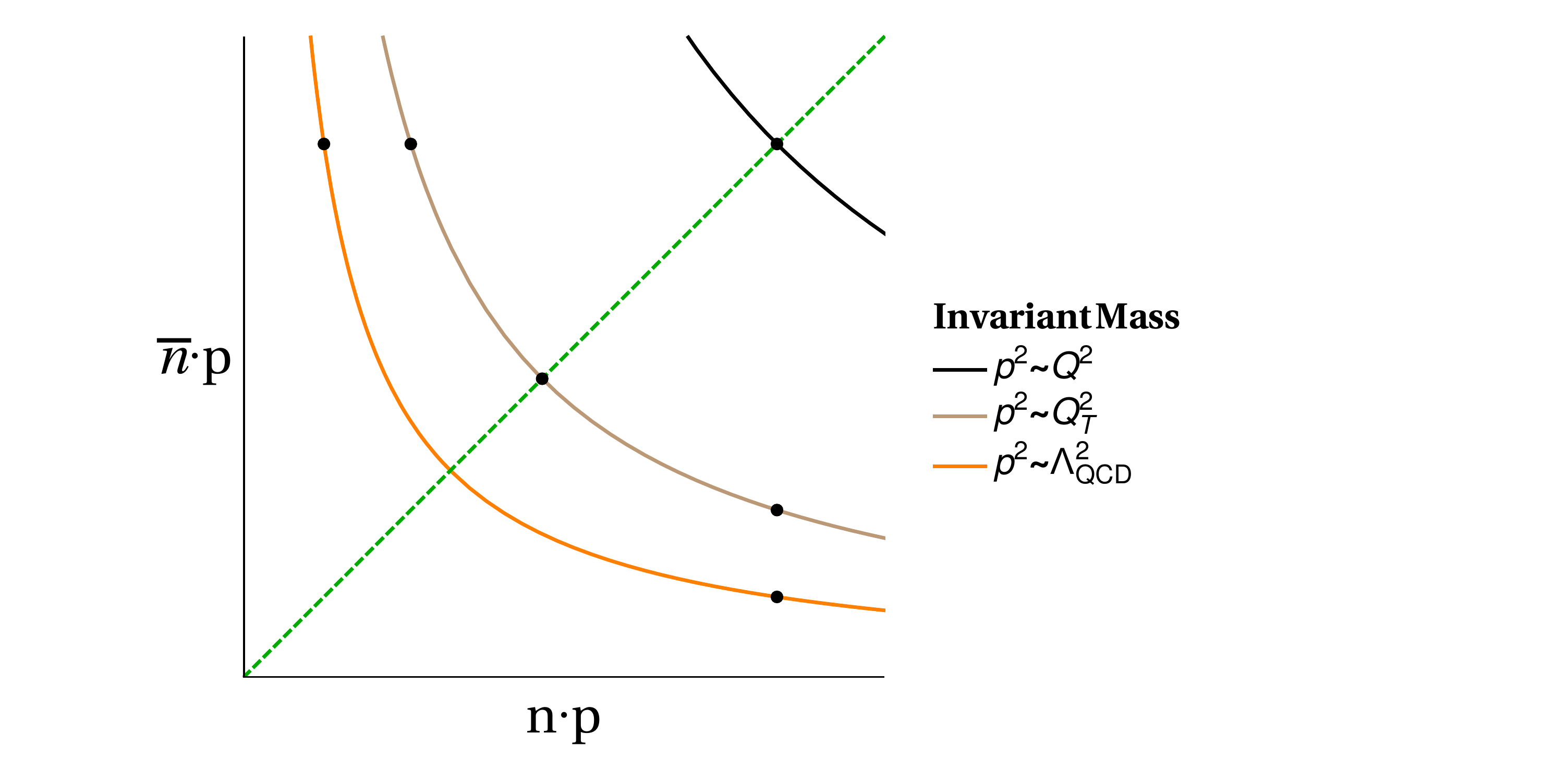}
\caption{\label{fig:tmd_modes} The invariant mass for the modes contributing to the transverse momentum spectrum of the Drell-Yan pair.}
\end{figure}

The factorization theorem for the Drell-Yan transverse momentum distribution takes the form\footnote{Several equivalent definitions can be found in the literature. They differ ultimately with regards to how the rapidity divergences are regulated, and the necessary subtractions that must take place given the form of the regulation.}:
 \begin{align}\label{eq:fact_theorem_t_spectrum}
 \frac{d\sigma}{dy d^2\vec{Q}_TdQ^2}&=\sigma_0 \int\frac{d^4q}{(2\pi)^3}\delta^+(n\cdot q\,\nbar\cdot q-Q^2)\delta\Big(y-\frac{1}{2}\ln\frac{n\cdot q}{\bar{n}\cdot q}\Big) \delta^{(2)}(\vec{Q}_T-\vec{q}_{\perp})\nonumber\\
 &\qquad\qquad\int d^4b e^{ib\cdot q}H_{q\bar{q}}(Q)B_{n,q/N_A}(0,n\cdot b,\vec{b}_{\perp})B_{\nbar, \bar{q}/N_B}(\bar{n}\cdot b,0,\vec{b}_{\perp})S(0,0,\vec{b}_{\perp})+\text{q}\leftrightarrow\bar{\text{q}}...\,.\\
 \label{eq:fact_theorem_t_spectrum_momentum_frac}&=\sigma_0 \int d^2b e^{i\vec{b}_{\perp}\cdot \vec{Q}_T}H_{q\bar{q}}(Q)B_{n,\perp,q/N_A}(x_A,\vec{b}_{\perp})B_{\nbar,\perp,\bar{q}/N_B}(x_B,\vec{b}_{\perp})S_{\perp}(\vec{b}_{\perp})+\text{q}\leftrightarrow\bar{\text{q}}...\,.
 \end{align}
$Q^2$ is the invariant mass of the Drell-Yan pair, and $\sigma_0$ is the leading-order (LO) cross section. The light-cone coordinates are defined with respect to the beam axis, and satisfy:
\begin{align}
n^2&=\bar{n}^2=0,\qquad n\cdot \bar{n}=2\,,\nonumber\\
q&=(\nbar\cdot q,n\cdot q,q_{\perp})=(q^+,q^-,q_{\perp})\,,
\end{align}
When switching to the $x_{A,B}$ momentum fractions of the partons in the hard collision, we intrepret the light-cone momentum components of the Drell-Yan pair as:
\begin{align}
x_{A,B}=e^{\pm y}\sqrt{\frac{Q^2}{s}}\,,\qquad q^+= x_A P_A^+\,,\qquad q^-= x_B P_B^-
\end{align} 
$s=P_A^+ \, P_B^-$ is the hadronic center of mass of the two colliding nucleons $N_A$ and $N_B$, the momenta of which can be written as $P_A = P_A^+ n/2$ and $P_B = P_B^- \nbar/2$ respectively.

The $H_{q\bar{q}}$ (which we will shorten to $H$) is the hard function that contains all the virtual correction to the LO contribution, and is obtained by matching from QCD to SCET. The $B_{n, q/N}$ (which we will shorten to just $B_{n}$) are beam functions encoding the energetic emissions along the beam axis, and $S$ is a soft factor encoding the contribution from soft states. In the fully differential form, these functions have the following operator definitions:
\begin{align}
B_{n}(b^+, b^- ,\vec{b}_{\perp})&=\text{tr}\langle N(P)|\bar{\chi}_n(b)\frac{\nbar\!\!\!\slash}{2}\chi_n(0)|N(P)\rangle\,,\\
S(b^+ , b^- ,\vec{b}_{\perp})&=\frac{1}{d_a}\text{tr}\langle 0|{\rm T}\{S_{\nbar}^\dagger(0)S_{n}(0)\} {\rm \bar{T}}\{S_{n}^\dagger(b)S_{\nbar}(b)\}|0\rangle\,,
\end{align}
where $b=(b^+,b^-,\vec{b}_{\perp})$ and $d_a=N_c=3$ for the Drell-Yan process. The $\chi_n$ field is a gauge invariant quark field operator dressed with a collinear wilson, and together with the soft and collinear wilson lines have the respective definitions:
\begin{align}
\chi_n(x)&=W^{\dagger}_n(-\infty,x)\psi_n(x)\\
W_{n}(x)&={\rm P}\,\text{exp}\Big(ig\int_{-\infty}^{0}ds\,\nbar\cdot A(x+s\nbar)\Big)\,,\\
S_{n}(x)&={\rm P}\,\text{exp}\Big(ig\int_{-\infty}^{0}ds\,n\cdot A(x+sn)\Big)\,.
\end{align}
In the factorization of the transverse momentum distribution, these functions compute the contribution to the observable from modes with momentum scaling:
\begin{align}\label{eq:SCET_II_scaling}
p_n\sim Q\Big(1,\lambda^2,\lambda\Big)  &\qquad p_{\nbar}\sim Q \Big(1,\lambda^2,\lambda\Big)\,,\nonumber\\
p_s\sim &Q\Big(\lambda,\lambda,\lambda\Big)\,,\\
\lambda=&\frac{\qt}{Q}\ll 1\,.
\end{align}
As a result, some light-cone coordinates need to be set to zero for proper power counting of the multipole expansion, and the relevant beam\footnote{The beam function under this momentum scaling is often called transverse momentum dependent parton distribution functions (TMD-PDF's)} and soft functions become,
\begin{align}\label{eq:qt_factorization_functions}
B_{n,\perp}(x ,\vec{b}_{\perp})=\int \frac{d \,b^-}{2\pi}e^{\frac{i}{2} (x P^+ b^-) }B_{n}(0, b^- ,\vec{b}_{\perp}) \,,\qquad S_{\perp}(\vec{b}_{\perp})=S(0, 0,\vec{b}_{\perp}) \,,
\end{align}
Often the momentum modes are called ``on-shell,'' since their dispersion relation satisfies $p_{n}^2=p_{\nbar}^2=p^2_s=Q^2\lambda^2$ homogenously, and as $\lambda\rightarrow 0$, they scale to exactly on-shell emissions. These modes have the important property that they are all at the same invariant mass-scale, as depicted in Fig. \ref{fig:tmd_modes}, so that the appropriate effective field theory is SCET$_{II}$, and are distinguished only with size of their relative energy scale or typical rapidity. Since dimensional regularization is invariant under boosts, one cannot distinguish these modes from each other in an integral with dimensional regularization alone. In so called  SCET$_{I}$ theories, modes are distinguished by their invariant masses, and since dimensional regularization breaks dilatation invariance, it suffices to regulate the theory completely and seperate the modes. A further regulator is needed when integrating over the light-cone variables in a typical diagram, and several have been proposed in the literature, each their various strengths and weaknesses. They may be classed into analytic style regulators \cite{Becher:2011dz,Chiu:2012ir}, deformations of the wilson line directions \cite{Ji:2004wu,Collins:2011zzd}, or finally a mass added to the eikonal propagator (the ``$\delta$'' regulator) \cite{Chiu:2009yx, Echevarria:2015byo}. Beyond the obvious requirement of ease of calculational use, one would also demand the regulator preserve gauge invariance, non-Abelian exponentiation \cite{Gatheral:1983cz,Frenkel:1984pz}, and a democratic treatment of sectors  (at least up to terms that vanish as the regulator is taken to its singular limit). For all regulators that have an explicit mass scale associated, like deformations of the wilson line direction or the $\delta$-regulator, the zero-bin subtraction will not be zero \cite{Manohar:2006nz, Chiu:2009yx}. 
  
It is important to note the origin of the light-cone singularities. In the factorization theorem of Eq. \eqref{eq:fact_theorem_t_spectrum}, the TMD-beam functions are localized at either $b^+=0$ or $b^-=0$, while the soft function is localized at both. This prevents momentum sharing in these small momentum components, since if we were to perform the fourier transform in Eq. \eqref{eq:fact_theorem_t_spectrum}, no recoil convolutions would appear in either the $n$ or $\nbar$ directions. The rapidity and mass of the Drell-Yan pair sets these momentum scales once and for all, to leading power. This is a direct consequence of the multipole expansion and the scaling of Eq. \eqref{eq:SCET_II_scaling}, and such an expansion is necessary to guarantee no large logarithms appear in the EFT.

Introducing a regulation scheme with the appropriate subtractions, one will have a generalized renormalized definition of the TMD-beam function and soft function (which can be combined together to form a TMDPDF). One removes systematically the light-cone and the ultra-violet divergences from each function. Removing these divergences will introduce a scale at which the divergences are subtracted, which we will generically call $\nu$ for the rapidity divergences, and $\mu$ for the ultra-violet. Since the physical cross-section is finite, these divergences will cancel between the various functions in the factorization formula, and the variation under the scale where the divergences are subtracted are controlled by generalized renormalization group equations:
\begin{align}\label{eq:RG_Generic}
\mu^2\frac{d}{d\mu^2} \ln B_{n,\perp}\Big(x_A,\vec{b}_{\perp};\mu,\nu\Big)&=\gamma_{B}\Big(\frac{\nu}{x_A P_A^+};\alpha_s(\mu)\Big) \,,\nonumber\\
\mu^2\frac{d}{d\mu^2} \ln B_{\nbar,\perp}\Big(x_B,\vec{b}_{\perp};\mu,\nu\Big)&=\gamma_{B}\Big(\frac{\nu}{x_B P_B^-};\alpha_s(\mu)\Big) \,,\nonumber\\
\mu^2\frac{d}{d\mu^2} \ln S_{\perp}\Big(\vec{b}_{\perp};\mu,\nu\Big)&=\gamma_{S}\Big(\frac{\mu}{\nu};\alpha_s(\mu)\Big) \,,\nonumber\\
\mu^2\frac{d}{d\mu^2} \ln H(Q;\mu)&=\gamma_{H}\Big(\frac{\mu}{Q};\alpha_s(\mu)\Big)\,.
\end{align}
and for the rapidity renormalization:
\begin{align}\label{eq:RRG_Generic}
\nu^2 \frac{d}{d\nu^2}\ln B_{n,\perp}\Big(x_A,\vec{b}_{\perp};\mu,\nu\Big)&=-\frac{1}{2} \gamma_{R}\Big(\frac{\mu |\vec{b}_{\perp}|}{b_0};\alpha_s(\mu)\Big),\nonumber\\
\nu^2 \frac{d}{d\nu^2}\ln B_{\nbar,\perp}\Big(x_B,\vec{b}_{\perp};\mu,\nu\Big)&=-\frac{1}{2} \gamma_{R}\Big(\frac{\mu |\vec{b}_{\perp}|}{b_0};\alpha_s(\mu)\Big),\nonumber\\
\nu^2 \frac{d}{d\nu^2}\ln S_{\perp}\Big(\vec{b}_{\perp};\mu,\nu\Big)&=\gamma_{R}\Big(\frac{\mu |\vec{b}_{\perp}|}{b_0};\alpha_s(\mu)\Big),
\end{align}
where $b_0=2e^{-\gamma_E}$. Since divergences cancel in the physical cross-section, the ultra-violet anomalous dimensions satisfy the constraint,
\begin{align}
&\gamma_{B}\Big(\frac{\nu}{x_A P_A^+};\alpha_s(\mu)\Big)+\gamma_{B}\Big(\frac{\nu}{x_B P_B^-};\alpha_s(\mu)\Big)+\gamma_{S}\Big(\frac{\mu}{\nu};\alpha_s(\mu)\Big)+\gamma_{H}\Big(\frac{\mu}{Q};\alpha_s(\mu)\Big)=0\,.
\end{align}
Similar constraint is manifestly written for the rapidity renormalization group. We have used the fact that the anomalous dimensions of the two TMD-PDFs should be the same up to relabeling $n$ and $\nbar$, given that the regularization procedure treats the two beam sectors identically. The arguments of these anomalous dimensions are dictated by the factorization structure in \eqref{eq:fact_theorem_t_spectrum}. The hard production scale $Q^2\sim x_A x_B P_A^+ P_B^-$ gets factorized into the large momentum components of the beams sectors, the $x_A P_A^+$ or $x_B P_B^-$ of the $n$ or $\nbar$ collinear sectors respectively. This hard production scale appears in the hard function $H$, including its anomalous dimension, and so to cancel it, it must reappear in the low scale EFT matrix elements. However, no propagator in the low scale matrix elements has virtuality at this hard scale by construction, so that in the beam sectors, the scale $Q$ can only appear associated with the large light-cone momentum component. Yet it is precisely integrals over these light-cone components that give rise to the rapidity divergence, so that the beam functions must depend ``anomalously'' on the ratio $\frac{\nu}{x_A P_A^+}$ or $\frac{\nu}{x_B P_B^-}$. The soft function's $\nu$-dependence is then constrained by the fact the anomalous dimensions sum to zero. Importantly, Lorentz invariance dictates that in the anomalous dimension, the logarithm of $x_A P_A^+$ must combine with the logarithm of $x_B P_B^-$ to form the scale $Q^2$, so that at most one logarithm of the $\nu$ scale can appear in the logarithm of the renormalized functions, see Refs. \cite{Manohar:2003vb,Chiu:2007dg}. Thus the rapidity scale $\nu$ does not appear in the anomalous dimension of Eq. \eqref{eq:RRG_Generic}, and the ultra-violet anomalous dimensions have the form:
\begin{align}
\gamma_{B}\Big(\frac{\nu}{x P^\pm};\alpha_s(\mu)\Big)&= \Gcusp(\alpha_s(\mu))\ln\Big(\frac{\nu}{x P^\pm}\Big)+\left[\gamma_{s}\Big(\alpha_s(\mu)\Big) - \gamma_{h}\Big(\alpha_s(\mu)\Big) \right]\\
\gamma_{S}\Big(\frac{\mu}{\nu};\alpha_s(\mu)\Big)&= \Gcusp(\alpha_s(\mu))\ln\Big(\frac{\mu^2}{\nu^2}\Big)-\gamma_{s}\Big(\alpha_s(\mu)\Big)\\
\gamma_{H}\Big(\frac{\mu}{Q};\alpha_s(\mu)\Big)&= -\Gcusp(\alpha_s(\mu))\ln\Big(\frac{\mu^2}{Q^2}\Big)+\gamma_{h}\Big(\alpha_s(\mu)\Big)
\end{align}
By consistency, the $\mu$ dependence of the rapidity anomalous dimension\footnote{It is Eq. \eqref{eq:RRG_Generic} which justifies the language of ``the rapidity anomalous dimension''. However, it is important to remember the rapidity anomalous dimension is in general process dependent.} is controlled by the cusp anomalous dimension for wilson lines:
\begin{align}\label{eq:RRG_mu}
\mu^2 \frac{d}{d \mu^2}\gamma_{R}\Big(\frac{\mu |\vec{b}_{\perp}|}{b_0};\alpha_s(\mu)\Big)&=\nu^2 \frac{d}{d \nu^2}\gamma_{S}\Big(\frac{\mu}{\nu};\alpha_s(\mu)\Big)=-\Gcusp\Big(\alpha_s(\mu)\Big)\,.
\end{align}
This leads to an all-orders form for the rapidity anomalous dimension:
\begin{align}
\gamma_{R}\Big(\frac{\mu |\vec{b}_{\perp}|}{b_0};\alpha_s(\mu)\Big)&=\int^{b_0^2/\vec{b}_\perp^2}_{\mu^2}\frac{d\mu'^2}{\mu'^2}\Gcusp\Big(\alpha_s(\mu')\Big)+\gamma_{r}\Big(\alpha_s\left(b_0 /|\vec{b}_{\perp}|\right)\Big)\,.
\end{align}

Lastly, we comment on the scheme dependence of how the rapidity divergences are isolated and removed from the bare functions. Schematically, the cross-section in coordinate space has the form:
\begin{align}
\frac{d\sigma}{dydQ^2d^2\vec{b}_{\perp}}&=\sigma_0H\Big(Q,\mu\Big)B_{n,\perp}\Big(x_A,\vec{b}_{\perp};\mu,\nu\Big)B_{\nbar,\perp}\Big(x_B,\vec{b}_{\perp};\mu,\nu\Big)S_\perp \Big(\vec{b}_{\perp};\mu,\nu\Big)+...\,.
\end{align}
Then we may consider the derivative:
\begin{align}\label{eq:Rap_Anom_Dim_from_Xsec_start}
Q^2\frac{d}{dQ^2}\ln\Big(\frac{1}{H(Q,\mu)}\frac{d\sigma}{dydQ^2d^2\vec{b}_{\perp}}\Big)&=\gamma_{R}\Big(\frac{\mu |\vec{b}_{\perp}| }{b_0}\Big)\,,
\end{align}
where we have used the fact that the total derivate with respect to $Q^2$ acts as partial derivative on the momentum components on the Drell-Yan pair:
\begin{align}
Q^2\frac{d}{dQ^2}&=\frac{1}{2} x_A P_A^+ \frac{\partial}{\partial(x_A P_A^+)}+\frac{1}{2} x_A P_B^-\frac{\partial}{\partial(x_B P_B^-)},
\end{align}
and $x_A P_A^+$ and $x_B P_B^-$ only appear in $B_{n,\perp}$ and $B_{\nbar,\perp}$ with $\nu$ in combinations of $\frac{\nu}{x_A P_A^+}$ and $\frac{\nu}{x_B P_B^-}$ according to the renormalization group equations. The left-hand side of Eq. \eqref{eq:Rap_Anom_Dim_from_Xsec_start} is independent of how the low-scale matrix elements are regulated. Indeed, the hard-function is the same in a wide variety of infra-red observables both with and without rapidity divergences, and thus can be calculated with or without intermediate rapidity regularization. Then the cancellation of rapidity divergences between the zero-bin subtracted collinear and soft functions allows us to conclude:
\begin{align}\label{eq:Rap_Anom_Dim_from_Xsec}
Q^2 \frac{d}{dQ^2}\ln\Big(\frac{1}{H(Q,\mu^2)}\frac{d\sigma}{dydQ^2d^2\vec{b}_{\perp}}\Big)&=\gamma_{R}\Big(\frac{\mu |\vec{b}_{\perp}| }{b_0};\alpha_s(\mu)\Big)\,.
\end{align}
Thus all scheme dependence of the rapidity anomalous dimension is directly controlled by the scheme dependence of the hard function, and the anomalous dimension is independent of the regularization procedure\footnote{To emphasize, this conclusion holds for any procedure for which the beam sectors are interchangeable under relabeling, and rapidity divergences cancel between low-scale sectors. This cancellation is dependent on the correct zero-bin subtraction being applied to the various sectors given the regulator. The constants of each sector cannot be constrained by this arguement, and thus constitute the scheme dependence of the renormalized function.}. There is a further scheme dependence involved with the decomposition into the cusp and non-cusp contributions to the rapidity anomalous dimension, however, this depedence is completely controlled as an initial condition to the solution of the differential equation \eqref{eq:RRG_mu}.

For all functions and parameters appeared here, we default to a fixed order expansion around $\alpha_S/4/\pi$ as:
\begin{align}
H\Big(Q;\mu\Big)&= \sum_{i=0}^{\infty}\Big(\frac{\alpha_s(\mu)}{4\pi}\Big)^{i} H_i\Big(Q;\mu\Big)\\
B_{n}\Big(b^+,b^-,\vec{b}_{\perp};\mu\Big)&= \sum_{i=0}^{\infty}\Big(\frac{\alpha_s(\mu)}{4\pi}\Big)^{i} B^{n}_i\Big(b^+,b^-,\vec{b}_{\perp};\mu\Big)\\
\ln S\Big(b^+,b^-,\vec{b}_{\perp};\mu \Big)&= \sum_{i=0}^{\infty}\Big(\frac{\alpha_s(\mu)}{4\pi}\Big)^{i} S_i\Big(b^+,b^-,\vec{b}_{\perp};\mu\Big)\\
B_{n,\perp}\Big(x,\vec{b}_{\perp};\mu,\nu\Big)&= \sum_{i=0}^{\infty}\Big(\frac{\alpha_s(\mu)}{4\pi}\Big)^{i} B^{n,\perp}_i\Big(x,\vec{b}_{\perp};\mu,\nu\Big)\\
\ln S_{\perp}\Big(\vec{b}_{\perp};\mu,\nu\Big)&= \sum_{i=0}^{\infty}\Big(\frac{\alpha_s(\mu)}{4\pi}\Big)^{i} S^{\perp}_i\Big(\vec{b}_{\perp};\mu,\nu\Big)\\
\Gcusp\Big(\alpha_s(\mu)\Big)&=\sum_{i=0}^{\infty}\Big(\frac{\alpha_s(\mu)}{4\pi}\Big)^{i+1}\gcusp_{i}\\
\gamma_{r,h,s}\Big(\alpha_s(\mu)\Big)&=\sum_{i=0}^{\infty}\Big(\frac{\alpha_s(\mu)}{4\pi}\Big)^{i+1}\gamma^{r,h,s}_i
\end{align}
Note that for soft functions, we have assumed non-Abelian exponentiation when performing the expansion \cite{Gatheral:1983cz,Frenkel:1984pz}. We gather in App. \ref{app:anom_dim} each of anomalous dimensions to the highest known perturbative order.

\section{The Exponential Regularization Procedure}
\label{sec:3}

\begin{figure}
  \centering
  \includegraphics[width=0.5\textwidth]{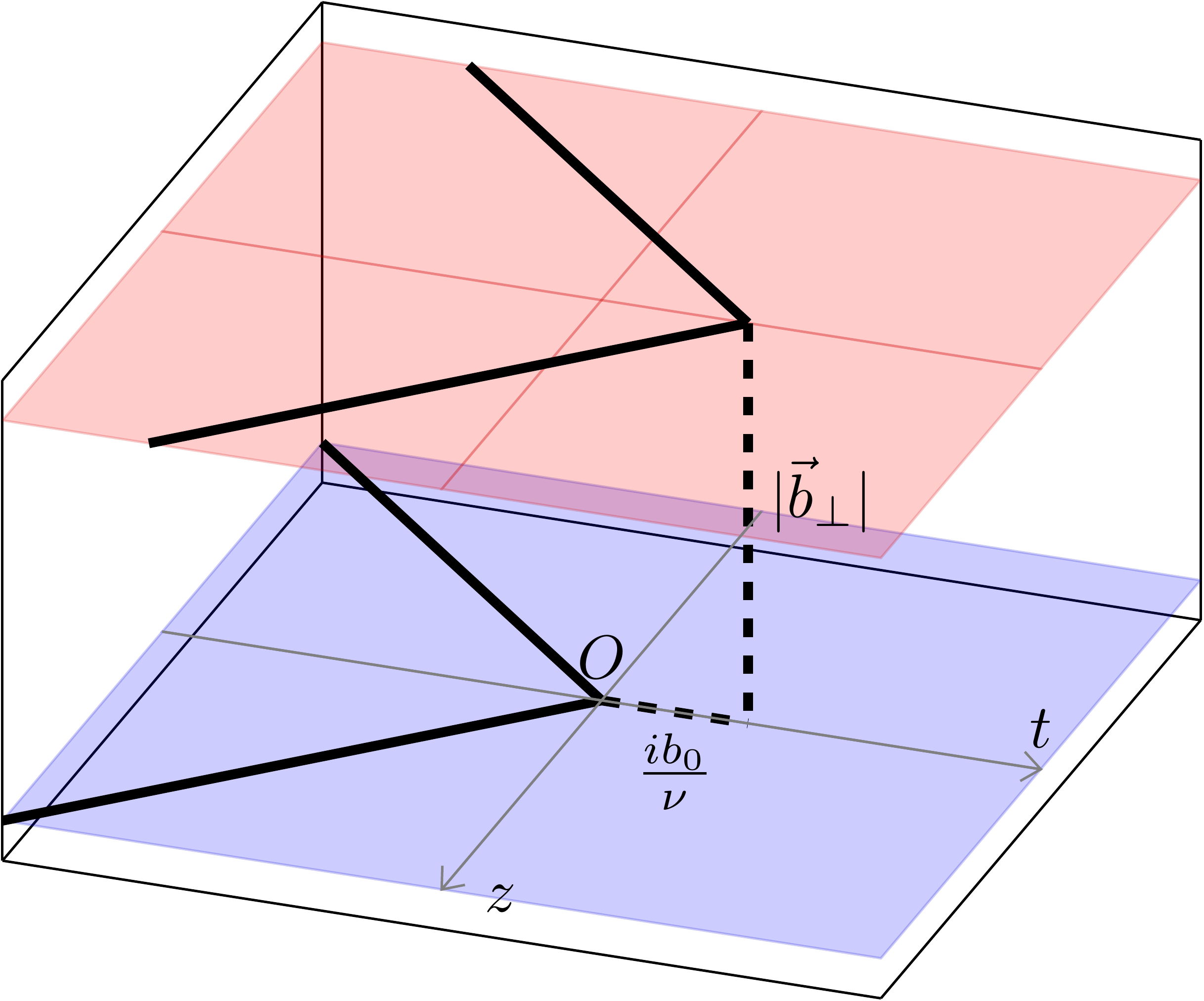}
  \caption{Coordinate space soft function.}
  \label{fig:1}
\end{figure}

We now explain how one can calculate the rapidity anomalous dimension of the soft function of Eq. \eqref{eq:qt_factorization_functions} through the exponential regulated soft function.
We first note that the origin of the divergences lay in the multipole expansion between the beam and soft sector's light-cone components in Eq. \eqref{eq:fact_theorem_t_spectrum_momentum_frac}. We are free to consider then not a strict expansion, but the limiting behavior of the functions as the light-cone components are localized. Specifically, we consider the soft function in coordinate space:
\begin{align}
  \label{eq:spectrum_reg_soft_function}
  S(\vecb, \tau) = S\left( ib_0\tau/2, ib_0\tau/2, \vecb\right).
\end{align} 
Since no information is lost by taking $b^+=b^-=ib_0\tau/2$ - as the fully-differential soft function is always a function of the product $b^+b^-$ by the RPI$_{III}$ transformations of the effective theory (see Ref. \cite{Manohar:2002fd}) - we use the same notation, {\it i.e.} $S$ with no subscript for the soft function here. A picture for the coordinate space soft function is depicted in
Fig.~\ref{fig:1} with $\tau=1/\nu$. We will show later that the $\nu=1/\tau$ is indeed the artificial scale appearing in the rapidity regularization once we take the limit of $\tau \to 0$.
By imposing the energy constraint on the momentum crossing the cut in the diagrammatic expansion, we regulate the integrals over the light-cone components of momenta. This can be seen more clearing in the momentum space, where the measurement of $\tau$ forms an exponential damping factor for the rapidity divergence. It is in this sense we call the function the exponential regulated soft function. This deformation of the transverse momentum dependent (TMD) soft function is particularly revealing, since it is directly relatable to the threshold soft function of Ref. \cite{Belitsky:1998tc}, which we define as:
\begin{align}\label{eq:spec_reg_soft_threshold_limit}
S_{\rm thr}(\tau)&= S\left( ib_0\tau/2, ib_0\tau/2, 0\right)=S(\vec{b}_{\perp}=0,\tau).
\end{align}
The limit $\vec{b}_{\perp}=0$ can be taken smoothly, both before and after renormalization, since no on-shell singularities are probed in this limit. One can perform the deformation to any SCET$_{II}$ soft functions, forming a general regularization procedure for these theories. The relation to the standard threshold soft function, and the fact that the limit is smooth implies several important features about the exponential regulated soft function. First, its UV anomalous dimension is the same as the threshold soft function:
\begin{align}
\label{eq:threshold_renorm_group}
\mu^2 \frac{d}{d \mu^2}\ln S_{\rm thr}(\tau;\mu)&=\mu^2 \frac{d}{d\mu^2}\ln S(\vec{b}_{\perp},\tau;\mu)=\gamma_{S}\Big(\tau \mu;\alpha_s(\mu)\Big)\,.
\end{align}
To qualify as a valid regularization scheme for the TMD soft function, it also has to satisfy the following condition in the $\tau\to 0$ limit (for a derivation of this result, we refer to Sec. \ref{app:FD_Functions}):
\begin{align}\label{eq:spec_reg_soft_function_rap_anom_dim}
\lim_{\tau\rightarrow 0}\tau^2 \frac{d}{d \tau^2}\ln S(\vec{b}_{\perp},\tau;\mu)&=-\gamma_{R}\Big(\frac{\mu |\vec{b}_{\perp}| }{b_0};\alpha_s(\mu)\Big)\,,
\end{align}
from which we can derive constraints on the function form of $S(\vec{b}_{\perp},\tau;\mu)$.
To make our statement explicit, let's first use Eq. \eqref{eq:threshold_renorm_group} to write the exponential regulated soft function as \footnote{We explicitly write the logarithm of the soft function, since this is most natural from the non-Abelian exponentiation theorem.}:
\begin{align}\label{eq:all_orders_form_spec_reg_soft_function}
\ln S(\vec{b}_{\perp},\tau;\mu)&=\int_{\frac{1}{\tau^2}}^{\mu^2} \frac{d\mu'^2}{\mu'^2} \gamma_{S}\Big(\tau \mu';\alpha_s(\mu')\Big) + \ln S\Big(\vec{b}_{\perp},\tau;\frac{1}{\tau}\Big)\,.
\end{align}
where the second term on the RHS is the $\mu$-independent part of the soft function, and has a well-behaved series expansion about $\vec{b}_{\perp}=0$. By demanding Eq. \eqref{eq:spec_reg_soft_function_rap_anom_dim} holds, we obtain the following equation:
\begin{align}
\lim_{\tau \rightarrow 0} \bigg\{ \int^{\frac{1}{\tau^2}}_{\mu^2}\frac{d\mu'^2}{\mu'^2} \Gcusp\Big(\alpha_s(\mu')\Big) & - \gamma_s\Big(\alpha_s(\frac{1}{\tau})\Big)+\tau^2\frac{d}{d\tau^2}\ln S\Big(\tau,\vec{b}_{\perp};\frac{1}{\tau}\Big) \bigg\} \nonumber\\
&=\int^{b_0^2/\vec{b}_{\perp}^{2}}_{\mu^2}\frac{d\mu'^2}{\mu'^2}\Gcusp\Big(\alpha_s(\mu')\Big)+\gamma_r\Big(\alpha_s(b_0/|\vec{b}_{\perp}|)\Big)\,.
\end{align}
This is a non-trivial constraint, since at each order in pertubation theory, the double logarithmic contribution to the $\tau \rightarrow 0$ behavior of the $\mu$-independent part must be fixed by the cusp anomalous dimension, and higher order logarithms are determined from the beta-function, both of which form important checks on any calculation of the function. The same regularization can be easily adapted to regulate the rapidity divergence in the TMD-PDFs through the use of fully differential beam function, which we will defer to future work.

This regularization of the TMD soft function has several features to commend it. Firstly, since it is defined via a measurement constraint on the final state radiation, it is manifestly gauge invariant. Non-abelian exponentiation also follows trivially, which we have used in writing down Eq. \eqref{eq:all_orders_form_spec_reg_soft_function}, since the measurement factorizes in its Laplace form to act on each final state parton. Lastly, as seen from Eq. \eqref{eq:spec_reg_soft_threshold_limit}, we can actually realize the exponential regulated soft function from its Taylor series expansion about the threshold limit, where all integrals will be reducible to known master integrals. As explained in Sec. \ref{sec:bootstrap}, this means by matching the Taylor series to an ansatz of special functions, we can deduce the full transverse-space dependence of the function from a finite number of terms. Being able to deduce the full transverse-space dependence is critical to being able to construct the rapidity anomalous dimension. In the all-orders form of the exponential regulated soft function in Eq. \eqref{eq:all_orders_form_spec_reg_soft_function}, transverse-space dependence is entirely controlled by its $\mu$-independent part, which depends on its arguments solely through the scaleless ratio of $x=- \vec{b}_\perp^2/b_0^2/\tau^2$ (neglecting the scale dependence in $\alpha_S$). It is the Taylor series about $x=0$ that is probed by the threshold limit, but it is the $x\rightarrow \infty$ that controls the rapidity anomalous dimension in Eq. \eqref{eq:spec_reg_soft_function_rap_anom_dim}. Technically, an infinite number of terms would be necessary, \emph{assuming} an infinite radius of convergence. However, the space of functions appearing in perturbative calculations is tightly constrained, allowing the full dependence to be deduced from only a finite number of terms even when the taylor series has a finite radius of convergence. It is fascinating that there is a mother function relating both threshold resummation to the transverse momentum resummation: both can be obtained by taking appropriate limits of a single function.

\begin{figure}[ht]
  \centering
  \subfigure[]{
\includegraphics[width=0.2\textwidth]{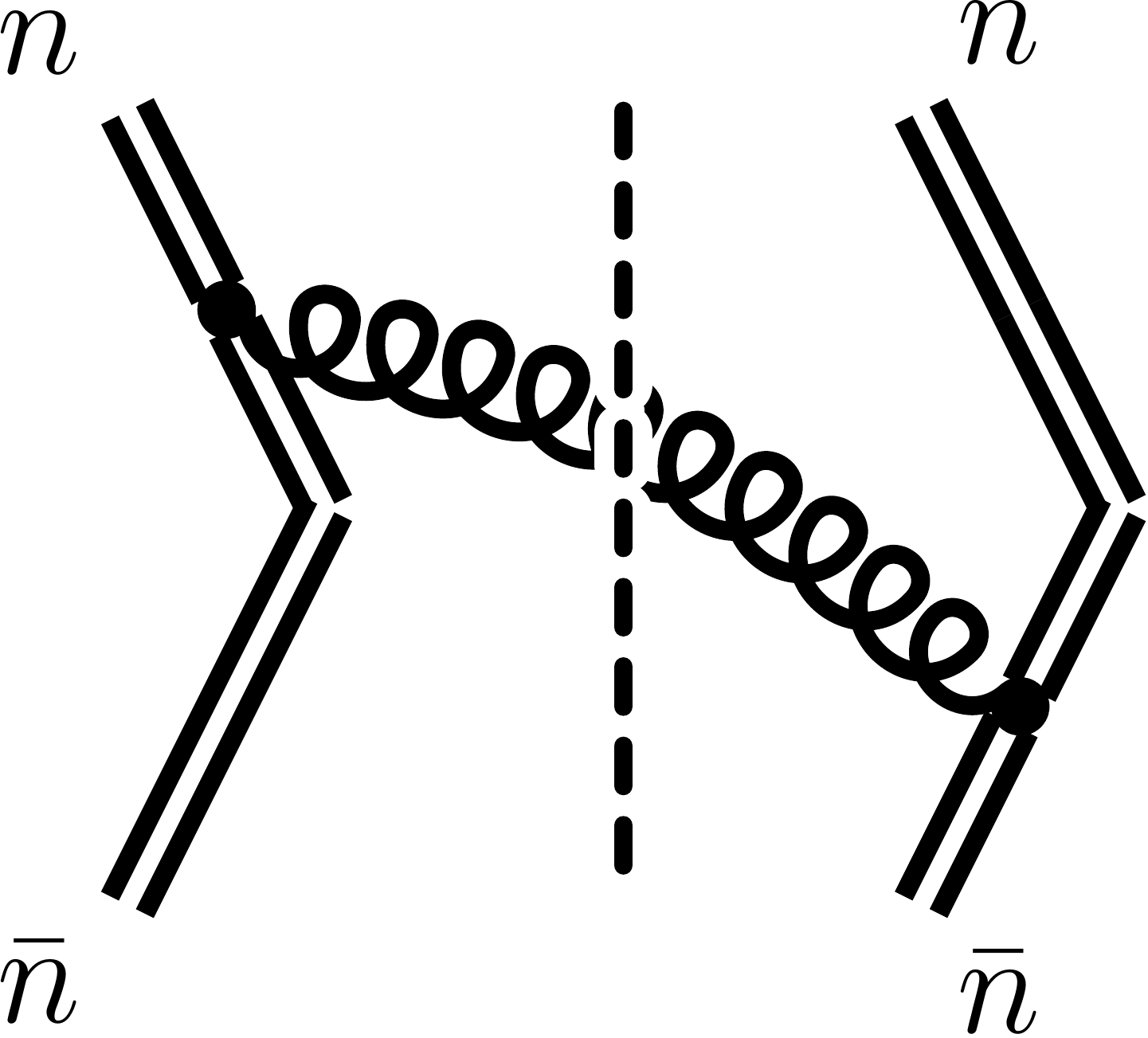}
\label{fig:sub1}}
\qquad
  \subfigure[]{
\includegraphics[width=0.2\textwidth]{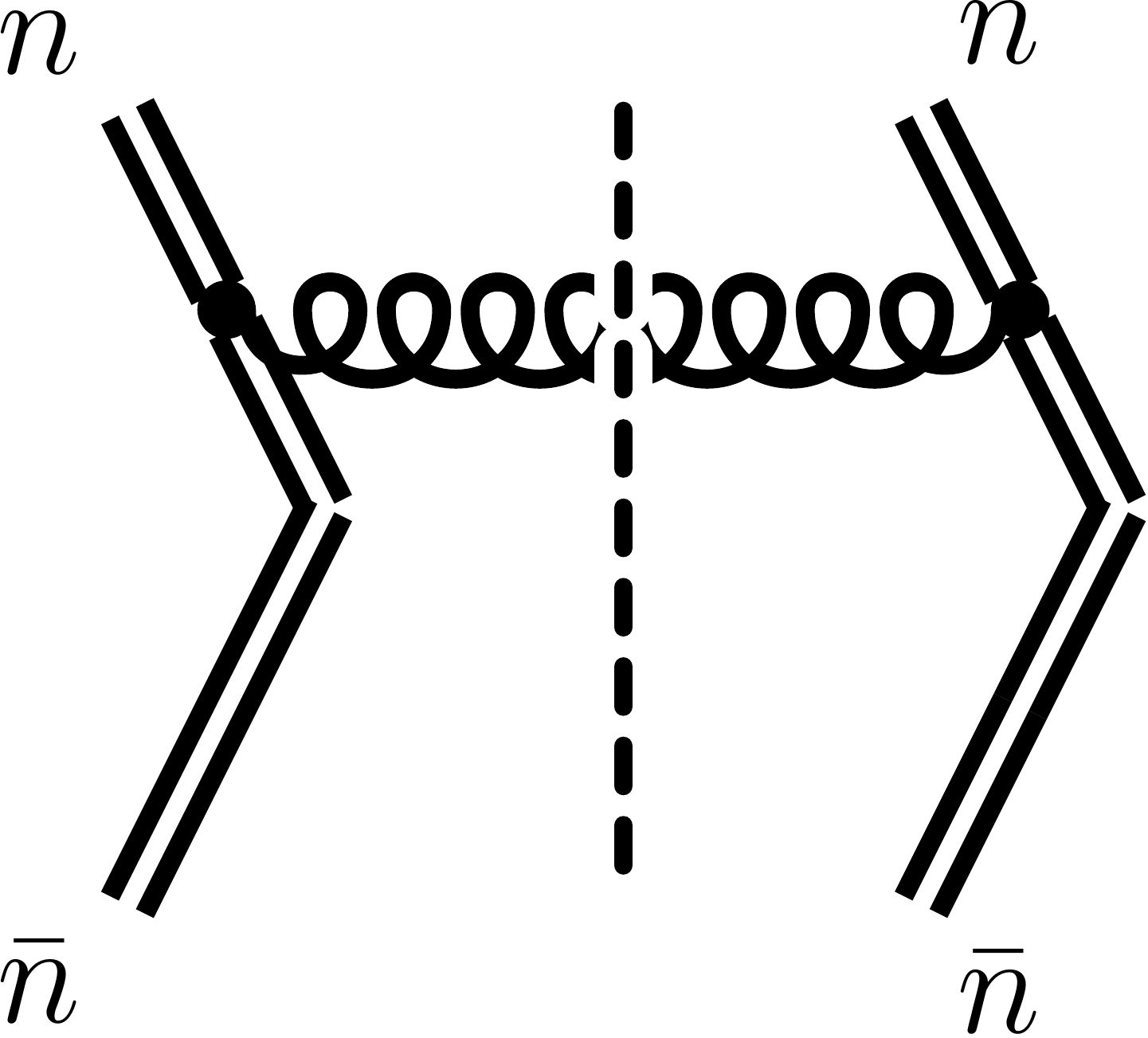}
\label{fig:sub2}}
  \caption{Cut diagrams for one-loop soft function. Double lines are
    Wilson lines.}
  \label{fig:2}
\end{figure}

To illustrate how the regulator actually works, we take the one-loop
calculation of the soft function as an example. The relevant diagrams
are depicted in Fig.~\ref{fig:2}. For light-like Wilson lines,
Fig.~\ref{fig:sub2} vanishes and we only need to consider
Fig.~\ref{fig:sub1} and its conjugate.

The bare exponential regulated soft function is given by the integral
\begin{align}
  \label{eq:2}
  \widetilde{S}_1 (\vecb, \tau) =&\, 2 (4\pi)^2 C_F \left(\frac{\mu^2
  e^{\gae}}{4\pi} \right)^{2-d/2} \int \frac{d^d k}{(2\pi)^{d-1}}
  \theta(k^0) 
  \delta(k^2)
\brk
\cdot \exp\left( - (k^++k^-)\tau e^{-\gae} - \im \vecb
  \mcdot \vec{k}_\perp \right)  \frac{ n \mcdot \bar{n}}{k^+ k^-}
\end{align}
where $d = 4 - 2 \e$. We work in the
$\overline{\textrm{MS}}$ scheme by a redefinition of the bare scale $\mu^2_0
= \mu^2 e^\gae/(4\pi) $.  Note that $\vecb$ is in two dimension, while $\vec{k}_\perp$ is
in $2 - 2 \e$ dimension. Due to rotation invariance  in the $\perp$
plane for Drell-Yan production, we let $\vecb = |\vec{b}_\perp| ( 0, 1)$.  Without loss of generality, we can
parameterize $\vec{k}_\perp$ as
\begin{align}
  \label{eq:4}
  \vec{k}_\perp = |\vec{k}_\perp| ( \sin\theta, \cos\theta, \vec{0}_{-2 \e})
\end{align}
It is also convenient to use light-cone coordinate for the integral measure,
\begin{align}
  \label{eq:3}
  \int \frac{d^{4-2\e} k}{(2\pi)^{d-1}} =  \frac{1}{2 (2\pi)^{3-2\e}}
 \Omega_{-2 \e} \int dk^+dk^- k^{1-2\e}_\perp dk_\perp \sin^{-2 \e}\theta d\theta 
\end{align}
where
\begin{align}
  \label{eq:5}
  \Omega_n  = \frac{2 \pi^{(n+1)/2}}{\Gamma((n+1)/2)}
\end{align}
is the area of unit sphere in $n$ dimension. Integrating out $\theta$, making the
following change of variables
\begin{align}
  \label{eq:7}
  r = k^+k^-, \quad v = \frac{k^+}{k^-}
\end{align}
with the Jacobian $1/(2v)$, and using the on-shell delta function, we arrive at
\begin{align}
  \label{eq:6}
   \widetilde{S}_1 (\vecb, \tau) =4C_F \left(\frac{b}{b_0} \mu^2\right)^\e
  \int^\infty_0 \frac{ dk_\perp}{k_\perp^{1 + \e}} J_{-\e}(bk_\perp)
  \int^\infty_0  \frac{dv}{v} \exp\left[ -\left(
  \frac{1}{\sqrt{v}} + \sqrt{v} \right)k_\perp \tau e^{-\gae} \right]
\end{align}
where $J_n(z)$ is the Bessel function of the first kind. The variable $v$ is related to the rapidity of soft gluon by $v
= \exp(-2 Y)$. It is clear from \Eqn{eq:6} that without the threshold
regulator factor, the $v$ integral diverges at the end points of
infinite rapidity. This is the so-called light-cone/rapidity
singularity. The exponential regulator
provides an exponential damping factor at infinite rapidity. The
resulting $v$ and $k_\perp$ integrals can be done in closed form,
giving
\begin{align}
  \label{eq:9}
  \widetilde{S}_1(\vecb,\tau) = C_F \frac{4}{\e^2} \mu^{2\e} e^{-\e
  \gae} \Gamma(1-\e)   \tau^{2\e} \,_2F_1\left(-\e,-\e;1-\e; - \frac{\vecbsq}{b^2_0
  \tau^2} \right)
\end{align}
It is straightforward to expand the above expression using, e.g.,
\texttt{HypExp}~\cite{Huber:2005yg} to arrive at
\begin{align}
  \label{eq:10}
  \widetilde{S}_1(\vecb,\tau) =C_F \left[ \frac{4}{\e^2} + \frac{4}{\e}
  \ln(\mu^2 \tau^2) + 2 \ln^2(\mu^2 \tau^2) + 4 \Li_2 \left(-
  \frac{\vecbsq}{b^2_0\tau^2} \right) + 2\zeta_2\right]
\end{align}
The renormalized fully differential soft function at one-loop is then
obtained by removing the $\e$ poles,
\begin{align}
  \label{eq:11}
  S_1(\vecb,\tau;\mu) = C_F\left[ 2 \ln^2(\mu^2 \tau^2) + 4 \Li_2 \left(-
  \frac{\vecbsq}{b^2_0\tau^2} \right) + 2\zeta_2\right]
\end{align}
The exponential regulated $\qt$ soft function is obtained by taking
the $\tau\to 0$ limit and keeping only the non-vanishing terms,
\begin{align}
  \label{eq:12}
  S^\perp_1(\vecb,\tau;\mu) = C_F \left[ 2 \ln^2 \left(
  \frac{\vecbsq \mu^2}{b^2_0} \right) - 4  \ln \left(
  \frac{\vecbsq \mu^2}{b^2_0} \right) \ln \left(
  \frac{\vecbsq }{b^2_0\tau^2} \right) - 2 \zeta_2 \right]
\end{align}
Once we identify $\nuw = \tau^{-1}$, we can make smooth connection with the rapidity RG
formalism, and check that \Eqn{eq:12} satisfy the $\mu$ and $\nu$ RG equation. 

\section{Transverse Momentum and Threshold Factorization}\label{app:FD_Functions}

\begin{figure}
\includegraphics[scale=.5]{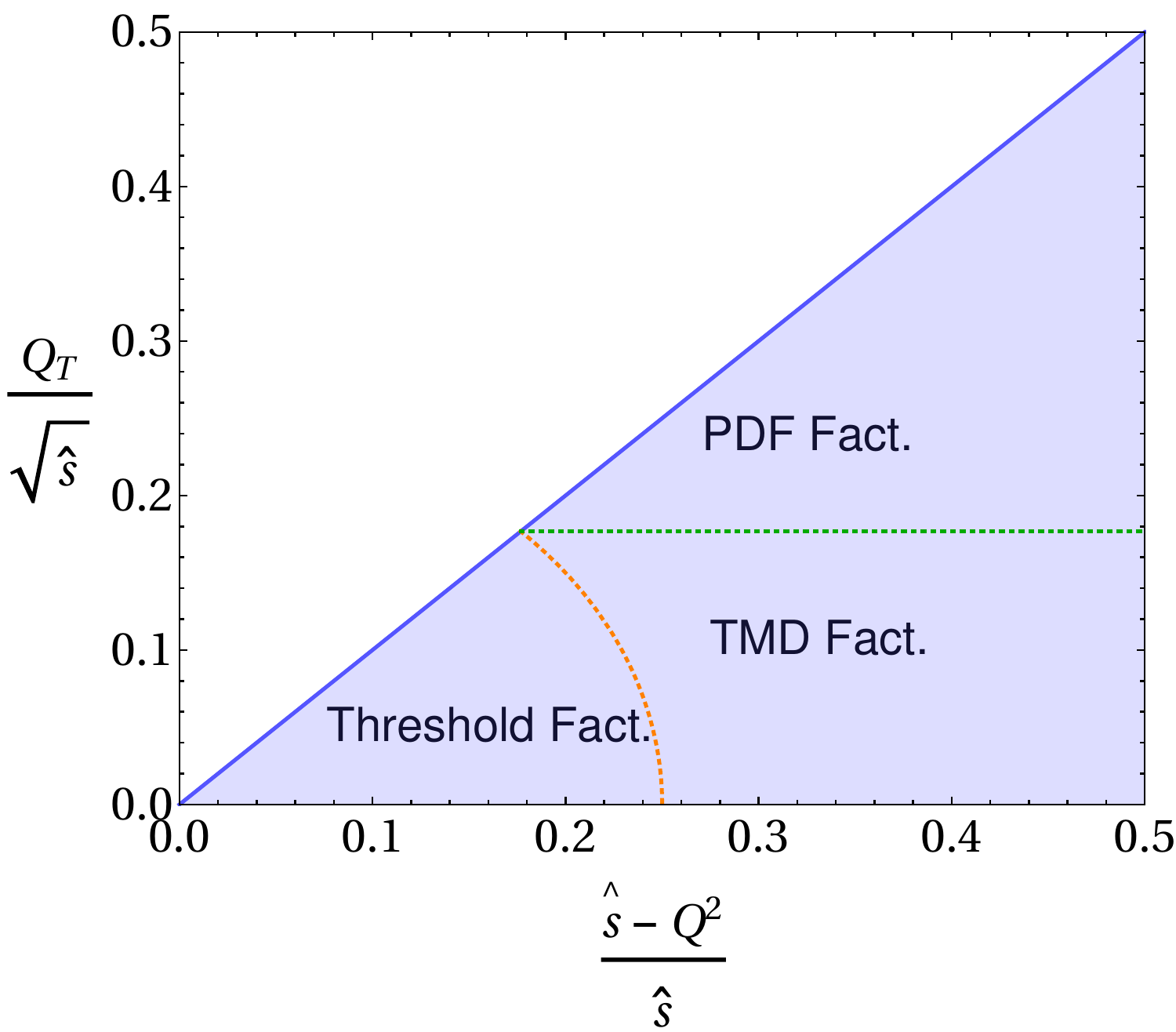}
\caption{\label{fig:threshold_pt_phase_space}The physical phase space for a Drell-Yan pair at zero rapidity. The shaded region is the allowed phase space, and different factorization regimes are indicated. The dotted lines indicate where approximate transition regions between the factorizations are found.}
\end{figure}

\begin{figure}
\includegraphics[scale=.5]{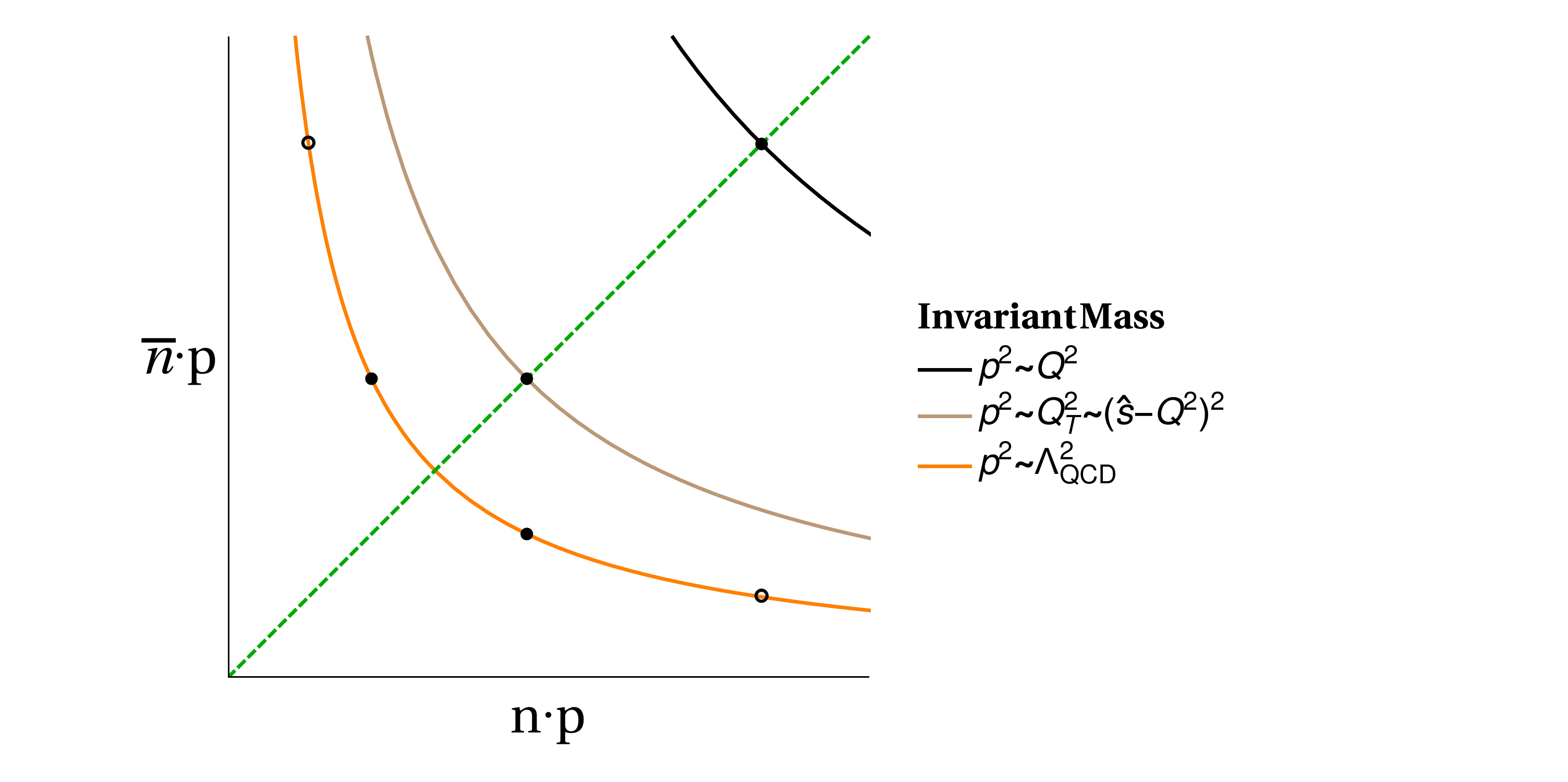}
\caption{\label{fig:threshold_modes}The invariant mass for the modes contributing to the threshold production of the Drell-Yan pair. The circles at large light-cone momenta are virtual fluctuations that feed directly to the hard interaction, but have no corresponding real emissions at that scale. Solid dots correspond to modes that can emit into the final state under the threshold constraint.}
\end{figure}

In contrast to the factorization of Eq. \eqref{eq:fact_theorem_t_spectrum}, a distinct formula was proposed in Ref. \cite{Mantry:2009qz}, which did not perform the multipole expansion:
\begin{align}\label{eq:fact_frank_sony}
\frac{d\sigma}{dy dQ^2d^2\vec{Q}_T}&=\sigma_0 \int\frac{d^4q}{(2\pi)^3}\delta^+(q^+ q^- - Q^2)\delta\Big(y-\frac{1}{2}\ln\frac{q^-}{q^+}\Big) \delta^{(2)}(\vec{Q}_T-\vec{q}_{\perp})\nonumber\\
&\qquad\qquad\int d^4b e^{ib\cdot q}H(Q) B_{n}(b^+,b^-,\vec{b}_{\perp})B_{\nbar}(b^+,b^-,\vec{b}_{\perp})S (b^+,b^-,\vec{b}_{\perp})+...\,.
\end{align}
This factorization utilized fully differential beam and soft functions, that are sensitive to the total momentum flow crossing the cut in the diagrammatic expansion of the functions. Since the multipole expansion was not performed, large logarithms may still remain in its perturbative expansion, even after renormalization group evolution. One can find consistent factorization theorems that utilize these fully differential functions in a multi-differential measurement of beam thrust and Drell-Yan transverse momentum, see Refs. \cite{Jain:2011iu, Larkoski:2014tva, Procura:2014cba}\footnote{For a further discussion of fully-differential beam functions, see also Refs. \cite{Collins:2007ph,Rogers:2008jk,Gaunt:2014xxa}.}.

The relative values of $b^+,b^-$ are unimportant, since they always appear in the product of $b^+b^-$ as explained earlier. Thus to examine the limit to the TMD soft function, we set $b^+=b^-= t$ and write:
\begin{align}
S(t,t,\vec{b}_{\perp})=S(\vec{b}_{\perp},t)\,.
\end{align}
Now the exponential regulated soft function is connected to the fully differential by the analytic continuation from $\tau\rightarrow -2 i t/b_0$. One can equally well use this Fourier transformed function as a definition of the exponential regulated soft function, instead of the Laplace. However, given that much work on soft threshold integrals has been done in Laplace space, as well as to avoid a proliferation of imaginary numbers, we found it convenient to adopt the Laplace space definition, {\it i.e.} using $\tau$ as the new argument.

To understand the origin of our central result, Eq. \eqref{eq:spec_reg_soft_function_rap_anom_dim}, we simply approach the factorization for the differential spectrum of the Drell-Yan pair from two different limits. For a Drell-Yan pair, the allowed phase space at zero rapidity in terms of the transverse momentum and the residual partonic energy scale is plotted in Fig. \ref{fig:threshold_pt_phase_space}. First we consider the factorization starting with the standard inclusive differential Drell-Yan cross-section, at large to moderate $\qt$, moving along the upper line in Fig. \ref{fig:threshold_pt_phase_space}. To avoid convolutions, it is simplest to work in position space, and the standard inclusive Drell-Yan cross-section admits a factorization into collinear PDFs as:
\begin{align}\label{eq:fact_inclusive_DY}
d\sigma &\sim \sum_{i,j} \hat{\sigma}_{ij}(Q;b^-,b^+,\vec{b}_{\perp})f_{n,i}(b^-)f_{\nbar,j}(b^+) + ...\,,
\end{align}
where $\hat{\sigma}$ is the inclusive hard coefficient, {\it i.e.} partonic cross section, and the standard collinear PDF is related to the fully differential beam functions by taking the transverse and small lightcone component to zero. This cross-section admits a further factorization in the threshold region, where the hard inclusive coefficient splits into the form factor derived hard function and a soft factor that is fully differential as in Ref. \cite{Mantry:2009qz}, and the PDFs are taken to their threshold expressions, that is, taking the Bjorken scale $x \to 1$ in the momentum space:
\begin{align}\label{eq:fact_inclusive_DY_threshold}
\hat{\sigma}_{ij}(Q;\nbar\cdot b,n\cdot b,\vec{b}_{\perp})&\to \delta_{iq}\delta_{j\bar{q}} H(Q)S(b^+,b^-,\vec{b}_{\perp})+...\\
f_{n}(b^-)=B_{n}(0,b^- b,0)&\to B_{n,{\rm thr}}(0,b^-,0)+...\\
f_{\nbar}(b^+)=B_{\nbar}(b^+,0,0)&\to B_{\nbar,{\rm thr}}(b^+,0,0)+...
\end{align}
Substituting these functions into Eq. \eqref{eq:fact_inclusive_DY}, we achieve the threshold factorization for the differential spectrum of Drell-Yan:
\begin{align}\label{eq:fact_inclusive_DY_threshold_spectrum}
d\sigma &\sim H(Q)S(b^+,b^-,\vec{b}_{\perp}) B_{n,{\rm thr}}(0,b^-,0) B_{\nbar,{\rm thr}}(b^+,0,0)
\end{align}
Thus the fully differential or exponential regulated soft function does appear in a factorization theorem with homogeneous power counting, when the modes are organized as in Fig.~\ref{fig:threshold_modes}.

\begin{figure}
\includegraphics[scale=.5]{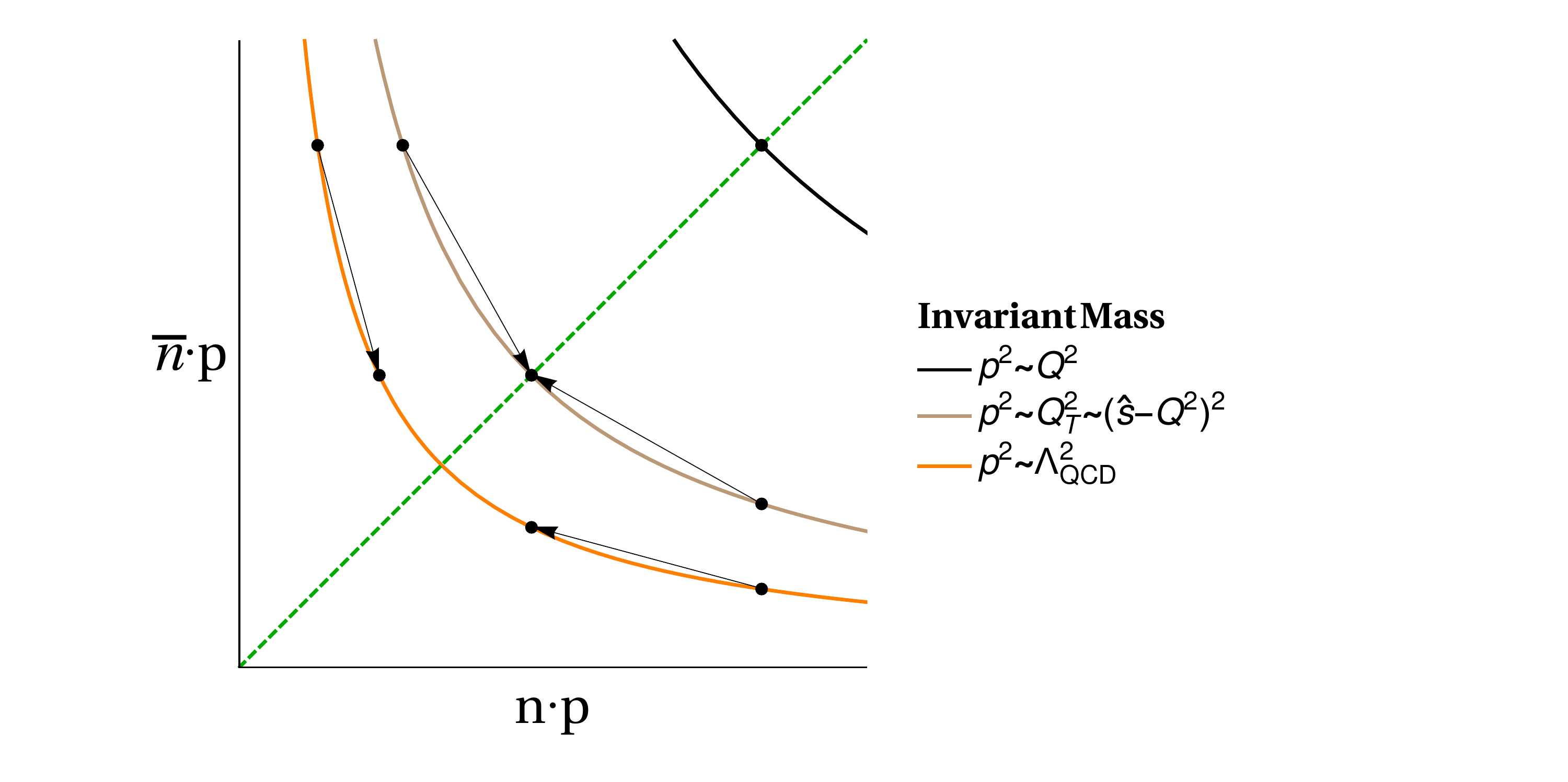}
\caption{\label{fig:tmd_to_threshold_modes} Refactorizing the TMD functions in the power counting of the threshold region. Both the PDF's and coefficients for TMD matching to PDF's get expanded.}
\end{figure}

Alternatively, we may approach the threshold regime already assuming small transverse momentum. Let's rewrite Eq. \eqref{eq:fact_theorem_t_spectrum} as,
\begin{align}\label{eq:fact_inclusive_DY_qt_spectrum}
d\sigma &\sim H(Q) B_{n}(0,b^-,\vec{b}_{\perp}) B_{\nbar}(b^+,0,\vec{b}_{\perp}) S(0,0,\vec{b}_{\perp})\,.
\end{align}
The TMD-beam functions can then be further factorized into an additional soft factor and the threshold PDF, reminicent of Ref. \cite{Korchemsky:1992xv}:
\begin{align}\label{eq:TMD_at_Threshold}
B_{n}(0,b^-,\vec{b}_{\perp})&\to B_{n,{\rm thr}}(0,b^-,0) S(0,b^-,\vec{b}_{\perp})+...\nonumber\\
B_{\nbar}(b^+,0,\vec{b}_{\perp})&\to B_{\nbar,{\rm thr}}(b^+,0,0) S(b^+,0,\vec{b}_{\perp})+...\,.
\end{align}
both sides of these equations have the same rapidity divergences, which on the right hand side are carried by the soft factor alone. This is the same soft factor appearing in the SCET$_+$ factorization of the multi-differential beam thrust and transverse momentum phase space, see Ref. \cite{Procura:2014cba}. By substituting Eq.~\eqref{eq:TMD_at_Threshold} in to Eq.~\eqref{eq:fact_inclusive_DY_qt_spectrum}, we again achieve another threshold factorization for the Drell-Yan process, where now all functions have been refactorized in the threshold power counting as in Fig. \ref{fig:tmd_to_threshold_modes}.

Demanding consistency between these two factorizations in their overlapping domain of validity, we conclude:
\begin{align}\label{eq:fully_differential_factorization}
S(b^+,b^-,\vec{b}_{\perp})&=S(\nbar\cdot b,0,\vec{b}_{\perp})S(0,n\cdot b,\vec{b}_{\perp})S(0,0,\vec{b}_{\perp})+\mathcal{O}\Big(\frac{b^+ b^-}{|\vec{b}_{\perp}|^2}\Big)
\end{align}
This equality holds at the level of \emph{renormalized} functions. The left-hand side is free from rapidity divergences, but in the limit $b^+, b^- \rightarrow 0$ (the small $\tau$ limit) has a large logarithm at each order in perturbation theory (the limit to zero light-cone position is not smooth). This corresponds to the fact that each factor on the right is naively rapidity divergent. With appropriate regularization and subtractions, these divergences will cancel, making way for the RRG. Following the arguments of the logarithms of the intermediate rapidity renormalization, we are then lead to Eq. \eqref{eq:spec_reg_soft_function_rap_anom_dim} similarly to how we concluded Eq. \eqref{eq:Rap_Anom_Dim_from_Xsec}. That is, since the rapidity divergences cancel between the three soft functions, we can interprete the fully differential soft function as a direct calculation of the rapidity renormalized soft function. Note that the expansion is very important. When we factorize the threshold region in the inclusive hard coefficient, we perform no expansion between the energy of the hadronic final state and its transverse momentum. In contrast, when further factorizing the small transverse momentum factorization in the threshold limit, an expansion between the energy of the final state and its transverse momentum has \emph{already} been performed to arrive at Eq. \eqref{eq:fact_theorem_t_spectrum}. The expansion in \eqref{eq:fully_differential_factorization} is the common region of validity between these two approaches to the threshold region\footnote{Though the appropriate threshold factorization is of the differential spectrum \eqref{eq:fact_inclusive_DY}.}.

The smoothness of the limit $|\vec{b}_{\perp}|\rightarrow 0$ is also seen from the threshold factorization of Eq. \eqref{eq:fact_inclusive_DY} using Eq. \eqref{eq:fact_inclusive_DY_threshold}. If we fourier transform Eq. \eqref{eq:fact_inclusive_DY} with respect to $\vec{Q}_T$, and take the limit $|\vec{b}_{\perp}|\rightarrow 0$, we recover the traditional factorization of the threshold Drell-Yan spectrum, see for instance Ref. \cite{Becher:2007ty}. Since this factorization has no singularities associated with its localization at zero impact parameter, we conclude the limit $|\vec{b}_{\perp}|\rightarrow 0$ is smooth to all orders, which is born out by explicit calculations up to and including three loops. Again, this is not surprising since the resummation structure driven by the renormalization group for the threshold factorization is resumming large logarithms associated with the light-cone variables $n\cdot b$ and $\nbar\cdot b$, not the transverse momentum.

Similar functions appear in joint resummation (see Refs. \cite{Laenen:2000ij,Kulesza:2002rh,Kulesza:2003wn}) that seeks to combine threshold and transverse momentum resummation. In particular, a similar refactorization to Eqs. \eqref{eq:TMD_at_Threshold} and \eqref{eq:fully_differential_factorization} was considered. There the authors sought to combine into a single formula the resummation for both the threshold logarithms and the transverse momentum spectrum. Our aim has been distinct, which was to provide a new method for calculating all quantities needed for resummation from a single fully differential function. However, the family of factorizations we have derived would also allow us to examine the structure for genuine joint resummation. We find that there are \emph{three} distinct factorization theorems, each of which is seperately consistent under ultraviolet and rapidity renormalization, Eqs. \eqref{eq:fact_theorem_t_spectrum}, \eqref{eq:fact_inclusive_DY_threshold_spectrum} and Eq. \eqref{eq:fact_inclusive_DY_threshold_spectrum} with the substitution of Eq. \eqref{eq:fully_differential_factorization}. One can consider a merging scheme as derived by \cite{Procura:2014cba} that would also attempt to combine both threshold and transverse momentum resummation, such that the scheme is accurate to $N^3LL$ in all limits. One could also include small-$x$ resummation following \cite{Forte:2015gve,Marzani:2015oyb}.

\section{Bootstrapping the fully differential soft function}\label{sec:bootstrap}

At first sight, the one-loop calculation using exponential regulator
in \sect{sec:3} doesn't seem to simplify the calculation. Even worse,
exponential regulator introduce an extra non-trivial scale
$\tau=1/\nu$ into the problem, which leads to the appearance of non-trivial analytic
function in the one-loop calculation. However, such seeming weaknesses will be shown to be strengths, once we examine the two-loop
calculation for fully differential soft function already performed
in Ref.~\cite{Li:2011zp}, where the results are given in terms of
polylogarithms up to weight four with rational coefficients. In this
section, we shall show that the simple structure of the results in
Ref.~\cite{Li:2011zp} allows us to calculate the fully differential
soft function without actually calculating the corresponding Feynman
integrals. 

As defined in Sec. \ref{sec:review}, we can expand the renormalized fully differential soft function in the following exponential, thanks to the on-Abelian exponentiation theorem:
\begin{align}
  \label{eq:14}
  S(\vecb, \tau; \mu) = \exp\Biggl[ \frac{\alpha_s(\mu)}{4\pi} S_1(\vecb,\tau; \mu) + 
\left(\frac{\alpha_s(\mu)}{4\pi}\right)^2 S_2(\vecb,\tau; \mu) + \left(\frac{\alpha_s(\mu)}{4\pi}\right)^3 S_3(\vecb,\tau; \mu) + \Ord(\alpha_s^4) \Biggr]
\end{align}
The results in Ref.~\cite{Li:2011zp} then can be rewritten in terms of Harmonic
Polylogarithms~(HPLs) of Remiddi and Vermaseren~\cite{Remiddi:1999ew},
taking into account the exponentiation in \Eqn{eq:14}:
\begin{align}
  S_1(\vecb, \tau; \mu = \tau^{-1}) = & c^s_1 + 4 C_a H_{2}
\nbrk
 S_2(\vecb, \tau; \mu = \tau^{-1}) = & c^s_2 + C_A C_a \Big(-8 \zeta_2 H_{2}+\frac{268}{9}
  H_{2}+\frac{44}{3} H_{3}
-8 H_{4}
-\frac{44}{3} H_{2,1}-8 H_{2,2}
\brk
-16 H_{3,1}-16 H_{2,1,1}\Big)
+C_a n_f \Big(-\frac{40}{9} H_{2}-\frac{8}{3} H_{3}+\frac{8}{3}
H_{2,1}\Big)
\label{eq:qcdc}
\end{align}
where we have only kept the scale independent part by setting $\mu =
\tau^{-1}$. $c^s_i$ are scale independent constant in threshold
resummation, whose explicit formula are collected in the appendix. We
use $C_a$ to denote the Casimir of the initial parton. $C_a = C_F$ for
Drell-Yan production, and $C_a = C_A$ for Higgs
production. $H_{\vec{w}} \equiv H_{\vec{w}}(x)$ are HPLs with weight
vector $\vec{w}$, while $x = - \vec{b}_\perp^2/(b^2_0 \tau^2)$. We have used
the shorthand notation for the weight vector of
HPLs~\cite{Remiddi:1999ew}~\footnote{In this notation, weight vector
  with $n$ trailing zeros to the left of a $1$ is written as $n+1$. For example,
$H_{0,0,0,1,0,1} \equiv H_{4,2}$.}. The exceedingly simplicity of
\Eqn{eq:qcdc} makes one wonder whether there is simpler way to obtain
them, instead of the brute-force calculation done in
Ref.~\cite{Li:2011zp}. Indeed, we found that the results in
\Eqn{eq:qcdc} can be obtained using bootstrap method, which we shall
explain below. 

The bootstrap program is extremely successful in calculating
scattering amplitudes in planar $\Neqfour$ SYM, in particular for
six-point amplitudes. Briefly speaking, for 
a $L$-loop planar amplitudes with the Bern-Dixon-Smirnov~\cite{Bern:2005iz}
factored out, one can make an ansatz consists of rational linear combination of
transcendental function of transcendental weight $2L$. In general the
ansatz contains a large number of unknown coefficients. Remarkably, in the case of
six-point planar amplitudes, they can be uniquely fixed by expanding
the ansatz in the boundaries of phase space, where prediction exist
thanks to knowledge of resummation and integrability. This approach is
so powerful that even planar five-loop NMHV amplitudes in $\Neqfour$ SYM can be obtained in
tis way~\cite{Dixon:2015iva}.

On the other hand, examples of application of bootstrap method in QCD
calculation are less common. The reason is that, in QCD the ansatz is
usually much more complicated than in $\Neqfour$ at given loop
order. For example, the transcendental functions in the ansatz can be
multiplied by non-trivial ratio function of kinematics variables, and
the transcendental weight can ranged from $1$ to $2L$ in an $L$-loop
amplitude. Furthermore, integrability is lost in QCD, therefore
the number of boundary data for fixing the ansatz are much smaller
than in $\Neqfour$ SYM.

Nevertheless, the simplicity of \Eqn{eq:qcdc} is hard to ignored:
\begin{itemize}
\item At one and two loops, the results are given solely in terms of HPLs with
rational coefficients. Furthermore, the indices of weight vector are
drawn from the set $\{0,1\}$ only.
\item The last entry of the weight vectors, or the first entry of the
  symbols~\cite{Goncharov:2010jf}, are always $1$. This is
  ensured by that the threshold limit $\vecbsq \to 0$ is a smooth
  limit. Fully differential soft function admits a simple Taylor
  series expansion over $\vecbsq$ in that limit. We will explain more
  about this in the following.
\item The first entry of the weight vectors are always $0$, at least
  through to two loops. 
\end{itemize}
These observations lead us to make the following ansatz for the scale
independent part of the fully differential soft function at $L$ loop:
\begin{align}
  \label{eq:15}
  S_L(\vecb,\tau;\mu=\tau^{-1}) \dot= c^s_L + \sum_i r_i F_i(x)
\end{align}
where $c^s_L$ is the $L$-loop scale independent constant of threshold
soft function. $r_i\in \mathbb{Q}$ are rational numbers. $F_i(x)$ are
transcendental function with transcendental weight $2\leq [F_i(x)]
\leq 2 L$. These can include single HPL $H_{0,\vec{w}_{n-2},1}$, where
$\vec{w}_{n-2}$ is a weight vector of length $n-2$ with entries drawn
from $\{0,1\}$. We also allow $F_i(x)$ to be product (multiple)
zeta value of weight $2\leq m \leq 2 L -2 $ and a HPL
$H_{0,\vec{w}_{n-m-2},1}$. The summation is over all possible
$F_i(x)$. With the ansatz at hand, what remains is to fixed the
rational coefficients $r_i$ using all possible constraints. We
identify two such constraints.

The first constraint comes the fact that rapidity divergence is only a single
logarithmic divergence at each order on the exponential, \Eqn{eq:14},
but the scale independent term in \Eqn{eq:15} can contains higher
order logarithmic divergence in $\tau \to 0$. Therefore, there must be
non-trivial cancellation for the higher order logarithmic divergence
between the scale dependent part and the scale independent part. As an
concrete example, consider the one-loop ansatz:
\begin{align}
  \label{eq:16}
  S(\vecb, \tau;\mu) = C_a \left[ 2 \ln^2 (\mu^2 \tau^2) + 2 \zeta_2 +
  r_1 H_{2} \right]
\end{align}
where the scale dependent part is uniquely fixed by RG equation, and
we have used $c^s_1  = 2 C_a \zeta_2$. The linearity of $\ln \tau$
divergence demands that the $\ln^2 \tau$ divergence in \Eqn{eq:16}
should cancel. Using 
\begin{align}
  \label{eq:17}
  \lim_{\tau \to 0} H_2(x) = \lim_{\tau \to 0} \Li_2 \left( -
  \frac{\vecbsq}{b^2_0 \tau^2} \right) = - \frac{1}{2} \ln^2\left(
  \frac{\vecbsq}{b^2_0 \tau^2}\right) - \zeta_2 + \Ord(\tau)
\end{align}
 we find that
 \begin{align}
   \label{eq:18}
   r_1 = 4 C_a
 \end{align}
which is in agreement with the result in \Eqn{eq:11}. In general, at
$L$-loop, the logarithmic divergent terms $\ln^k \tau$ with $1< k \leq
2 L$ must cancel between the scale independent part and the scale
dependent part.

Beyond one-loop, the linearity in the logarithmic rapidity divergence
is not enough to completely fix the unknown coefficients. The
remaining degrees of freedom have to be fixed using the second constraint,
which comes from the expansion in the threshold limit $\vecbsq \to
0$. 

Following \Eqn{eq:spectrum_reg_soft_function}, we notice that an expansion in $\vecbsq$ is
possible at the integrand level by expanding the exponential, 
\begin{align}
  \label{eq:19}
  S(\vecb,\tau;\mu) &= \int \frac{d^d k}{(2 \pi)^d} \theta(k^0)
  \theta(k^2) \exp( - k^0 b_0\tau -i \vecb \mcdot \vec{k}_\perp )
 \hat{S}(k,\mu) \nonumber\\
  &= \int \frac{d^d k}{(2 \pi)^d} \theta(k^0)
  \theta(k^2) \exp( - k^0 b_0\tau )
  \sum_{n=0}^\infty \frac{ (-i \vecb \mcdot \vec{k}_\perp)^n}{n!} \hat{S}(k,\mu),
\end{align}
where $\hat{S}(k,\mu)$ is the fully differential soft
function in the momentum space, with $k$ being the total momentum of
radiation crossing the cut. We recognize that the first
term in Eq.~\eqref{eq:19} is simply the threshold soft function,
\begin{align}
  \label{eq:20}
  S(\vecb,\tau;\mu)\Big|_{n=0} = \int \frac{d^d k}{(2
  \pi)^d}\theta(k^0) \theta(k^2)  \exp\left( - k^0 b_0
  \tau \right)
   \hat S(k,\mu)
\end{align}
Starting from $n=1$, we encounter terms of the form
\begin{align}
  \label{eq:22}
 (- i  \vecb \mcdot \vec{k}_\perp)^n = (i b_\perp \mcdot k)^n
\end{align}
Note that $S(k,\mu)$ is a function of $k^+k^-$ and $k^2$. By Lorentz
invariance, $k^\mu$ becomes $n^\mu$ or $\bar{n}^\mu$
after the $d^dk$ integral. Using $b_\perp \mcdot n = b_\perp \mcdot
\bar{n} = 0$, we obtain that only even $n$ survives in \Eqn{eq:19}.
 The first non-vanishing term start at $n=2$,
\begin{align}
  \label{eq:23}
  ( - i \vecb \mcdot \vec{k}_\perp)^2 = - \vecb^\alpha \vecb^\beta
  \vec{k}_\perp^\alpha \vec{k}_\perp^\beta \dot= - \frac{1}{d-2} \vecbsq
  \vec{k}_\perp^{\, 2} =
-  \frac{\vecbsq}{d-2} ( k^+ k^- - k^2 )
\end{align}
where $\dot =$ means that the equality only holds after integrating
over $k$. For arbitrary positive integer $m$, we have
\begin{align}
  \label{eq:24}
  ( -i \vecb \mcdot \vec{k}_\perp)^{2m} = f(2m) \big(\vecbsq\big)^m (k^+ k^- - k^2)^{m}
\end{align}
where the function $f(2m)$ is given by
\begin{align}
  \label{eq:25}
f(2m) =  (-1)^{m} \frac{1  \cdot 3 \cdot 5 
  \ldots  (2m-1)}{d_\perp \cdot  (d_\perp + 2)\cdot (d_\perp + 4) \ldots
   (d_\perp + 2m-2)}
\end{align}
with $d_\perp = d-2$, and we also define $f(0) = 1$.
We therefore obtain that the fully differential soft function is given
by the expansion
\begin{align}
  \label{eq:26}
\boxed{S(\vecb,\tau;\mu)= \sum_{m=0}^\infty \frac{f(2m)}{(2m)!}\big(\vecbsq\big)^m \int \frac{d^d k}{(2
  \pi)^d}\theta(k^0) \theta(k^2)  \exp\left( - k^0 b_0
  \tau \right) (k^+k^- - k^2)^{m}
   \hat{S}(k,\mu) }
\end{align}
that is, by insertion of numerator $(k^+k^- - k^2)^{m}$ into the integrand for
threshold soft function. By expressing $k$ as the sum of soft momentum
crossing the cut $k^\mu = \sum_{i \in \rm cut} k^\mu_i$, the numerator
insertion can be reduced to scalar master integrals relevant for the
threshold soft function via Integration-By-Parts identities~(IBP)~\cite{Chetyrkin:1981qh}. At high power of $m$, the reduction can be very computational-demanding.
Fortunately, some trick can be used to ease the effort, which is explained in the
App. \ref{app:integral_reduction}. We note that these integrals have been
computed to three loops~\cite{Anastasiou:2013srw,Li:2013lsa,Duhr:2013msa,Anastasiou:2015yha,Li:2014bfa,Zhu:2014fma}, in an effort to
obtain the soft-virtual corrections for Higgs production at N$^3$LO~\cite{Anastasiou:2014vaa,Li:2014afw}.
\Eqn{eq:26} is one of our main result, which in principle provides
any number of boundary conditions for fixing the unknown
coefficients $r_i$~\footnote{In reality, the number of boundary
  condition is limited by the computation resource demanded by the IBP
reduction.}.

As a concrete example, we present below the expansion in $x = -
\vecbsq /b^2_0 \tau^2$ through to $x^5$ at one and two loops for the
scale independent part:
\begin{align}
  \label{eq:27}
  S(\vecb,\tau;\mu=\tau^{-1}) = &\, C_a \left(\frac{4
                                  x^5}{25}+\frac{x^4}{4}+\frac{4
                                  x^3}{9}+x^2+4 x+2 \zeta_2\right) +
                                  \Ord ( x^6)
\nbrk
  S(\vecb,\tau;\mu=\tau^{-1}) = &\, 
\beta^{\rm gen}_0 C_a \left(\frac{113 x^5}{375}+\frac{19
                                  x^4}{48}+\frac{14
                                  x^3}{27}+\frac{x^2}{2}-4 x-\frac{4
                                  \zeta_3}{3}\right)
\brk
+C_A C_a \Biggl[\delta_R \left(\frac{4 x^5}{75}+\frac{x^4}{12}+\frac{4
                                  x^3}{27}+\frac{x^2}{3}+\frac{4
                                  x}{3}+\frac{\zeta_2}{3}+\frac{52}{27}\right)
\brk
-\frac{8}{25} x^5 \zeta_2-\frac{30829 x^5}{11250}-\frac{1}{2} x^4
                                  \zeta_2-\frac{949
                                  x^4}{288}-\frac{8}{9} x^3
                                  \zeta_2-\frac{98 x^3}{27}-2 x^2
                                  \zeta_2
\brk
-\frac{19 x^2}{18}-8 x \zeta_2+\frac{448 x}{9}+\frac{64
                                  \zeta_2}{9}+\frac{22 \zeta_3}{9}-30
                                  \zeta_4+\frac{2272}{81}\Biggr]
\brk
+2 C_a n_f T_f \left(\frac{28 x^5}{125}+\frac{x^4}{4}+\frac{16
                                  x^3}{81}-\frac{4 x^2}{9}-\frac{88
                                  x}{9}-\frac{10 \zeta_2}{9}-\frac{4
                                  \zeta_3}{9}-\frac{328}{81}\right)
\brk
+2 C_a n_s T_s \left(\frac{11 x^5}{375}+\frac{x^4}{48}-\frac{2 x^3}{81}-\frac{5 x^2}{18}-\frac{28 x}{9}-\frac{4 \zeta_2}{9}-\frac{\zeta_3}{9}-\frac{160}{81}\right)
\brk
+ \Ord(x^6)
\end{align}
where we have given the results for SU($N_c$) gauge theory with $n_f$
number of Dirac Fermion, $n_s$ number of complex scalar, both in
fundamental representation. $\delta_R$ is a parameter choosing the
specific regularization scheme. $\delta_R=0$ for
Four-Dimensional-Helicity~(FDH) scheme~\cite{Bern:2002zk}, and $\delta_R=1$ for 't
Hooft-Veltman scheme. $\beta^{\rm gen}_0$ is the corresponding
beta function:
\begin{align}
  \label{eq:28}
  \beta^{\rm gen}_0 = \frac{11}{3} C_A - \frac{4}{3} n_f T_f -
  \frac{1}{3} n_s T_s
\end{align}
Using a generic two-loop ansatz from \Eqn{eq:15}, it is
straightforward to determine all the coefficients by combining the
linearity constraint for the $\ln\tau$ divergence, and the data from
threshold expansion. 
For example, the one and two-loop result in FDH scheme for $\Neqfour$ SYM can be
bootstrapped 
from these constraints by setting $n_f \to 2$, $n_s \to 3$, $ C_A,
T_f = T_s \to N_c$, and $\delta_R = 0$:
\begin{align}
  S^{\Neqfour}_1(\tau,\vecb;\mu=\tau^{-1}) = & c^{s,\Neqfour}_{1} + 4 N_c H_{2}
\nbrk
S^{\Neqfour}_2(\tau,\vecb;\mu=\tau^{-1}) = & c^{s,\Neqfour}_2 + N_c^2 \Big(-8 \zeta_2 H_{2}-8 H_{4}
-8 H_{2,2}
-16 H_{3,1}-16 H_{2,1,1}\Big)
\label{eq:neq4c}
\end{align}
where the constants $c^{s,\Neqfour}_i$ are given in the App. \ref{app:anom_dim}. By
setting $n_s = 0$ and $\delta_R = 1$, we also reproduce the QCD result
in \Eqn{eq:qcdc}. We note that the result in $\Neqfour$ SYM agrees the
leading transcendental part of the QCD result~\footnote{The
  transcendental weight of HPLs are defined as the length of its
  weight vector, and the transcendental weight of $\zeta_n$ is
  $n$. Furthermore, transcendental weight is additive under
  multiplication.}, while there is no lower transcendental weight part
in $\Neqfour$ SYM. Such relation between $\Neqfour$ SYM and QCD was
first observed for anomalous dimension of twist two
operator~\cite{Kotikov:2003fb}. In the context of soft function in
SCET, this is true for threshold soft function through to three loops~\cite{Li:2014afw}.
 
\begin{figure}[ht]
  \centering
  \includegraphics[scale=0.3]{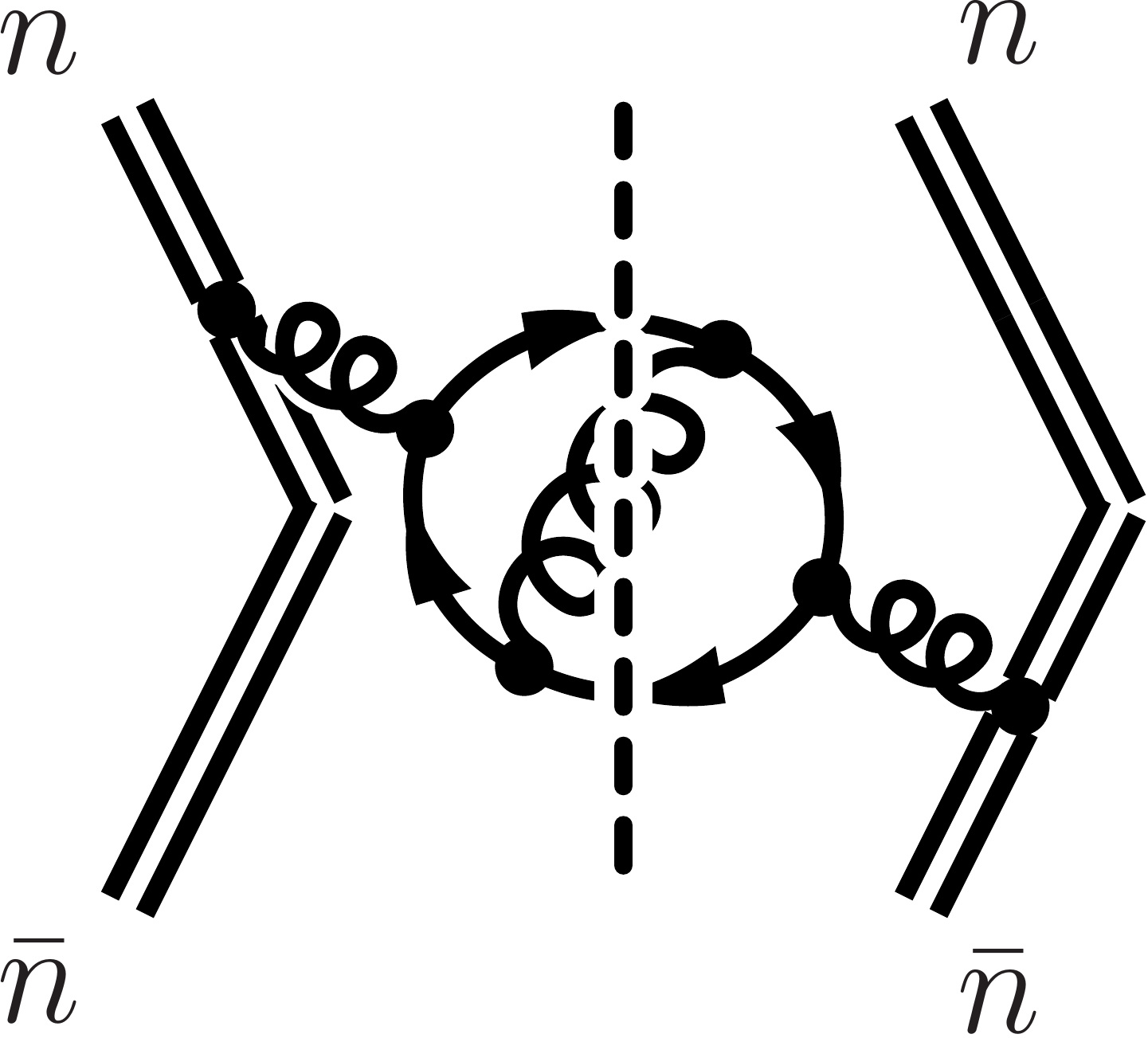}
  \caption{Representative diagram for color structure $C_a C_F n_f$ at
    three loop in QCD.}
  \label{fig:cacfnf}
\end{figure}

As outlined above, the calculation of fully differential soft function
doesn't require calculating a single Feynman diagram. At least through
to two loops, it only relies
on the constraints imposed by linearity of $\ln\tau$ divergence, and
more importantly boundary data from threshold soft
function. At higher orders, it is limited by the availability of
threshold data, the amount computation resource required for IBP
reduction for \Eqn{eq:26}, and potentially also the completeness of
the ansatz in \Eqn{eq:15}~\footnote{For example, function that do not
  follow the pattern observed at one and two loops might appear at
  higher loops.}. To illustrate the power of this approach, we
consider the calculation of fully differential
soft function in QCD at three loops for two relatively simple color
structure, $C_a n^2_f$ and $C_a C_F n_f$. These contributions are
simple because they correspond to self-energy insertion of one-loop
diagram. A representative diagram which contribute to $C_a C_F n_f$
color structure is depicted in Fig.~\ref{fig:cacfnf}. We find that the
ansatz in \Eqn{eq:15} is sufficient for bootstrapping these color
structure at three loops. The results reads
\begin{align}
  \label{eq:29}
  S_3(\tau,\vecb;\mu) = &\,  C_a n_f^2 \Biggl(\frac{400}{81} H_{2}+\frac{160}{27}
H_{3}+\frac{32}{9} H_{4}-\frac{160}{27} H_{2,1}
-\frac{32}{9} H_{3,1}+\frac{32}{9} H_{2,1,1} \Biggr)
\brk
+  C_a C_F n_f \Biggl( 32 \zeta_3 H_{2}-\frac{110}{3}
  H_{2}-8 H_{3}+8 H_{2,1} \Biggr) + S_{3, C_a C_A^2} + S_{3, C_a C_A n_f}
\end{align}
The full QCD results require significant more work and we defer them
to a second publication. 

As explain in \sect{sec:3}, the fully differential soft function acts
as a mother function to both the TMD soft function and the threshold soft
function. It is an easy exercise to obtain TMD soft function by taking
the limit $\tau\to 0$, and keeping the non-vanishing terms only. Then
one can identify $\nu = \tau^{-1}$. We find
that as expected, the TMD soft function satisfy the $\mu$ and $\nu$
evolution equation. The corresponding rapidity anomalous dimension is
extracted to be
\begin{align}
  \label{eq:30}
    \grap_0 = & \, 0
\nbrk
\grap_1 = &\, C_A C_a \left(28 \zeta_3-\frac{808}{27}\right)+\frac{112
  C_a n_f}{27}
\end{align}
for QCD, and
\begin{align}
  \label{eq:31}
      \gamma_{0}^{r,\Neqfour} = & \, 0
\nbrk
\gamma_{1}^{r,\Neqfour} = &\, 28 N_c^2 \zeta_3
\end{align}
for $\Neqfour$ SYM.

The QCD result in \Eqn{eq:30} agrees with those obtained using
different methods~\cite{Gehrmann:2014yya,Echevarria:2015byo,Luebbert:2016itl}.
The corresponding result for the scale independent part are given by
\begin{align}
  \label{eq:32}
      c^\perp_{1} = & \, -2 C_a \zeta_2
\nbrk
  c^\perp_{2} = & \, C_A C_a \left(-\frac{67 \zeta_2}{3}-\frac{154
      \zeta_3}{9}+10 \zeta_4+\frac{2428}{81}\right)
\brk
+C_a n_f \left(\frac{10 \zeta_2}{3}+\frac{28 \zeta_3}{9}-\frac{328}{81}\right)
\end{align}
for QCD, and
\begin{align}
  \label{eq:33}
    c^{\perp,\Neqfour}_{1} = & \, -2 N_c \zeta_2
\nbrk
  c^{\perp,\Neqfour}_{2} = & \, 10 N_c^2 \zeta_4
\end{align}
for $\Neqfour$ SYM.

\section{Conclusions}

We have introduced a new method to calculate naively rapidity divergent soft functions, by deforming the soft function's measurement into one that is calculable in dimensional regularization, and can be reconnected to the naive rapidity divergent soft function. The most practical benefit of this method will be the three-loop non-cusp anomalous dimension needed for transverse momentum resummation for Drell-Yan and Higgs production. This is the largest source of certainty on the $N^3LL$ analysis of these spectra, given that the exact value of four loop cusp anomalous dimension has been found to have negligible impact. This anomalous dimension for QCD processes and the $N^3LL+NNLO$ transverse momentum spectrum for Higgs production will be presented in companion papers.

Ultimately underpinning this procedure is the factorization of multi-differential cross-section, where the same cross-section in different singular regions experiences different factorizations. These factorizations must be consistent with each other, even after resumming logarithms, allowing us to derive results on the other factorizations from ones more amiable to calculation. Other observables sensitive to rapidity divergences can also be treated this way, perhaps simplifying certain two or even three-loop calculations. Beyond perturbation theory, what is particularly appealing about this type of multi-differential factorization is that it gives an in-principle non-perturbative regularization of the rapidity divergences, something that hitherto was only feasible by tilting wilson-lines off the light-cone. For our specific processes, it suggests that the non-perturbative corrections to both transverse momentum resummation and threshold resummation coming from soft radiation are intimately connected, and work (see for instance \cite{Becher:2013iya,D'Alesio:2014vja,Korchemsky:1994is}) on such corrections in transverse momentum resummation should be revisited in this light. Indeed, given the rapid development of integrability technology (\cite{Beisert:2010jr}) in planar $N=4$ SYM for null polygonal wilson loops, it is reasonable to suppose that the fully differential soft function can be calculated \emph{exactly} in that theory. In this example, the typical model for non-perturbative corrections - being inspired from renormalon related loop corrections probed by the running coupling \cite{Mueller:1984vh,Zakharov:1992bx,Beneke:1993yn,Korchemsky:1994is} - should fail, since the beta-function vanishes in that theory. The theory being conformal has no new structures arising in the deep infra-red.

Another benefit of this multiple region factorization is that it clarifies the structure of the transverse momentum spectrum in the threshold region, and in particular, how one can perform joint resummation using the techniques of \cite{Larkoski:2014tva, Procura:2014cba}. Again we have two boundary theories, the threshold region and the TMD-PDF region, that are connected by an intermediate theory where functions refactorize. Evolving this intermediate theory to its natural scales would give a natural joint resummation formula that reduces to transverse momentum resummation or threshold resummation in the correct phase-space regions. Going beyond \cite{Laenen:2000ij}, we can also naturally include the full low-scale matrix elements consistently, giving a $N^3LL$ formula uniformly valid across phase-space. It would be interesting to see how such a resummation changes the transverse momentum spectrum, particularly at colliders with lower center of mass energies.

\begin{acknowledgments}
This work was supported by the Office of Nuclear Physics of the U.S. Department of Energy
(DOE) under Contract DE-SC0011090. Fermilab is operated by Fermi Research Alliance, LLC under Contract No. DE-AC02-07CH11359 with the United States Department of Energy.
\end{acknowledgments}

\appendix

\section{Anomalous Dimension and Matching Coefficients}
\label{app:anom_dim}

We collect all the anomalous dimension and matching coefficients needed in this work here. We present the results for $\Neqfour$ SYM case first.

The cusp anomalous dimension in $\Neqfour$ SYM have been perturbatively computed through to four loops in large $N_c$ limit~\cite{Bern:2006ew}. The first three-loop expansion is given by
\begin{align}
  \Gamma^{{\rm cusp},\Neqfour}_{0} = & \, 4 N_c
\nbrk
  \Gamma^{{\rm cusp},\Neqfour}_{1} = & \, -8 N_c^2 \zeta_2
\nbrk
  \Gamma^{{\rm cusp},\Neqfour}_{2} = & \, 88 N_c^3 \zeta_4
\end{align}

The threshold anomalous dimension in $\Neqfour$ through to three loops
has been computed to be~\cite{Li:2014afw}
\begin{align}
  \gamma^{s,\Neqfour}_{0} = & \, 0
\nbrk
  \gamma^{s,\Neqfour}_{1} = & \, 28 N_c^2 \zeta_3
\nbrk
  \gamma^{s,\Neqfour}_{2} = & \, N_c^3 \left(-\frac{176}{3} \zeta_2 \zeta_3-192 \zeta_5\right)
\end{align}
The rapidity anomalous dimension is the same as threshold anomalous
dimension, at least through to two loops in $\Neqfour$ SYM,
\begin{align}
  \gamma_{i}^{s,\Neqfour} = \gamma_{i}^{r,\Neqfour} , \quad i=0,1
\end{align}
The threshold soft function constants in $\Neqfour$ SYM through to three loops are given by~\cite{Li:2014afw}
\begin{align}
  c_{0}^{s,\Neqfour} = & \, 2 N_c \zeta_2
\nbrk
  c_{1}^{s,\Neqfour} = & \, -30 N_c^2 \zeta_4
\nbrk
  c_{2}^{s,\Neqfour} = & \, N_c^3 \left(\frac{1072 \zeta_3^2}{9}+\frac{8506 \zeta_6}{27}\right)
\end{align}
And the constants for the TMD soft function with exponential regulator in $\Neqfour$ SYM through to two loops are given by
\begin{align}
  c^{\perp,\Neqfour}_{0} = & \, -2 N_c \zeta_2
\nbrk
  c^{\perp,\Neqfour}_{1} = & \, 10 N_c^2 \zeta_4
\end{align}

The QCD cusp anomalous dimension through to three loops have been
computed in ref.~\cite{Moch:2004pa}. The results are
\begin{align}
  \gcusp_{0} = & \, 4 C_a
\nbrk
  \gcusp_{1} = & \, C_A C_a \left(\frac{268}{9}-8 \zeta_2\right)-\frac{40 C_a n_f}{9}
\nbrk
  \gcusp_{2} = & \, C_A^2 C_a \left(-\frac{1072 \zeta_2}{9}+\frac{88
      \zeta_3}{3}+88 \zeta_4+\frac{490}{3}\right)
\brk
+C_A C_a n_f \left(\frac{160 \zeta_2}{9}-\frac{112
    \zeta_3}{3}-\frac{836}{27}\right)
\brk
+C_a C_F n_f \left(32 \zeta_3-\frac{110}{3}\right)-\frac{16 C_a n_f^2}{27}
\end{align}

The threshold soft anomalous dimension can be extracted from quark or
gluon form factor~\cite{Baikov:2009bg,Lee:2010cga,Gehrmann:2010ue} and the $\delta(1-x)$ part of the QCD splitting function~\cite{Moch:2004pa}:
\begin{align}
   \gsoft_{0} = & \, 0
\nbrk
  \gsoft_{1} = & \, C_A C_a \left(\frac{22 \zeta_2}{3}+28 \zeta_3-\frac{808}{27}\right)+C_a n_f \left(\frac{112}{27}-\frac{4 \zeta_2}{3}\right)
\nbrk
  \gsoft_{2} = & \, C_A^2 C_a \left(-\frac{176}{3} \zeta_3
    \zeta_2+\frac{12650 \zeta_2}{81}+\frac{1316 \zeta_3}{3}-176
    \zeta_4-192 \zeta_5-\frac{136781}{729}\right)\nn\\
&+C_A C_a n_f \left(-\frac{2828
  \zeta_2}{81}-\frac{728 \zeta_3}{27}
+48 \zeta_4+\frac{11842}{729}\right)\nn\\
&+C_a C_F n_f \left(-4
\zeta_2-\frac{304 \zeta_3}{9} 
-16 \zeta_4+\frac{1711}{27}\right)+C_a n_f^2 \left(\frac{40 \zeta_2}{27}-\frac{112 \zeta_3}{27}+\frac{2080}{729}\right)
\end{align}

The threshold soft function constants in QCD through to three loops
are~\cite{Li:2014afw}:
\begin{align}
  c^s_{0} = & \, 2 C_a \zeta_2
\nbrk
  c^s_{1} = & \, C_A C_a \left(\frac{67 \zeta_2}{9}-\frac{22
      \zeta_3}{9}-30 \zeta_4+\frac{2428}{81}\right)
+C_a n_f \left(-\frac{10 \zeta_2}{9}+\frac{4 \zeta_3}{9}-\frac{328}{81}\right)
\nbrk
  c^s_{2} = & \, C_A^2 C_a \left(\frac{1072
      \zeta_3^2}{9}-\frac{220}{9} \zeta_2 \zeta_3-\frac{87052
      \zeta_3}{243}-\frac{20371 \zeta_2}{729}
-\frac{9527 \zeta_4}{27}-\frac{968 \zeta_5}{9}+\frac{8506
  \zeta_6}{27}+\frac{5211949}{13122}\right)
\brk
+C_A C_a n_f \left(-\frac{8}{9} \zeta_3 \zeta_2+\frac{2638
    \zeta_2}{729}+\frac{1216 \zeta_3}{81}+\frac{928 \zeta_4}{27}
-\frac{16 \zeta_5}{3}-\frac{412765}{6561}\right)
\brk
+C_a C_F n_f
\left(\frac{16}{3} \zeta_3 \zeta_2-\frac{55 \zeta_2}{9}
+\frac{2840 \zeta_3}{81}+\frac{152 \zeta_4}{9}+\frac{224
  \zeta_5}{9}-\frac{42727}{486}\right)
\brk
+C_a n_f^2 \left(-\frac{8 \zeta_2}{81}+\frac{880 \zeta_3}{243}+\frac{52 \zeta_4}{27}-\frac{256}{6561}\right)
\end{align}

The rapidity anomalous dimension in QCD up to two loops are:
\begin{align}
\grap_{0}&= \,0
\nbrk
\grap_{1}&= \, C_a C_A \left(28 \zeta_3 - \frac{808}{27}\right) + \frac{112 C_a n_f}{27}
\end{align}
And the constants for the TMD soft function with the exponential
regulator in QCD
$c^{\perp}_i=S^\perp_i(\vec{b};\mu=\frac{b_0}{|\vec{b}_\perp|},\mu=\frac{b_0}{|\vec{b}_\perp|})$
through to two loops are: 
\begin{align}
    c^{\perp}_{1} = & \, -2 C_a \zeta_2
\nbrk
  c^{\perp}_{2} = & \, C_a C_A \left(-\frac{67 \zeta_2}{3}-\frac{154
      \zeta_3}{9}+10 \zeta_4+\frac{2428}{81}\right)
+C_a n_f \left(\frac{10 \zeta_2}{3}+\frac{28 \zeta_3}{9}-\frac{328}{81}\right)
\end{align}

\section{Integral Reduction \label{app:integral_reduction}}

The integration to obtain fully-different soft function in the momentum space can be written as follows,
\begin{align}
   & \int [dq][dk] \;\delta^d(k - \sum_{i} k_i) = \int \frac{d^d q_1}{(2\pi)^d} \frac{d^d q_2}{(2\pi)^d} \dots \; \frac{d^d k_1}{(2\pi)^{d-2}}\delta(k_1^2) \frac{d^d k_2}{(2\pi)^{d-2}}\delta(k_2^2) \dots \; \delta^d(k -  \sum_{i} k_i ),
\end{align}
where $q_{1,2,\dots}$ are loop momenta, $k_{1,2,\dots}$ are momenta of soft radiations, and $k$ denotes the total momentum of emissions $\sum_i k_i$. Combined with integration of Eq.~\ref{eq:26}, we get
\begin{align}
  & \sum_{m=0}^\infty \frac{f(2m)}{(2m)!}\big(\vecbsq\big)^m \int \frac{d^d k}{(2
  \pi)^d}\theta(k^0) \theta(k^2) e^{- k^0 b_0\tau}  (k^+k^- - k^2)^{m} \int [dq][dk] \;\delta^d(k - \sum_{i} k_i) \nonumber\\
  &= \sum_{m=0}^\infty \frac{f(2m)}{(2m)!} \big(\vecbsq\big)^m \int \frac{d k^0}{2 \pi} \theta(k^0) e^{- k^0 b_0\tau}  \int [dq][dk] (k^+k^- - k^2)^{m} \;\delta(k^0 - \frac{k^- + k^+}{2} ),
\end{align}
Here the integration of our concern is only the first part,
\begin{align}
  \int [dq][dk] (k^+k^- - k^2)^{m} \;\delta(k^0 - \frac{k^- + k^+}{2} ),
\end{align}
since this is where the integral reduction can be applied. Note that at $m=0$, it is exactly the integration to obtain the threshold soft function in the momentum space.

The integrand generally consist of propagators of three types, 1) eikonal propagator along $n$ in the form of $1/P^-$, 2) eikonal propagator along $\bar{n}$ in the form of $1/P^+$, and 3) regular propagator in the form of $1/P^2$, where $P$ denote some linear combination of loop and emission momenta $q_i, k_i$ in the relevant Feynman diagrams. Because we can write the insertion of $(k^+k^- - k^2)^{m}$ as polynomials of inverse propagators, we will treat it as part of the integrand and stop writing it explicitly in the integration from now on.

The insertion of $(k^+ k^- -k^2)^m$ at large $m$ often leads to very high rank integrals due to the fact that each of $k^+$ and $k^-$ corresponds to at least one inverse propagator. And the fast raising rank from increasing $m$ makes reduction through IBP identities very challenging, which becomes nearly impossible when $m$ approaches 15 or larger. Fortunately, the reduction of $k^+$ and $k^-$ to arbitrary powers can be done without resorting to IBP identities. To see it, we first insert some delta functions to separate the above integration into two steps,
\begin{eqnarray}
      && \int [dq][dk] \;\delta(k^0 - \frac{k^- + k^+}{2} ) \nonumber\\
      &=& \int d k^- d k^+ \delta(k^0 - \frac{k^- + k^+}{2} ) \int [dq][dk] \; \delta(k^- - \sum_{i}  k_i^-) \delta(k^+ - \sum_{i}  k_i^+).
\end{eqnarray}
For the first integration, let's consider the following integral which is schematical representation of the actual integrand,
\begin{eqnarray}
      && \int [dq][dk] \; \delta(k^- - \sum_{i} k_i^-) \delta(k^+ - \sum_{i} k_i^+) \frac{(k^-)^a (k^+)^b}{(P_{1}^-)^{i_1} (P_{2}^-)^{i_2} \dots (Q_{1}^+)^{j_1} (Q_{2}^+)^{j_2} \dots (K_{1}^2)^{k_1} (K_{2}^2)^{k_2} \dots} \nonumber\\
      &=& {\rm F}(d) \; (k^-)^{a+l-1-i_1-i_2-\dots-k_1-k_2-\dots} (k^+)^{b+l-1-j_1-j_2-\dots-k_1-k_2-\dots},
\end{eqnarray}
where $P_{1,2,\dots}$, $Q_{1,2,\dots}$, $K_{1,2,\dots}$ are linear combinations of loop and emission momenta, and $i_{1,2,\dots}$, $j_{1,2,\dots}$, $k_{1,2,\dots}$ are the indices of propagators. The second line follows from dimensional analysis and rescaling dependence on $n$ and $\nbar$. ${\rm F}(d)$ is a function of the space-time dimension $d$ that can only be obtained through direct calculation; $l$ is given by the number of loops $n_l$ and soft emissions $n_e$, $l = (n_e (d-2) + n_l d)/2$ \footnote{In practice, inverse unitarity allows us to treat $\delta(k_{1,2,\dots}^2)$ in the phase space integration as propagators $1/k_{1,2,\dots}^2$, and the definition of $m$ becomes $l = (n_e + n_l) d/2$ accordingly.}. It is straightforward to write down the result of the second integration,
\begin{eqnarray}
      && \int [dq][dk] \;\delta(k^0 - \frac{k^- + k^+}{2} ) \frac{(k^-)^a (k^+)^b}{(P_{1}^-)^{i_1} (P_{2}^-)^{i_2} \dots (Q_{1}^+)^{j_1} (Q_{2}^+)^{j_2} \dots (K_{1}^2)^{k_1} (K_{2}^2)^{k_2} \dots} \nonumber\\
      &=& \int d k^- d k^+ \delta(k^0 - \frac{k^- + k^+}{2} ) F(d) \; (k^-)^{a+l-1-(i_1+i_2+\dots)-(k_1+k_2+\dots)} (k^+)^{b+l-1-(j_1+j_2+\dots)-(k_1+k_2+\dots)} \nonumber\\
      &=& 2\, {\rm F}(d)\; (2 k^0)^{a+b+2 l-1-(i_1+i_2+\dots)-(j_1+j_2+\dots)-2(k_1+ k_2+\dots)} \nonumber\\
      && {\rm B}(a+l-(i_1+i_2+\dots)-(k_1+k_2+\dots), b+l-(j_1+j_2+\dots)-(k_1+k_2+\dots)),
\end{eqnarray}
where ${\rm B}$ is the standard Beta function, ${\rm B}(x,y)=\Gamma(x)\Gamma(y)/\Gamma(x+y)$.
So it becomes clear that we can write any integral with non-zero $a$ or $b$ in term of the corresponding integral with zero $a$ and $b$ by multiplying it the proper factor,
\begin{eqnarray}
       && \int [dq][dk] \;\delta(k^0 - \frac{k^- + k^+}{2} ) \frac{(k^-)^a (k^+)^b}{(P_{1}^-)^{i_1} (P_{2}^-)^{i_2} \dots (Q_{1}^+)^{j_1} (Q_{2}^+)^{j_2} \dots (K_{1}^2)^{k_1} (K_{2}^2)^{k_2} \dots}  \nonumber\\
       &=& (2k^0)^{a+b} \frac{{\rm B}(a+l-(i_1+i_2+\dots)-(k_1+k_2+\dots), b+l-(j_1+j_2+\dots)-(k_1+k_2+\dots))}{{\rm B}(l-(i_1+i_2+\dots)-(k_1+k_2+\dots), l-(j_1+j_2+\dots)-(k_1+k_2+\dots))} \nonumber\\
       &&  \times \int [dq][dk] \;\delta(k^0 - \frac{k^- + k^+}{2} ) \frac{1}{(P_{1}^-)^{i_1} (P_{2}^-)^{i_2} \dots (Q_{1}^+)^{j_1} (Q_{2}^+)^{j_2} \dots (K_{1}^2)^{k_1} (K_{2}^2)^{k_2} \dots}.
\end{eqnarray}
which essentially realizes the reduction of insertions of $k^+$ and $k^-$ of arbitrary powers. As for the reduction of $k^2$ to arbitrary powers, we find no alternative but relying on the full IBP reduction machinery~\cite{Lee:2012cn,Smirnov:2008iw}.

\bibliography{new_rap_reg_bib}

\begin{thebibliography}{98}%
\makeatletter
\providecommand \@ifxundefined [1]{%
 \@ifx{#1\undefined}
}%
\providecommand \@ifnum [1]{%
 \ifnum #1\expandafter \@firstoftwo
 \else \expandafter \@secondoftwo
 \fi
}%
\providecommand \@ifx [1]{%
 \ifx #1\expandafter \@firstoftwo
 \else \expandafter \@secondoftwo
 \fi
}%
\providecommand \natexlab [1]{#1}%
\providecommand \enquote  [1]{``#1''}%
\providecommand \bibnamefont  [1]{#1}%
\providecommand \bibfnamefont [1]{#1}%
\providecommand \citenamefont [1]{#1}%
\providecommand \href@noop [0]{\@secondoftwo}%
\providecommand \href [0]{\begingroup \@sanitize@url \@href}%
\providecommand \@href[1]{\@@startlink{#1}\@@href}%
\providecommand \@@href[1]{\endgroup#1\@@endlink}%
\providecommand \@sanitize@url [0]{\catcode `\\12\catcode `\$12\catcode
  `\&12\catcode `\#12\catcode `\^12\catcode `\_12\catcode `\%12\relax}%
\providecommand \@@startlink[1]{}%
\providecommand \@@endlink[0]{}%
\providecommand \url  [0]{\begingroup\@sanitize@url \@url }%
\providecommand \@url [1]{\endgroup\@href {#1}{\urlprefix }}%
\providecommand \urlprefix  [0]{URL }%
\providecommand \Eprint [0]{\href }%
\providecommand \doibase [0]{http://dx.doi.org/}%
\providecommand \selectlanguage [0]{\@gobble}%
\providecommand \bibinfo  [0]{\@secondoftwo}%
\providecommand \bibfield  [0]{\@secondoftwo}%
\providecommand \translation [1]{[#1]}%
\providecommand \BibitemOpen [0]{}%
\providecommand \bibitemStop [0]{}%
\providecommand \bibitemNoStop [0]{.\EOS\space}%
\providecommand \EOS [0]{\spacefactor3000\relax}%
\providecommand \BibitemShut  [1]{\csname bibitem#1\endcsname}%
\let\auto@bib@innerbib\@empty
\bibitem [{\citenamefont {Chiu}\ \emph
  {et~al.}(2012{\natexlab{a}})\citenamefont {Chiu}, \citenamefont {Jain},
  \citenamefont {Neill},\ and\ \citenamefont {Rothstein}}]{Chiu:2011qc}%
  \BibitemOpen
  \bibfield  {author} {\bibinfo {author} {\bibfnamefont {J.-y.}\ \bibnamefont
  {Chiu}}, \bibinfo {author} {\bibfnamefont {A.}~\bibnamefont {Jain}}, \bibinfo
  {author} {\bibfnamefont {D.}~\bibnamefont {Neill}}, \ and\ \bibinfo {author}
  {\bibfnamefont {I.~Z.}\ \bibnamefont {Rothstein}},\ }\href {\doibase
  10.1103/PhysRevLett.108.151601} {\bibfield  {journal} {\bibinfo  {journal}
  {Phys. Rev. Lett.}\ }\textbf {\bibinfo {volume} {108}},\ \bibinfo {pages}
  {151601} (\bibinfo {year} {2012}{\natexlab{a}})},\ \Eprint
  {http://arxiv.org/abs/1104.0881} {arXiv:1104.0881 [hep-ph]} \BibitemShut
  {NoStop}%
\bibitem [{\citenamefont {Chiu}\ \emph
  {et~al.}(2012{\natexlab{b}})\citenamefont {Chiu}, \citenamefont {Jain},
  \citenamefont {Neill},\ and\ \citenamefont {Rothstein}}]{Chiu:2012ir}%
  \BibitemOpen
  \bibfield  {author} {\bibinfo {author} {\bibfnamefont {J.-Y.}\ \bibnamefont
  {Chiu}}, \bibinfo {author} {\bibfnamefont {A.}~\bibnamefont {Jain}}, \bibinfo
  {author} {\bibfnamefont {D.}~\bibnamefont {Neill}}, \ and\ \bibinfo {author}
  {\bibfnamefont {I.~Z.}\ \bibnamefont {Rothstein}},\ }\href {\doibase
  10.1007/JHEP05(2012)084} {\bibfield  {journal} {\bibinfo  {journal} {JHEP}\
  }\textbf {\bibinfo {volume} {05}},\ \bibinfo {pages} {084} (\bibinfo {year}
  {2012}{\natexlab{b}})},\ \Eprint {http://arxiv.org/abs/1202.0814}
  {arXiv:1202.0814 [hep-ph]} \BibitemShut {NoStop}%
\bibitem [{\citenamefont {Becher}\ \emph {et~al.}(2011)\citenamefont {Becher},
  \citenamefont {Bell},\ and\ \citenamefont {Neubert}}]{Becher:2011pf}%
  \BibitemOpen
  \bibfield  {author} {\bibinfo {author} {\bibfnamefont {T.}~\bibnamefont
  {Becher}}, \bibinfo {author} {\bibfnamefont {G.}~\bibnamefont {Bell}}, \ and\
  \bibinfo {author} {\bibfnamefont {M.}~\bibnamefont {Neubert}},\ }\href
  {\doibase 10.1016/j.physletb.2011.09.005} {\bibfield  {journal} {\bibinfo
  {journal} {Phys. Lett.}\ }\textbf {\bibinfo {volume} {B704}},\ \bibinfo
  {pages} {276} (\bibinfo {year} {2011})},\ \Eprint
  {http://arxiv.org/abs/1104.4108} {arXiv:1104.4108 [hep-ph]} \BibitemShut
  {NoStop}%
\bibitem [{\citenamefont {Becher}\ and\ \citenamefont
  {Bell}(2012{\natexlab{a}})}]{Becher:2012qc}%
  \BibitemOpen
  \bibfield  {author} {\bibinfo {author} {\bibfnamefont {T.}~\bibnamefont
  {Becher}}\ and\ \bibinfo {author} {\bibfnamefont {G.}~\bibnamefont {Bell}},\
  }\href {\doibase 10.1007/JHEP11(2012)126} {\bibfield  {journal} {\bibinfo
  {journal} {JHEP}\ }\textbf {\bibinfo {volume} {11}},\ \bibinfo {pages} {126}
  (\bibinfo {year} {2012}{\natexlab{a}})},\ \Eprint
  {http://arxiv.org/abs/1210.0580} {arXiv:1210.0580 [hep-ph]} \BibitemShut
  {NoStop}%
\bibitem [{\citenamefont {Tackmann}\ \emph {et~al.}(2012)\citenamefont
  {Tackmann}, \citenamefont {Walsh},\ and\ \citenamefont
  {Zuberi}}]{Tackmann:2012bt}%
  \BibitemOpen
  \bibfield  {author} {\bibinfo {author} {\bibfnamefont {F.~J.}\ \bibnamefont
  {Tackmann}}, \bibinfo {author} {\bibfnamefont {J.~R.}\ \bibnamefont {Walsh}},
  \ and\ \bibinfo {author} {\bibfnamefont {S.}~\bibnamefont {Zuberi}},\ }\href
  {\doibase 10.1103/PhysRevD.86.053011} {\bibfield  {journal} {\bibinfo
  {journal} {Phys. Rev.}\ }\textbf {\bibinfo {volume} {D86}},\ \bibinfo {pages}
  {053011} (\bibinfo {year} {2012})},\ \Eprint {http://arxiv.org/abs/1206.4312}
  {arXiv:1206.4312 [hep-ph]} \BibitemShut {NoStop}%
\bibitem [{\citenamefont {Banfi}\ \emph
  {et~al.}(2012{\natexlab{a}})\citenamefont {Banfi}, \citenamefont {Monni},
  \citenamefont {Salam},\ and\ \citenamefont {Zanderighi}}]{Banfi:2012jm}%
  \BibitemOpen
  \bibfield  {author} {\bibinfo {author} {\bibfnamefont {A.}~\bibnamefont
  {Banfi}}, \bibinfo {author} {\bibfnamefont {P.~F.}\ \bibnamefont {Monni}},
  \bibinfo {author} {\bibfnamefont {G.~P.}\ \bibnamefont {Salam}}, \ and\
  \bibinfo {author} {\bibfnamefont {G.}~\bibnamefont {Zanderighi}},\ }\href
  {\doibase 10.1103/PhysRevLett.109.202001} {\bibfield  {journal} {\bibinfo
  {journal} {Phys. Rev. Lett.}\ }\textbf {\bibinfo {volume} {109}},\ \bibinfo
  {pages} {202001} (\bibinfo {year} {2012}{\natexlab{a}})},\ \Eprint
  {http://arxiv.org/abs/1206.4998} {arXiv:1206.4998 [hep-ph]} \BibitemShut
  {NoStop}%
\bibitem [{\citenamefont {Banfi}\ \emph {et~al.}(2015)\citenamefont {Banfi},
  \citenamefont {Caola}, \citenamefont {Dreyer}, \citenamefont {Monni},
  \citenamefont {Salam}, \citenamefont {Zanderighi},\ and\ \citenamefont
  {Dulat}}]{Banfi:2015pju}%
  \BibitemOpen
  \bibfield  {author} {\bibinfo {author} {\bibfnamefont {A.}~\bibnamefont
  {Banfi}}, \bibinfo {author} {\bibfnamefont {F.}~\bibnamefont {Caola}},
  \bibinfo {author} {\bibfnamefont {F.~A.}\ \bibnamefont {Dreyer}}, \bibinfo
  {author} {\bibfnamefont {P.~F.}\ \bibnamefont {Monni}}, \bibinfo {author}
  {\bibfnamefont {G.~P.}\ \bibnamefont {Salam}}, \bibinfo {author}
  {\bibfnamefont {G.}~\bibnamefont {Zanderighi}}, \ and\ \bibinfo {author}
  {\bibfnamefont {F.}~\bibnamefont {Dulat}},\ }\href@noop {} {\  (\bibinfo
  {year} {2015})},\ \Eprint {http://arxiv.org/abs/1511.02886} {arXiv:1511.02886
  [hep-ph]} \BibitemShut {NoStop}%
\bibitem [{\citenamefont {Becher}\ and\ \citenamefont
  {Neubert}(2012)}]{Becher:2012qa}%
  \BibitemOpen
  \bibfield  {author} {\bibinfo {author} {\bibfnamefont {T.}~\bibnamefont
  {Becher}}\ and\ \bibinfo {author} {\bibfnamefont {M.}~\bibnamefont
  {Neubert}},\ }\href {\doibase 10.1007/JHEP07(2012)108} {\bibfield  {journal}
  {\bibinfo  {journal} {JHEP}\ }\textbf {\bibinfo {volume} {07}},\ \bibinfo
  {pages} {108} (\bibinfo {year} {2012})},\ \Eprint
  {http://arxiv.org/abs/1205.3806} {arXiv:1205.3806 [hep-ph]} \BibitemShut
  {NoStop}%
\bibitem [{\citenamefont {Becher}\ \emph
  {et~al.}(2013{\natexlab{a}})\citenamefont {Becher}, \citenamefont {Neubert},\
  and\ \citenamefont {Rothen}}]{Becher:2013xia}%
  \BibitemOpen
  \bibfield  {author} {\bibinfo {author} {\bibfnamefont {T.}~\bibnamefont
  {Becher}}, \bibinfo {author} {\bibfnamefont {M.}~\bibnamefont {Neubert}}, \
  and\ \bibinfo {author} {\bibfnamefont {L.}~\bibnamefont {Rothen}},\ }\href
  {\doibase 10.1007/JHEP10(2013)125} {\bibfield  {journal} {\bibinfo  {journal}
  {JHEP}\ }\textbf {\bibinfo {volume} {10}},\ \bibinfo {pages} {125} (\bibinfo
  {year} {2013}{\natexlab{a}})},\ \Eprint {http://arxiv.org/abs/1307.0025}
  {arXiv:1307.0025 [hep-ph]} \BibitemShut {NoStop}%
\bibitem [{\citenamefont {Stewart}\ \emph {et~al.}(2014)\citenamefont
  {Stewart}, \citenamefont {Tackmann}, \citenamefont {Walsh},\ and\
  \citenamefont {Zuberi}}]{Stewart:2013faa}%
  \BibitemOpen
  \bibfield  {author} {\bibinfo {author} {\bibfnamefont {I.~W.}\ \bibnamefont
  {Stewart}}, \bibinfo {author} {\bibfnamefont {F.~J.}\ \bibnamefont
  {Tackmann}}, \bibinfo {author} {\bibfnamefont {J.~R.}\ \bibnamefont {Walsh}},
  \ and\ \bibinfo {author} {\bibfnamefont {S.}~\bibnamefont {Zuberi}},\ }\href
  {\doibase 10.1103/PhysRevD.89.054001} {\bibfield  {journal} {\bibinfo
  {journal} {Phys. Rev.}\ }\textbf {\bibinfo {volume} {D89}},\ \bibinfo {pages}
  {054001} (\bibinfo {year} {2014})},\ \Eprint {http://arxiv.org/abs/1307.1808}
  {arXiv:1307.1808} \BibitemShut {NoStop}%
\bibitem [{\citenamefont {Stewart}\ \emph {et~al.}(2010)\citenamefont
  {Stewart}, \citenamefont {Tackmann},\ and\ \citenamefont
  {Waalewijn}}]{Stewart:2009yx}%
  \BibitemOpen
  \bibfield  {author} {\bibinfo {author} {\bibfnamefont {I.~W.}\ \bibnamefont
  {Stewart}}, \bibinfo {author} {\bibfnamefont {F.~J.}\ \bibnamefont
  {Tackmann}}, \ and\ \bibinfo {author} {\bibfnamefont {W.~J.}\ \bibnamefont
  {Waalewijn}},\ }\href {\doibase 10.1103/PhysRevD.81.094035} {\bibfield
  {journal} {\bibinfo  {journal} {Phys. Rev.}\ }\textbf {\bibinfo {volume}
  {D81}},\ \bibinfo {pages} {094035} (\bibinfo {year} {2010})},\ \Eprint
  {http://arxiv.org/abs/0910.0467} {arXiv:0910.0467 [hep-ph]} \BibitemShut
  {NoStop}%
\bibitem [{\citenamefont {Stewart}\ \emph {et~al.}(2011)\citenamefont
  {Stewart}, \citenamefont {Tackmann},\ and\ \citenamefont
  {Waalewijn}}]{Stewart:2010pd}%
  \BibitemOpen
  \bibfield  {author} {\bibinfo {author} {\bibfnamefont {I.~W.}\ \bibnamefont
  {Stewart}}, \bibinfo {author} {\bibfnamefont {F.~J.}\ \bibnamefont
  {Tackmann}}, \ and\ \bibinfo {author} {\bibfnamefont {W.~J.}\ \bibnamefont
  {Waalewijn}},\ }\href {\doibase 10.1103/PhysRevLett.106.032001} {\bibfield
  {journal} {\bibinfo  {journal} {Phys. Rev. Lett.}\ }\textbf {\bibinfo
  {volume} {106}},\ \bibinfo {pages} {032001} (\bibinfo {year} {2011})},\
  \Eprint {http://arxiv.org/abs/1005.4060} {arXiv:1005.4060 [hep-ph]}
  \BibitemShut {NoStop}%
\bibitem [{\citenamefont {Berger}\ \emph {et~al.}(2011)\citenamefont {Berger},
  \citenamefont {Marcantonini}, \citenamefont {Stewart}, \citenamefont
  {Tackmann},\ and\ \citenamefont {Waalewijn}}]{Berger:2010xi}%
  \BibitemOpen
  \bibfield  {author} {\bibinfo {author} {\bibfnamefont {C.~F.}\ \bibnamefont
  {Berger}}, \bibinfo {author} {\bibfnamefont {C.}~\bibnamefont
  {Marcantonini}}, \bibinfo {author} {\bibfnamefont {I.~W.}\ \bibnamefont
  {Stewart}}, \bibinfo {author} {\bibfnamefont {F.~J.}\ \bibnamefont
  {Tackmann}}, \ and\ \bibinfo {author} {\bibfnamefont {W.~J.}\ \bibnamefont
  {Waalewijn}},\ }\href {\doibase 10.1007/JHEP04(2011)092} {\bibfield
  {journal} {\bibinfo  {journal} {JHEP}\ }\textbf {\bibinfo {volume} {04}},\
  \bibinfo {pages} {092} (\bibinfo {year} {2011})},\ \Eprint
  {http://arxiv.org/abs/1012.4480} {arXiv:1012.4480 [hep-ph]} \BibitemShut
  {NoStop}%
\bibitem [{\citenamefont {Forshaw}\ \emph {et~al.}(2006)\citenamefont
  {Forshaw}, \citenamefont {Kyrieleis},\ and\ \citenamefont
  {Seymour}}]{Forshaw:2006fk}%
  \BibitemOpen
  \bibfield  {author} {\bibinfo {author} {\bibfnamefont {J.~R.}\ \bibnamefont
  {Forshaw}}, \bibinfo {author} {\bibfnamefont {A.}~\bibnamefont {Kyrieleis}},
  \ and\ \bibinfo {author} {\bibfnamefont {M.~H.}\ \bibnamefont {Seymour}},\
  }\href {\doibase 10.1088/1126-6708/2006/08/059} {\bibfield  {journal}
  {\bibinfo  {journal} {JHEP}\ }\textbf {\bibinfo {volume} {08}},\ \bibinfo
  {pages} {059} (\bibinfo {year} {2006})},\ \Eprint
  {http://arxiv.org/abs/hep-ph/0604094} {arXiv:hep-ph/0604094 [hep-ph]}
  \BibitemShut {NoStop}%
\bibitem [{\citenamefont {Rogers}\ and\ \citenamefont
  {Mulders}(2010)}]{Rogers:2010dm}%
  \BibitemOpen
  \bibfield  {author} {\bibinfo {author} {\bibfnamefont {T.~C.}\ \bibnamefont
  {Rogers}}\ and\ \bibinfo {author} {\bibfnamefont {P.~J.}\ \bibnamefont
  {Mulders}},\ }\href {\doibase 10.1103/PhysRevD.81.094006} {\bibfield
  {journal} {\bibinfo  {journal} {Phys. Rev.}\ }\textbf {\bibinfo {volume}
  {D81}},\ \bibinfo {pages} {094006} (\bibinfo {year} {2010})},\ \Eprint
  {http://arxiv.org/abs/1001.2977} {arXiv:1001.2977 [hep-ph]} \BibitemShut
  {NoStop}%
\bibitem [{\citenamefont {Catani}\ \emph {et~al.}(2012)\citenamefont {Catani},
  \citenamefont {de~Florian},\ and\ \citenamefont {Rodrigo}}]{Catani:2011st}%
  \BibitemOpen
  \bibfield  {author} {\bibinfo {author} {\bibfnamefont {S.}~\bibnamefont
  {Catani}}, \bibinfo {author} {\bibfnamefont {D.}~\bibnamefont {de~Florian}},
  \ and\ \bibinfo {author} {\bibfnamefont {G.}~\bibnamefont {Rodrigo}},\ }\href
  {\doibase 10.1007/JHEP07(2012)026} {\bibfield  {journal} {\bibinfo  {journal}
  {JHEP}\ }\textbf {\bibinfo {volume} {07}},\ \bibinfo {pages} {026} (\bibinfo
  {year} {2012})},\ \Eprint {http://arxiv.org/abs/1112.4405} {arXiv:1112.4405
  [hep-ph]} \BibitemShut {NoStop}%
\bibitem [{\citenamefont {Forshaw}\ \emph {et~al.}(2012)\citenamefont
  {Forshaw}, \citenamefont {Seymour},\ and\ \citenamefont
  {Siodmok}}]{Forshaw:2012bi}%
  \BibitemOpen
  \bibfield  {author} {\bibinfo {author} {\bibfnamefont {J.~R.}\ \bibnamefont
  {Forshaw}}, \bibinfo {author} {\bibfnamefont {M.~H.}\ \bibnamefont
  {Seymour}}, \ and\ \bibinfo {author} {\bibfnamefont {A.}~\bibnamefont
  {Siodmok}},\ }\href {\doibase 10.1007/JHEP11(2012)066} {\bibfield  {journal}
  {\bibinfo  {journal} {JHEP}\ }\textbf {\bibinfo {volume} {11}},\ \bibinfo
  {pages} {066} (\bibinfo {year} {2012})},\ \Eprint
  {http://arxiv.org/abs/1206.6363} {arXiv:1206.6363 [hep-ph]} \BibitemShut
  {NoStop}%
\bibitem [{\citenamefont {Rothstein}\ and\ \citenamefont
  {Stewart}(2016)}]{Rothstein:2016bsq}%
  \BibitemOpen
  \bibfield  {author} {\bibinfo {author} {\bibfnamefont {I.~Z.}\ \bibnamefont
  {Rothstein}}\ and\ \bibinfo {author} {\bibfnamefont {I.~W.}\ \bibnamefont
  {Stewart}},\ }\href@noop {} {\  (\bibinfo {year} {2016})},\ \Eprint
  {http://arxiv.org/abs/1601.04695} {arXiv:1601.04695 [hep-ph]} \BibitemShut
  {NoStop}%
\bibitem [{\citenamefont {Collins}\ \emph {et~al.}(1985)\citenamefont
  {Collins}, \citenamefont {Soper},\ and\ \citenamefont
  {Sterman}}]{Collins:1984kg}%
  \BibitemOpen
  \bibfield  {author} {\bibinfo {author} {\bibfnamefont {J.~C.}\ \bibnamefont
  {Collins}}, \bibinfo {author} {\bibfnamefont {D.~E.}\ \bibnamefont {Soper}},
  \ and\ \bibinfo {author} {\bibfnamefont {G.~F.}\ \bibnamefont {Sterman}},\
  }\href {\doibase 10.1016/0550-3213(85)90479-1} {\bibfield  {journal}
  {\bibinfo  {journal} {Nucl. Phys.}\ }\textbf {\bibinfo {volume} {B250}},\
  \bibinfo {pages} {199} (\bibinfo {year} {1985})}\BibitemShut {NoStop}%
\bibitem [{\citenamefont {Gaunt}(2014)}]{Gaunt:2014ska}%
  \BibitemOpen
  \bibfield  {author} {\bibinfo {author} {\bibfnamefont {J.~R.}\ \bibnamefont
  {Gaunt}},\ }\href {\doibase 10.1007/JHEP07(2014)110} {\bibfield  {journal}
  {\bibinfo  {journal} {JHEP}\ }\textbf {\bibinfo {volume} {07}},\ \bibinfo
  {pages} {110} (\bibinfo {year} {2014})},\ \Eprint
  {http://arxiv.org/abs/1405.2080} {arXiv:1405.2080 [hep-ph]} \BibitemShut
  {NoStop}%
\bibitem [{\citenamefont {Bozzi}\ \emph {et~al.}(2011)\citenamefont {Bozzi},
  \citenamefont {Catani}, \citenamefont {Ferrera}, \citenamefont {de~Florian},\
  and\ \citenamefont {Grazzini}}]{Bozzi:2010xn}%
  \BibitemOpen
  \bibfield  {author} {\bibinfo {author} {\bibfnamefont {G.}~\bibnamefont
  {Bozzi}}, \bibinfo {author} {\bibfnamefont {S.}~\bibnamefont {Catani}},
  \bibinfo {author} {\bibfnamefont {G.}~\bibnamefont {Ferrera}}, \bibinfo
  {author} {\bibfnamefont {D.}~\bibnamefont {de~Florian}}, \ and\ \bibinfo
  {author} {\bibfnamefont {M.}~\bibnamefont {Grazzini}},\ }\href {\doibase
  10.1016/j.physletb.2010.12.024} {\bibfield  {journal} {\bibinfo  {journal}
  {Phys. Lett.}\ }\textbf {\bibinfo {volume} {B696}},\ \bibinfo {pages} {207}
  (\bibinfo {year} {2011})},\ \Eprint {http://arxiv.org/abs/1007.2351}
  {arXiv:1007.2351 [hep-ph]} \BibitemShut {NoStop}%
\bibitem [{\citenamefont {de~Florian}\ \emph {et~al.}(2011)\citenamefont
  {de~Florian}, \citenamefont {Ferrera}, \citenamefont {Grazzini},\ and\
  \citenamefont {Tommasini}}]{deFlorian:2011xf}%
  \BibitemOpen
  \bibfield  {author} {\bibinfo {author} {\bibfnamefont {D.}~\bibnamefont
  {de~Florian}}, \bibinfo {author} {\bibfnamefont {G.}~\bibnamefont {Ferrera}},
  \bibinfo {author} {\bibfnamefont {M.}~\bibnamefont {Grazzini}}, \ and\
  \bibinfo {author} {\bibfnamefont {D.}~\bibnamefont {Tommasini}},\ }\href
  {\doibase 10.1007/JHEP11(2011)064} {\bibfield  {journal} {\bibinfo  {journal}
  {JHEP}\ }\textbf {\bibinfo {volume} {11}},\ \bibinfo {pages} {064} (\bibinfo
  {year} {2011})},\ \Eprint {http://arxiv.org/abs/1109.2109} {arXiv:1109.2109
  [hep-ph]} \BibitemShut {NoStop}%
\bibitem [{\citenamefont {Banfi}\ \emph
  {et~al.}(2012{\natexlab{b}})\citenamefont {Banfi}, \citenamefont {Dasgupta},
  \citenamefont {Marzani},\ and\ \citenamefont {Tomlinson}}]{Banfi:2012du}%
  \BibitemOpen
  \bibfield  {author} {\bibinfo {author} {\bibfnamefont {A.}~\bibnamefont
  {Banfi}}, \bibinfo {author} {\bibfnamefont {M.}~\bibnamefont {Dasgupta}},
  \bibinfo {author} {\bibfnamefont {S.}~\bibnamefont {Marzani}}, \ and\
  \bibinfo {author} {\bibfnamefont {L.}~\bibnamefont {Tomlinson}},\ }\href
  {\doibase 10.1016/j.physletb.2012.07.035} {\bibfield  {journal} {\bibinfo
  {journal} {Phys. Lett.}\ }\textbf {\bibinfo {volume} {B715}},\ \bibinfo
  {pages} {152} (\bibinfo {year} {2012}{\natexlab{b}})},\ \Eprint
  {http://arxiv.org/abs/1205.4760} {arXiv:1205.4760 [hep-ph]} \BibitemShut
  {NoStop}%
\bibitem [{\citenamefont {Becher}\ \emph
  {et~al.}(2013{\natexlab{b}})\citenamefont {Becher}, \citenamefont {Neubert},\
  and\ \citenamefont {Wilhelm}}]{Becher:2012yn}%
  \BibitemOpen
  \bibfield  {author} {\bibinfo {author} {\bibfnamefont {T.}~\bibnamefont
  {Becher}}, \bibinfo {author} {\bibfnamefont {M.}~\bibnamefont {Neubert}}, \
  and\ \bibinfo {author} {\bibfnamefont {D.}~\bibnamefont {Wilhelm}},\ }\href
  {\doibase 10.1007/JHEP05(2013)110} {\bibfield  {journal} {\bibinfo  {journal}
  {JHEP}\ }\textbf {\bibinfo {volume} {05}},\ \bibinfo {pages} {110} (\bibinfo
  {year} {2013}{\natexlab{b}})},\ \Eprint {http://arxiv.org/abs/1212.2621}
  {arXiv:1212.2621 [hep-ph]} \BibitemShut {NoStop}%
\bibitem [{\citenamefont {Echevarria}\ \emph
  {et~al.}(2015{\natexlab{a}})\citenamefont {Echevarria}, \citenamefont
  {Kasemets}, \citenamefont {Mulders},\ and\ \citenamefont
  {Pisano}}]{Echevarria:2015uaa}%
  \BibitemOpen
  \bibfield  {author} {\bibinfo {author} {\bibfnamefont {M.~G.}\ \bibnamefont
  {Echevarria}}, \bibinfo {author} {\bibfnamefont {T.}~\bibnamefont
  {Kasemets}}, \bibinfo {author} {\bibfnamefont {P.~J.}\ \bibnamefont
  {Mulders}}, \ and\ \bibinfo {author} {\bibfnamefont {C.}~\bibnamefont
  {Pisano}},\ }\href {\doibase 10.1007/JHEP07(2015)158} {\bibfield  {journal}
  {\bibinfo  {journal} {JHEP}\ }\textbf {\bibinfo {volume} {07}},\ \bibinfo
  {pages} {158} (\bibinfo {year} {2015}{\natexlab{a}})},\ \Eprint
  {http://arxiv.org/abs/1502.05354} {arXiv:1502.05354 [hep-ph]} \BibitemShut
  {NoStop}%
\bibitem [{\citenamefont {Neill}\ \emph {et~al.}(2015)\citenamefont {Neill},
  \citenamefont {Rothstein},\ and\ \citenamefont {Vaidya}}]{Neill:2015roa}%
  \BibitemOpen
  \bibfield  {author} {\bibinfo {author} {\bibfnamefont {D.}~\bibnamefont
  {Neill}}, \bibinfo {author} {\bibfnamefont {I.~Z.}\ \bibnamefont
  {Rothstein}}, \ and\ \bibinfo {author} {\bibfnamefont {V.}~\bibnamefont
  {Vaidya}},\ }\href {\doibase 10.1007/JHEP12(2015)097} {\bibfield  {journal}
  {\bibinfo  {journal} {JHEP}\ }\textbf {\bibinfo {volume} {12}},\ \bibinfo
  {pages} {097} (\bibinfo {year} {2015})},\ \Eprint
  {http://arxiv.org/abs/1503.00005} {arXiv:1503.00005 [hep-ph]} \BibitemShut
  {NoStop}%
\bibitem [{\citenamefont {Bagnaschi}\ \emph {et~al.}(2016)\citenamefont
  {Bagnaschi}, \citenamefont {Harlander}, \citenamefont {Mantler},
  \citenamefont {Vicini},\ and\ \citenamefont {Wiesemann}}]{Bagnaschi:2015bop}%
  \BibitemOpen
  \bibfield  {author} {\bibinfo {author} {\bibfnamefont {E.}~\bibnamefont
  {Bagnaschi}}, \bibinfo {author} {\bibfnamefont {R.~V.}\ \bibnamefont
  {Harlander}}, \bibinfo {author} {\bibfnamefont {H.}~\bibnamefont {Mantler}},
  \bibinfo {author} {\bibfnamefont {A.}~\bibnamefont {Vicini}}, \ and\ \bibinfo
  {author} {\bibfnamefont {M.}~\bibnamefont {Wiesemann}},\ }\href {\doibase
  10.1007/JHEP01(2016)090} {\bibfield  {journal} {\bibinfo  {journal} {JHEP}\
  }\textbf {\bibinfo {volume} {01}},\ \bibinfo {pages} {090} (\bibinfo {year}
  {2016})},\ \Eprint {http://arxiv.org/abs/1510.08850} {arXiv:1510.08850
  [hep-ph]} \BibitemShut {NoStop}%
\bibitem [{\citenamefont {de~Florian}\ and\ \citenamefont
  {Grazzini}(2000)}]{deFlorian:2000pr}%
  \BibitemOpen
  \bibfield  {author} {\bibinfo {author} {\bibfnamefont {D.}~\bibnamefont
  {de~Florian}}\ and\ \bibinfo {author} {\bibfnamefont {M.}~\bibnamefont
  {Grazzini}},\ }\href {\doibase 10.1103/PhysRevLett.85.4678} {\bibfield
  {journal} {\bibinfo  {journal} {Phys. Rev. Lett.}\ }\textbf {\bibinfo
  {volume} {85}},\ \bibinfo {pages} {4678} (\bibinfo {year} {2000})},\ \Eprint
  {http://arxiv.org/abs/hep-ph/0008152} {arXiv:hep-ph/0008152 [hep-ph]}
  \BibitemShut {NoStop}%
\bibitem [{\citenamefont {de~Florian}\ and\ \citenamefont
  {Grazzini}(2001)}]{deFlorian:2001zd}%
  \BibitemOpen
  \bibfield  {author} {\bibinfo {author} {\bibfnamefont {D.}~\bibnamefont
  {de~Florian}}\ and\ \bibinfo {author} {\bibfnamefont {M.}~\bibnamefont
  {Grazzini}},\ }\href {\doibase 10.1016/S0550-3213(01)00460-6} {\bibfield
  {journal} {\bibinfo  {journal} {Nucl. Phys.}\ }\textbf {\bibinfo {volume}
  {B616}},\ \bibinfo {pages} {247} (\bibinfo {year} {2001})},\ \Eprint
  {http://arxiv.org/abs/hep-ph/0108273} {arXiv:hep-ph/0108273 [hep-ph]}
  \BibitemShut {NoStop}%
\bibitem [{\citenamefont {Gehrmann}\ \emph {et~al.}(2014)\citenamefont
  {Gehrmann}, \citenamefont {Luebbert},\ and\ \citenamefont
  {Yang}}]{Gehrmann:2014yya}%
  \BibitemOpen
  \bibfield  {author} {\bibinfo {author} {\bibfnamefont {T.}~\bibnamefont
  {Gehrmann}}, \bibinfo {author} {\bibfnamefont {T.}~\bibnamefont {Luebbert}},
  \ and\ \bibinfo {author} {\bibfnamefont {L.~L.}\ \bibnamefont {Yang}},\
  }\href {\doibase 10.1007/JHEP06(2014)155} {\bibfield  {journal} {\bibinfo
  {journal} {JHEP}\ }\textbf {\bibinfo {volume} {06}},\ \bibinfo {pages} {155}
  (\bibinfo {year} {2014})},\ \Eprint {http://arxiv.org/abs/1403.6451}
  {arXiv:1403.6451 [hep-ph]} \BibitemShut {NoStop}%
\bibitem [{\citenamefont {Echevarria}\ \emph
  {et~al.}(2015{\natexlab{b}})\citenamefont {Echevarria}, \citenamefont
  {Scimemi},\ and\ \citenamefont {Vladimirov}}]{Echevarria:2015byo}%
  \BibitemOpen
  \bibfield  {author} {\bibinfo {author} {\bibfnamefont {M.~G.}\ \bibnamefont
  {Echevarria}}, \bibinfo {author} {\bibfnamefont {I.}~\bibnamefont {Scimemi}},
  \ and\ \bibinfo {author} {\bibfnamefont {A.}~\bibnamefont {Vladimirov}},\
  }\href@noop {} {\  (\bibinfo {year} {2015}{\natexlab{b}})},\ \Eprint
  {http://arxiv.org/abs/1511.05590} {arXiv:1511.05590 [hep-ph]} \BibitemShut
  {NoStop}%
\bibitem [{\citenamefont {Luebbert}\ \emph {et~al.}(2016)\citenamefont
  {Luebbert}, \citenamefont {Oredsson},\ and\ \citenamefont
  {Stahlhofen}}]{Luebbert:2016itl}%
  \BibitemOpen
  \bibfield  {author} {\bibinfo {author} {\bibfnamefont {T.}~\bibnamefont
  {Luebbert}}, \bibinfo {author} {\bibfnamefont {J.}~\bibnamefont {Oredsson}},
  \ and\ \bibinfo {author} {\bibfnamefont {M.}~\bibnamefont {Stahlhofen}},\
  }\href@noop {} {\  (\bibinfo {year} {2016})},\ \Eprint
  {http://arxiv.org/abs/1602.01829} {arXiv:1602.01829 [hep-ph]} \BibitemShut
  {NoStop}%
\bibitem [{\citenamefont {Collins}(2013)}]{Collins:2011zzd}%
  \BibitemOpen
  \bibfield  {author} {\bibinfo {author} {\bibfnamefont {J.}~\bibnamefont
  {Collins}},\ }\href {http://www.cambridge.org/de/knowledge/isbn/item5756723}
  {\emph {\bibinfo {title} {{Foundations of perturbative QCD}}}}\ (\bibinfo
  {publisher} {Cambridge University Press},\ \bibinfo {year}
  {2013})\BibitemShut {NoStop}%
\bibitem [{\citenamefont {Aybat}\ and\ \citenamefont
  {Rogers}(2011)}]{Aybat:2011zv}%
  \BibitemOpen
  \bibfield  {author} {\bibinfo {author} {\bibfnamefont {S.~M.}\ \bibnamefont
  {Aybat}}\ and\ \bibinfo {author} {\bibfnamefont {T.~C.}\ \bibnamefont
  {Rogers}},\ }\href {\doibase 10.1103/PhysRevD.83.114042} {\bibfield
  {journal} {\bibinfo  {journal} {Phys. Rev.}\ }\textbf {\bibinfo {volume}
  {D83}},\ \bibinfo {pages} {114042} (\bibinfo {year} {2011})},\ \Eprint
  {http://arxiv.org/abs/1101.5057} {arXiv:1101.5057 [hep-ph]} \BibitemShut
  {NoStop}%
\bibitem [{\citenamefont {Larkoski}\ \emph {et~al.}(2014)\citenamefont
  {Larkoski}, \citenamefont {Moult},\ and\ \citenamefont
  {Neill}}]{Larkoski:2014tva}%
  \BibitemOpen
  \bibfield  {author} {\bibinfo {author} {\bibfnamefont {A.~J.}\ \bibnamefont
  {Larkoski}}, \bibinfo {author} {\bibfnamefont {I.}~\bibnamefont {Moult}}, \
  and\ \bibinfo {author} {\bibfnamefont {D.}~\bibnamefont {Neill}},\ }\href
  {\doibase 10.1007/JHEP09(2014)046} {\bibfield  {journal} {\bibinfo  {journal}
  {JHEP}\ }\textbf {\bibinfo {volume} {1409}},\ \bibinfo {pages} {046}
  (\bibinfo {year} {2014})},\ \Eprint {http://arxiv.org/abs/1401.4458}
  {arXiv:1401.4458 [hep-ph]} \BibitemShut {NoStop}%
\bibitem [{\citenamefont {Anastasiou}\ \emph {et~al.}(2013)\citenamefont
  {Anastasiou}, \citenamefont {Duhr}, \citenamefont {Dulat},\ and\
  \citenamefont {Mistlberger}}]{Anastasiou:2013srw}%
  \BibitemOpen
  \bibfield  {author} {\bibinfo {author} {\bibfnamefont {C.}~\bibnamefont
  {Anastasiou}}, \bibinfo {author} {\bibfnamefont {C.}~\bibnamefont {Duhr}},
  \bibinfo {author} {\bibfnamefont {F.}~\bibnamefont {Dulat}}, \ and\ \bibinfo
  {author} {\bibfnamefont {B.}~\bibnamefont {Mistlberger}},\ }\href {\doibase
  10.1007/JHEP07(2013)003} {\bibfield  {journal} {\bibinfo  {journal} {JHEP}\
  }\textbf {\bibinfo {volume} {07}},\ \bibinfo {pages} {003} (\bibinfo {year}
  {2013})},\ \Eprint {http://arxiv.org/abs/1302.4379} {arXiv:1302.4379
  [hep-ph]} \BibitemShut {NoStop}%
\bibitem [{\citenamefont {Li}\ and\ \citenamefont {Zhu}(2013)}]{Li:2013lsa}%
  \BibitemOpen
  \bibfield  {author} {\bibinfo {author} {\bibfnamefont {Y.}~\bibnamefont
  {Li}}\ and\ \bibinfo {author} {\bibfnamefont {H.~X.}\ \bibnamefont {Zhu}},\
  }\href {\doibase 10.1007/JHEP11(2013)080} {\bibfield  {journal} {\bibinfo
  {journal} {JHEP}\ }\textbf {\bibinfo {volume} {11}},\ \bibinfo {pages} {080}
  (\bibinfo {year} {2013})},\ \Eprint {http://arxiv.org/abs/1309.4391}
  {arXiv:1309.4391 [hep-ph]} \BibitemShut {NoStop}%
\bibitem [{\citenamefont {Duhr}\ and\ \citenamefont
  {Gehrmann}(2013)}]{Duhr:2013msa}%
  \BibitemOpen
  \bibfield  {author} {\bibinfo {author} {\bibfnamefont {C.}~\bibnamefont
  {Duhr}}\ and\ \bibinfo {author} {\bibfnamefont {T.}~\bibnamefont
  {Gehrmann}},\ }\href {\doibase 10.1016/j.physletb.2013.10.063} {\bibfield
  {journal} {\bibinfo  {journal} {Phys. Lett.}\ }\textbf {\bibinfo {volume}
  {B727}},\ \bibinfo {pages} {452} (\bibinfo {year} {2013})},\ \Eprint
  {http://arxiv.org/abs/1309.4393} {arXiv:1309.4393 [hep-ph]} \BibitemShut
  {NoStop}%
\bibitem [{\citenamefont {Li}\ \emph {et~al.}(2014)\citenamefont {Li},
  \citenamefont {von Manteuffel}, \citenamefont {Schabinger},\ and\
  \citenamefont {Zhu}}]{Li:2014bfa}%
  \BibitemOpen
  \bibfield  {author} {\bibinfo {author} {\bibfnamefont {Y.}~\bibnamefont
  {Li}}, \bibinfo {author} {\bibfnamefont {A.}~\bibnamefont {von Manteuffel}},
  \bibinfo {author} {\bibfnamefont {R.~M.}\ \bibnamefont {Schabinger}}, \ and\
  \bibinfo {author} {\bibfnamefont {H.~X.}\ \bibnamefont {Zhu}},\ }\href
  {\doibase 10.1103/PhysRevD.90.053006} {\bibfield  {journal} {\bibinfo
  {journal} {Phys. Rev.}\ }\textbf {\bibinfo {volume} {D90}},\ \bibinfo {pages}
  {053006} (\bibinfo {year} {2014})},\ \Eprint {http://arxiv.org/abs/1404.5839}
  {arXiv:1404.5839 [hep-ph]} \BibitemShut {NoStop}%
\bibitem [{\citenamefont {Anastasiou}\ \emph {et~al.}(2014)\citenamefont
  {Anastasiou}, \citenamefont {Duhr}, \citenamefont {Dulat}, \citenamefont
  {Furlan}, \citenamefont {Gehrmann}, \citenamefont {Herzog},\ and\
  \citenamefont {Mistlberger}}]{Anastasiou:2014vaa}%
  \BibitemOpen
  \bibfield  {author} {\bibinfo {author} {\bibfnamefont {C.}~\bibnamefont
  {Anastasiou}}, \bibinfo {author} {\bibfnamefont {C.}~\bibnamefont {Duhr}},
  \bibinfo {author} {\bibfnamefont {F.}~\bibnamefont {Dulat}}, \bibinfo
  {author} {\bibfnamefont {E.}~\bibnamefont {Furlan}}, \bibinfo {author}
  {\bibfnamefont {T.}~\bibnamefont {Gehrmann}}, \bibinfo {author}
  {\bibfnamefont {F.}~\bibnamefont {Herzog}}, \ and\ \bibinfo {author}
  {\bibfnamefont {B.}~\bibnamefont {Mistlberger}},\ }\href {\doibase
  10.1016/j.physletb.2014.08.067} {\bibfield  {journal} {\bibinfo  {journal}
  {Phys. Lett.}\ }\textbf {\bibinfo {volume} {B737}},\ \bibinfo {pages} {325}
  (\bibinfo {year} {2014})},\ \Eprint {http://arxiv.org/abs/1403.4616}
  {arXiv:1403.4616 [hep-ph]} \BibitemShut {NoStop}%
\bibitem [{\citenamefont {Zhu}(2015)}]{Zhu:2014fma}%
  \BibitemOpen
  \bibfield  {author} {\bibinfo {author} {\bibfnamefont {H.~X.}\ \bibnamefont
  {Zhu}},\ }\href {\doibase 10.1007/JHEP02(2015)155} {\bibfield  {journal}
  {\bibinfo  {journal} {JHEP}\ }\textbf {\bibinfo {volume} {02}},\ \bibinfo
  {pages} {155} (\bibinfo {year} {2015})},\ \Eprint
  {http://arxiv.org/abs/1501.00236} {arXiv:1501.00236 [hep-ph]} \BibitemShut
  {NoStop}%
\bibitem [{\citenamefont {Anastasiou}\ \emph {et~al.}(2015)\citenamefont
  {Anastasiou}, \citenamefont {Duhr}, \citenamefont {Dulat}, \citenamefont
  {Furlan}, \citenamefont {Herzog},\ and\ \citenamefont
  {Mistlberger}}]{Anastasiou:2015yha}%
  \BibitemOpen
  \bibfield  {author} {\bibinfo {author} {\bibfnamefont {C.}~\bibnamefont
  {Anastasiou}}, \bibinfo {author} {\bibfnamefont {C.}~\bibnamefont {Duhr}},
  \bibinfo {author} {\bibfnamefont {F.}~\bibnamefont {Dulat}}, \bibinfo
  {author} {\bibfnamefont {E.}~\bibnamefont {Furlan}}, \bibinfo {author}
  {\bibfnamefont {F.}~\bibnamefont {Herzog}}, \ and\ \bibinfo {author}
  {\bibfnamefont {B.}~\bibnamefont {Mistlberger}},\ }\href {\doibase
  10.1007/JHEP08(2015)051} {\bibfield  {journal} {\bibinfo  {journal} {JHEP}\
  }\textbf {\bibinfo {volume} {08}},\ \bibinfo {pages} {051} (\bibinfo {year}
  {2015})},\ \Eprint {http://arxiv.org/abs/1505.04110} {arXiv:1505.04110
  [hep-ph]} \BibitemShut {NoStop}%
\bibitem [{\citenamefont {Bauer}\ \emph {et~al.}(2001)\citenamefont {Bauer},
  \citenamefont {Fleming}, \citenamefont {Pirjol},\ and\ \citenamefont
  {Stewart}}]{Bauer:2000yr}%
  \BibitemOpen
  \bibfield  {author} {\bibinfo {author} {\bibfnamefont {C.~W.}\ \bibnamefont
  {Bauer}}, \bibinfo {author} {\bibfnamefont {S.}~\bibnamefont {Fleming}},
  \bibinfo {author} {\bibfnamefont {D.}~\bibnamefont {Pirjol}}, \ and\ \bibinfo
  {author} {\bibfnamefont {I.~W.}\ \bibnamefont {Stewart}},\ }\href {\doibase
  10.1103/PhysRevD.63.114020} {\bibfield  {journal} {\bibinfo  {journal}
  {Phys.Rev.}\ }\textbf {\bibinfo {volume} {D63}},\ \bibinfo {pages} {114020}
  (\bibinfo {year} {2001})},\ \Eprint {http://arxiv.org/abs/hep-ph/0011336}
  {arXiv:hep-ph/0011336 [hep-ph]} \BibitemShut {NoStop}%
\bibitem [{\citenamefont {Bauer}\ \emph {et~al.}(2000)\citenamefont {Bauer},
  \citenamefont {Fleming},\ and\ \citenamefont {Luke}}]{Bauer:2000ew}%
  \BibitemOpen
  \bibfield  {author} {\bibinfo {author} {\bibfnamefont {C.~W.}\ \bibnamefont
  {Bauer}}, \bibinfo {author} {\bibfnamefont {S.}~\bibnamefont {Fleming}}, \
  and\ \bibinfo {author} {\bibfnamefont {M.~E.}\ \bibnamefont {Luke}},\ }\href
  {\doibase 10.1103/PhysRevD.63.014006} {\bibfield  {journal} {\bibinfo
  {journal} {Phys.Rev.}\ }\textbf {\bibinfo {volume} {D63}},\ \bibinfo {pages}
  {014006} (\bibinfo {year} {2000})},\ \Eprint
  {http://arxiv.org/abs/hep-ph/0005275} {arXiv:hep-ph/0005275 [hep-ph]}
  \BibitemShut {NoStop}%
\bibitem [{\citenamefont {Bauer}\ and\ \citenamefont
  {Stewart}(2001)}]{Bauer:2001ct}%
  \BibitemOpen
  \bibfield  {author} {\bibinfo {author} {\bibfnamefont {C.~W.}\ \bibnamefont
  {Bauer}}\ and\ \bibinfo {author} {\bibfnamefont {I.~W.}\ \bibnamefont
  {Stewart}},\ }\href {\doibase 10.1016/S0370-2693(01)00902-9} {\bibfield
  {journal} {\bibinfo  {journal} {Phys.Lett.}\ }\textbf {\bibinfo {volume}
  {B516}},\ \bibinfo {pages} {134} (\bibinfo {year} {2001})},\ \Eprint
  {http://arxiv.org/abs/hep-ph/0107001} {arXiv:hep-ph/0107001 [hep-ph]}
  \BibitemShut {NoStop}%
\bibitem [{\citenamefont {Grinstein}\ and\ \citenamefont
  {Rothstein}(1998)}]{Grinstein:1997gv}%
  \BibitemOpen
  \bibfield  {author} {\bibinfo {author} {\bibfnamefont {B.}~\bibnamefont
  {Grinstein}}\ and\ \bibinfo {author} {\bibfnamefont {I.~Z.}\ \bibnamefont
  {Rothstein}},\ }\href {\doibase 10.1103/PhysRevD.57.78} {\bibfield  {journal}
  {\bibinfo  {journal} {Phys. Rev.}\ }\textbf {\bibinfo {volume} {D57}},\
  \bibinfo {pages} {78} (\bibinfo {year} {1998})},\ \Eprint
  {http://arxiv.org/abs/hep-ph/9703298} {arXiv:hep-ph/9703298 [hep-ph]}
  \BibitemShut {NoStop}%
\bibitem [{\citenamefont {Beneke}\ \emph {et~al.}(2002)\citenamefont {Beneke},
  \citenamefont {Chapovsky}, \citenamefont {Diehl},\ and\ \citenamefont
  {Feldmann}}]{Beneke:2002ph}%
  \BibitemOpen
  \bibfield  {author} {\bibinfo {author} {\bibfnamefont {M.}~\bibnamefont
  {Beneke}}, \bibinfo {author} {\bibfnamefont {A.~P.}\ \bibnamefont
  {Chapovsky}}, \bibinfo {author} {\bibfnamefont {M.}~\bibnamefont {Diehl}}, \
  and\ \bibinfo {author} {\bibfnamefont {T.}~\bibnamefont {Feldmann}},\ }\href
  {\doibase 10.1016/S0550-3213(02)00687-9} {\bibfield  {journal} {\bibinfo
  {journal} {Nucl. Phys.}\ }\textbf {\bibinfo {volume} {B643}},\ \bibinfo
  {pages} {431} (\bibinfo {year} {2002})},\ \Eprint
  {http://arxiv.org/abs/hep-ph/0206152} {arXiv:hep-ph/0206152 [hep-ph]}
  \BibitemShut {NoStop}%
\bibitem [{\citenamefont {Becher}\ and\ \citenamefont
  {Neubert}(2011)}]{Becher:2010tm}%
  \BibitemOpen
  \bibfield  {author} {\bibinfo {author} {\bibfnamefont {T.}~\bibnamefont
  {Becher}}\ and\ \bibinfo {author} {\bibfnamefont {M.}~\bibnamefont
  {Neubert}},\ }\href {\doibase 10.1140/epjc/s10052-011-1665-7} {\bibfield
  {journal} {\bibinfo  {journal} {Eur. Phys. J.}\ }\textbf {\bibinfo {volume}
  {C71}},\ \bibinfo {pages} {1665} (\bibinfo {year} {2011})},\ \Eprint
  {http://arxiv.org/abs/1007.4005} {arXiv:1007.4005 [hep-ph]} \BibitemShut
  {NoStop}%
\bibitem [{\citenamefont {Echevarria}\ \emph {et~al.}(2012)\citenamefont
  {Echevarria}, \citenamefont {Idilbi},\ and\ \citenamefont
  {Scimemi}}]{GarciaEchevarria:2011rb}%
  \BibitemOpen
  \bibfield  {author} {\bibinfo {author} {\bibfnamefont {M.~G.}\ \bibnamefont
  {Echevarria}}, \bibinfo {author} {\bibfnamefont {A.}~\bibnamefont {Idilbi}},
  \ and\ \bibinfo {author} {\bibfnamefont {I.}~\bibnamefont {Scimemi}},\ }\href
  {\doibase 10.1007/JHEP07(2012)002} {\bibfield  {journal} {\bibinfo  {journal}
  {JHEP}\ }\textbf {\bibinfo {volume} {07}},\ \bibinfo {pages} {002} (\bibinfo
  {year} {2012})},\ \Eprint {http://arxiv.org/abs/1111.4996} {arXiv:1111.4996
  [hep-ph]} \BibitemShut {NoStop}%
\bibitem [{\citenamefont {Echevarría}\ \emph {et~al.}(2013)\citenamefont
  {Echevarría}, \citenamefont {Idilbi},\ and\ \citenamefont
  {Scimemi}}]{Echevarria:2012js}%
  \BibitemOpen
  \bibfield  {author} {\bibinfo {author} {\bibfnamefont {M.~G.}\ \bibnamefont
  {Echevarría}}, \bibinfo {author} {\bibfnamefont {A.}~\bibnamefont {Idilbi}},
  \ and\ \bibinfo {author} {\bibfnamefont {I.}~\bibnamefont {Scimemi}},\ }\href
  {\doibase 10.1016/j.physletb.2013.09.003} {\bibfield  {journal} {\bibinfo
  {journal} {Phys. Lett.}\ }\textbf {\bibinfo {volume} {B726}},\ \bibinfo
  {pages} {795} (\bibinfo {year} {2013})},\ \Eprint
  {http://arxiv.org/abs/1211.1947} {arXiv:1211.1947 [hep-ph]} \BibitemShut
  {NoStop}%
\bibitem [{\citenamefont {Collins}\ and\ \citenamefont
  {Soper}(1981)}]{Collins:1981uk}%
  \BibitemOpen
  \bibfield  {author} {\bibinfo {author} {\bibfnamefont {J.~C.}\ \bibnamefont
  {Collins}}\ and\ \bibinfo {author} {\bibfnamefont {D.~E.}\ \bibnamefont
  {Soper}},\ }\href {\doibase 10.1016/0550-3213(81)90339-4} {\bibfield
  {journal} {\bibinfo  {journal} {Nucl. Phys.}\ }\textbf {\bibinfo {volume}
  {B193}},\ \bibinfo {pages} {381} (\bibinfo {year} {1981})},\ \bibinfo {note}
  {[Erratum: Nucl. Phys.B213,545(1983)]}\BibitemShut {NoStop}%
\bibitem [{\citenamefont {Becher}\ and\ \citenamefont
  {Bell}(2012{\natexlab{b}})}]{Becher:2011dz}%
  \BibitemOpen
  \bibfield  {author} {\bibinfo {author} {\bibfnamefont {T.}~\bibnamefont
  {Becher}}\ and\ \bibinfo {author} {\bibfnamefont {G.}~\bibnamefont {Bell}},\
  }\href {\doibase 10.1016/j.physletb.2012.05.016} {\bibfield  {journal}
  {\bibinfo  {journal} {Phys. Lett.}\ }\textbf {\bibinfo {volume} {B713}},\
  \bibinfo {pages} {41} (\bibinfo {year} {2012}{\natexlab{b}})},\ \Eprint
  {http://arxiv.org/abs/1112.3907} {arXiv:1112.3907 [hep-ph]} \BibitemShut
  {NoStop}%
\bibitem [{\citenamefont {Ji}\ \emph {et~al.}(2005)\citenamefont {Ji},
  \citenamefont {Ma},\ and\ \citenamefont {Yuan}}]{Ji:2004wu}%
  \BibitemOpen
  \bibfield  {author} {\bibinfo {author} {\bibfnamefont {X.-d.}\ \bibnamefont
  {Ji}}, \bibinfo {author} {\bibfnamefont {J.-p.}\ \bibnamefont {Ma}}, \ and\
  \bibinfo {author} {\bibfnamefont {F.}~\bibnamefont {Yuan}},\ }\href {\doibase
  10.1103/PhysRevD.71.034005} {\bibfield  {journal} {\bibinfo  {journal} {Phys.
  Rev.}\ }\textbf {\bibinfo {volume} {D71}},\ \bibinfo {pages} {034005}
  (\bibinfo {year} {2005})},\ \Eprint {http://arxiv.org/abs/hep-ph/0404183}
  {arXiv:hep-ph/0404183 [hep-ph]} \BibitemShut {NoStop}%
\bibitem [{\citenamefont {Chiu}\ \emph {et~al.}(2009)\citenamefont {Chiu},
  \citenamefont {Fuhrer}, \citenamefont {Hoang}, \citenamefont {Kelley},\ and\
  \citenamefont {Manohar}}]{Chiu:2009yx}%
  \BibitemOpen
  \bibfield  {author} {\bibinfo {author} {\bibfnamefont {J.-y.}\ \bibnamefont
  {Chiu}}, \bibinfo {author} {\bibfnamefont {A.}~\bibnamefont {Fuhrer}},
  \bibinfo {author} {\bibfnamefont {A.~H.}\ \bibnamefont {Hoang}}, \bibinfo
  {author} {\bibfnamefont {R.}~\bibnamefont {Kelley}}, \ and\ \bibinfo {author}
  {\bibfnamefont {A.~V.}\ \bibnamefont {Manohar}},\ }\href {\doibase
  10.1103/PhysRevD.79.053007} {\bibfield  {journal} {\bibinfo  {journal} {Phys.
  Rev.}\ }\textbf {\bibinfo {volume} {D79}},\ \bibinfo {pages} {053007}
  (\bibinfo {year} {2009})},\ \Eprint {http://arxiv.org/abs/0901.1332}
  {arXiv:0901.1332 [hep-ph]} \BibitemShut {NoStop}%
\bibitem [{\citenamefont {Gatheral}(1983)}]{Gatheral:1983cz}%
  \BibitemOpen
  \bibfield  {author} {\bibinfo {author} {\bibfnamefont {J.~G.~M.}\
  \bibnamefont {Gatheral}},\ }\href {\doibase 10.1016/0370-2693(83)90112-0}
  {\bibfield  {journal} {\bibinfo  {journal} {Phys. Lett.}\ }\textbf {\bibinfo
  {volume} {B133}},\ \bibinfo {pages} {90} (\bibinfo {year}
  {1983})}\BibitemShut {NoStop}%
\bibitem [{\citenamefont {Frenkel}\ and\ \citenamefont
  {Taylor}(1984)}]{Frenkel:1984pz}%
  \BibitemOpen
  \bibfield  {author} {\bibinfo {author} {\bibfnamefont {J.}~\bibnamefont
  {Frenkel}}\ and\ \bibinfo {author} {\bibfnamefont {J.~C.}\ \bibnamefont
  {Taylor}},\ }\href {\doibase 10.1016/0550-3213(84)90294-3} {\bibfield
  {journal} {\bibinfo  {journal} {Nucl. Phys.}\ }\textbf {\bibinfo {volume}
  {B246}},\ \bibinfo {pages} {231} (\bibinfo {year} {1984})}\BibitemShut
  {NoStop}%
\bibitem [{\citenamefont {Manohar}\ and\ \citenamefont
  {Stewart}(2007)}]{Manohar:2006nz}%
  \BibitemOpen
  \bibfield  {author} {\bibinfo {author} {\bibfnamefont {A.~V.}\ \bibnamefont
  {Manohar}}\ and\ \bibinfo {author} {\bibfnamefont {I.~W.}\ \bibnamefont
  {Stewart}},\ }\href {\doibase 10.1103/PhysRevD.76.074002} {\bibfield
  {journal} {\bibinfo  {journal} {Phys.Rev.}\ }\textbf {\bibinfo {volume}
  {D76}},\ \bibinfo {pages} {074002} (\bibinfo {year} {2007})},\ \Eprint
  {http://arxiv.org/abs/hep-ph/0605001} {arXiv:hep-ph/0605001 [hep-ph]}
  \BibitemShut {NoStop}%
\bibitem [{\citenamefont {Manohar}(2003)}]{Manohar:2003vb}%
  \BibitemOpen
  \bibfield  {author} {\bibinfo {author} {\bibfnamefont {A.~V.}\ \bibnamefont
  {Manohar}},\ }\href {\doibase 10.1103/PhysRevD.68.114019} {\bibfield
  {journal} {\bibinfo  {journal} {Phys. Rev.}\ }\textbf {\bibinfo {volume}
  {D68}},\ \bibinfo {pages} {114019} (\bibinfo {year} {2003})},\ \Eprint
  {http://arxiv.org/abs/hep-ph/0309176} {arXiv:hep-ph/0309176 [hep-ph]}
  \BibitemShut {NoStop}%
\bibitem [{\citenamefont {Chiu}\ \emph {et~al.}(2008)\citenamefont {Chiu},
  \citenamefont {Golf}, \citenamefont {Kelley},\ and\ \citenamefont
  {Manohar}}]{Chiu:2007dg}%
  \BibitemOpen
  \bibfield  {author} {\bibinfo {author} {\bibfnamefont {J.-y.}\ \bibnamefont
  {Chiu}}, \bibinfo {author} {\bibfnamefont {F.}~\bibnamefont {Golf}}, \bibinfo
  {author} {\bibfnamefont {R.}~\bibnamefont {Kelley}}, \ and\ \bibinfo {author}
  {\bibfnamefont {A.~V.}\ \bibnamefont {Manohar}},\ }\href {\doibase
  10.1103/PhysRevD.77.053004} {\bibfield  {journal} {\bibinfo  {journal} {Phys.
  Rev.}\ }\textbf {\bibinfo {volume} {D77}},\ \bibinfo {pages} {053004}
  (\bibinfo {year} {2008})},\ \Eprint {http://arxiv.org/abs/0712.0396}
  {arXiv:0712.0396 [hep-ph]} \BibitemShut {NoStop}%
\bibitem [{\citenamefont {Manohar}\ \emph {et~al.}(2002)\citenamefont
  {Manohar}, \citenamefont {Mehen}, \citenamefont {Pirjol},\ and\ \citenamefont
  {Stewart}}]{Manohar:2002fd}%
  \BibitemOpen
  \bibfield  {author} {\bibinfo {author} {\bibfnamefont {A.~V.}\ \bibnamefont
  {Manohar}}, \bibinfo {author} {\bibfnamefont {T.}~\bibnamefont {Mehen}},
  \bibinfo {author} {\bibfnamefont {D.}~\bibnamefont {Pirjol}}, \ and\ \bibinfo
  {author} {\bibfnamefont {I.~W.}\ \bibnamefont {Stewart}},\ }\href {\doibase
  10.1016/S0370-2693(02)02029-4} {\bibfield  {journal} {\bibinfo  {journal}
  {Phys. Lett.}\ }\textbf {\bibinfo {volume} {B539}},\ \bibinfo {pages} {59}
  (\bibinfo {year} {2002})},\ \Eprint {http://arxiv.org/abs/hep-ph/0204229}
  {arXiv:hep-ph/0204229 [hep-ph]} \BibitemShut {NoStop}%
\bibitem [{\citenamefont {Belitsky}(1998)}]{Belitsky:1998tc}%
  \BibitemOpen
  \bibfield  {author} {\bibinfo {author} {\bibfnamefont {A.~V.}\ \bibnamefont
  {Belitsky}},\ }\href {\doibase 10.1016/S0370-2693(98)01249-0} {\bibfield
  {journal} {\bibinfo  {journal} {Phys. Lett.}\ }\textbf {\bibinfo {volume}
  {B442}},\ \bibinfo {pages} {307} (\bibinfo {year} {1998})},\ \Eprint
  {http://arxiv.org/abs/hep-ph/9808389} {arXiv:hep-ph/9808389 [hep-ph]}
  \BibitemShut {NoStop}%
\bibitem [{\citenamefont {Huber}\ and\ \citenamefont
  {Maitre}(2006)}]{Huber:2005yg}%
  \BibitemOpen
  \bibfield  {author} {\bibinfo {author} {\bibfnamefont {T.}~\bibnamefont
  {Huber}}\ and\ \bibinfo {author} {\bibfnamefont {D.}~\bibnamefont {Maitre}},\
  }\href {\doibase 10.1016/j.cpc.2006.01.007} {\bibfield  {journal} {\bibinfo
  {journal} {Comput.Phys.Commun.}\ }\textbf {\bibinfo {volume} {175}},\
  \bibinfo {pages} {122} (\bibinfo {year} {2006})},\ \Eprint
  {http://arxiv.org/abs/hep-ph/0507094} {arXiv:hep-ph/0507094 [hep-ph]}
  \BibitemShut {NoStop}%
\bibitem [{\citenamefont {Mantry}\ and\ \citenamefont
  {Petriello}(2010)}]{Mantry:2009qz}%
  \BibitemOpen
  \bibfield  {author} {\bibinfo {author} {\bibfnamefont {S.}~\bibnamefont
  {Mantry}}\ and\ \bibinfo {author} {\bibfnamefont {F.}~\bibnamefont
  {Petriello}},\ }\href {\doibase 10.1103/PhysRevD.81.093007} {\bibfield
  {journal} {\bibinfo  {journal} {Phys. Rev.}\ }\textbf {\bibinfo {volume}
  {D81}},\ \bibinfo {pages} {093007} (\bibinfo {year} {2010})},\ \Eprint
  {http://arxiv.org/abs/0911.4135} {arXiv:0911.4135 [hep-ph]} \BibitemShut
  {NoStop}%
\bibitem [{\citenamefont {Jain}\ \emph {et~al.}(2012)\citenamefont {Jain},
  \citenamefont {Procura},\ and\ \citenamefont {Waalewijn}}]{Jain:2011iu}%
  \BibitemOpen
  \bibfield  {author} {\bibinfo {author} {\bibfnamefont {A.}~\bibnamefont
  {Jain}}, \bibinfo {author} {\bibfnamefont {M.}~\bibnamefont {Procura}}, \
  and\ \bibinfo {author} {\bibfnamefont {W.~J.}\ \bibnamefont {Waalewijn}},\
  }\href {\doibase 10.1007/JHEP04(2012)132} {\bibfield  {journal} {\bibinfo
  {journal} {JHEP}\ }\textbf {\bibinfo {volume} {04}},\ \bibinfo {pages} {132}
  (\bibinfo {year} {2012})},\ \Eprint {http://arxiv.org/abs/1110.0839}
  {arXiv:1110.0839 [hep-ph]} \BibitemShut {NoStop}%
\bibitem [{\citenamefont {Procura}\ \emph {et~al.}(2015)\citenamefont
  {Procura}, \citenamefont {Waalewijn},\ and\ \citenamefont
  {Zeune}}]{Procura:2014cba}%
  \BibitemOpen
  \bibfield  {author} {\bibinfo {author} {\bibfnamefont {M.}~\bibnamefont
  {Procura}}, \bibinfo {author} {\bibfnamefont {W.~J.}\ \bibnamefont
  {Waalewijn}}, \ and\ \bibinfo {author} {\bibfnamefont {L.}~\bibnamefont
  {Zeune}},\ }\href {\doibase 10.1007/JHEP02(2015)117} {\bibfield  {journal}
  {\bibinfo  {journal} {JHEP}\ }\textbf {\bibinfo {volume} {1502}},\ \bibinfo
  {pages} {117} (\bibinfo {year} {2015})},\ \Eprint
  {http://arxiv.org/abs/1410.6483} {arXiv:1410.6483 [hep-ph]} \BibitemShut
  {NoStop}%
\bibitem [{\citenamefont {Collins}\ \emph {et~al.}(2008)\citenamefont
  {Collins}, \citenamefont {Rogers},\ and\ \citenamefont
  {Stasto}}]{Collins:2007ph}%
  \BibitemOpen
  \bibfield  {author} {\bibinfo {author} {\bibfnamefont {J.~C.}\ \bibnamefont
  {Collins}}, \bibinfo {author} {\bibfnamefont {T.~C.}\ \bibnamefont {Rogers}},
  \ and\ \bibinfo {author} {\bibfnamefont {A.~M.}\ \bibnamefont {Stasto}},\
  }\href {\doibase 10.1103/PhysRevD.77.085009} {\bibfield  {journal} {\bibinfo
  {journal} {Phys. Rev.}\ }\textbf {\bibinfo {volume} {D77}},\ \bibinfo {pages}
  {085009} (\bibinfo {year} {2008})},\ \Eprint {http://arxiv.org/abs/0708.2833}
  {arXiv:0708.2833 [hep-ph]} \BibitemShut {NoStop}%
\bibitem [{\citenamefont {Rogers}(2008)}]{Rogers:2008jk}%
  \BibitemOpen
  \bibfield  {author} {\bibinfo {author} {\bibfnamefont {T.~C.}\ \bibnamefont
  {Rogers}},\ }\href {\doibase 10.1103/PhysRevD.78.074018} {\bibfield
  {journal} {\bibinfo  {journal} {Phys. Rev.}\ }\textbf {\bibinfo {volume}
  {D78}},\ \bibinfo {pages} {074018} (\bibinfo {year} {2008})},\ \Eprint
  {http://arxiv.org/abs/0807.2430} {arXiv:0807.2430 [hep-ph]} \BibitemShut
  {NoStop}%
\bibitem [{\citenamefont {Gaunt}\ and\ \citenamefont
  {Stahlhofen}(2014)}]{Gaunt:2014xxa}%
  \BibitemOpen
  \bibfield  {author} {\bibinfo {author} {\bibfnamefont {J.~R.}\ \bibnamefont
  {Gaunt}}\ and\ \bibinfo {author} {\bibfnamefont {M.}~\bibnamefont
  {Stahlhofen}},\ }\href {\doibase 10.1007/JHEP12(2014)146} {\bibfield
  {journal} {\bibinfo  {journal} {JHEP}\ }\textbf {\bibinfo {volume} {12}},\
  \bibinfo {pages} {146} (\bibinfo {year} {2014})},\ \Eprint
  {http://arxiv.org/abs/1409.8281} {arXiv:1409.8281 [hep-ph]} \BibitemShut
  {NoStop}%
\bibitem [{\citenamefont {Korchemsky}\ and\ \citenamefont
  {Marchesini}(1993)}]{Korchemsky:1992xv}%
  \BibitemOpen
  \bibfield  {author} {\bibinfo {author} {\bibfnamefont {G.~P.}\ \bibnamefont
  {Korchemsky}}\ and\ \bibinfo {author} {\bibfnamefont {G.}~\bibnamefont
  {Marchesini}},\ }\href {\doibase 10.1016/0550-3213(93)90167-N} {\bibfield
  {journal} {\bibinfo  {journal} {Nucl. Phys.}\ }\textbf {\bibinfo {volume}
  {B406}},\ \bibinfo {pages} {225} (\bibinfo {year} {1993})},\ \Eprint
  {http://arxiv.org/abs/hep-ph/9210281} {arXiv:hep-ph/9210281 [hep-ph]}
  \BibitemShut {NoStop}%
\bibitem [{\citenamefont {Becher}\ \emph {et~al.}(2008)\citenamefont {Becher},
  \citenamefont {Neubert},\ and\ \citenamefont {Xu}}]{Becher:2007ty}%
  \BibitemOpen
  \bibfield  {author} {\bibinfo {author} {\bibfnamefont {T.}~\bibnamefont
  {Becher}}, \bibinfo {author} {\bibfnamefont {M.}~\bibnamefont {Neubert}}, \
  and\ \bibinfo {author} {\bibfnamefont {G.}~\bibnamefont {Xu}},\ }\href
  {\doibase 10.1088/1126-6708/2008/07/030} {\bibfield  {journal} {\bibinfo
  {journal} {JHEP}\ }\textbf {\bibinfo {volume} {07}},\ \bibinfo {pages} {030}
  (\bibinfo {year} {2008})},\ \Eprint {http://arxiv.org/abs/0710.0680}
  {arXiv:0710.0680 [hep-ph]} \BibitemShut {NoStop}%
\bibitem [{\citenamefont {Laenen}\ \emph {et~al.}(2001)\citenamefont {Laenen},
  \citenamefont {Sterman},\ and\ \citenamefont {Vogelsang}}]{Laenen:2000ij}%
  \BibitemOpen
  \bibfield  {author} {\bibinfo {author} {\bibfnamefont {E.}~\bibnamefont
  {Laenen}}, \bibinfo {author} {\bibfnamefont {G.~F.}\ \bibnamefont {Sterman}},
  \ and\ \bibinfo {author} {\bibfnamefont {W.}~\bibnamefont {Vogelsang}},\
  }\href {\doibase 10.1103/PhysRevD.63.114018} {\bibfield  {journal} {\bibinfo
  {journal} {Phys. Rev.}\ }\textbf {\bibinfo {volume} {D63}},\ \bibinfo {pages}
  {114018} (\bibinfo {year} {2001})},\ \Eprint
  {http://arxiv.org/abs/hep-ph/0010080} {arXiv:hep-ph/0010080 [hep-ph]}
  \BibitemShut {NoStop}%
\bibitem [{\citenamefont {Kulesza}\ \emph {et~al.}(2002)\citenamefont
  {Kulesza}, \citenamefont {Sterman},\ and\ \citenamefont
  {Vogelsang}}]{Kulesza:2002rh}%
  \BibitemOpen
  \bibfield  {author} {\bibinfo {author} {\bibfnamefont {A.}~\bibnamefont
  {Kulesza}}, \bibinfo {author} {\bibfnamefont {G.~F.}\ \bibnamefont
  {Sterman}}, \ and\ \bibinfo {author} {\bibfnamefont {W.}~\bibnamefont
  {Vogelsang}},\ }\href {\doibase 10.1103/PhysRevD.66.014011} {\bibfield
  {journal} {\bibinfo  {journal} {Phys. Rev.}\ }\textbf {\bibinfo {volume}
  {D66}},\ \bibinfo {pages} {014011} (\bibinfo {year} {2002})},\ \Eprint
  {http://arxiv.org/abs/hep-ph/0202251} {arXiv:hep-ph/0202251 [hep-ph]}
  \BibitemShut {NoStop}%
\bibitem [{\citenamefont {Kulesza}\ \emph {et~al.}(2004)\citenamefont
  {Kulesza}, \citenamefont {Sterman},\ and\ \citenamefont
  {Vogelsang}}]{Kulesza:2003wn}%
  \BibitemOpen
  \bibfield  {author} {\bibinfo {author} {\bibfnamefont {A.}~\bibnamefont
  {Kulesza}}, \bibinfo {author} {\bibfnamefont {G.~F.}\ \bibnamefont
  {Sterman}}, \ and\ \bibinfo {author} {\bibfnamefont {W.}~\bibnamefont
  {Vogelsang}},\ }\href {\doibase 10.1103/PhysRevD.69.014012} {\bibfield
  {journal} {\bibinfo  {journal} {Phys. Rev.}\ }\textbf {\bibinfo {volume}
  {D69}},\ \bibinfo {pages} {014012} (\bibinfo {year} {2004})},\ \Eprint
  {http://arxiv.org/abs/hep-ph/0309264} {arXiv:hep-ph/0309264 [hep-ph]}
  \BibitemShut {NoStop}%
\bibitem [{\citenamefont {Forte}\ and\ \citenamefont
  {Muselli}(2016)}]{Forte:2015gve}%
  \BibitemOpen
  \bibfield  {author} {\bibinfo {author} {\bibfnamefont {S.}~\bibnamefont
  {Forte}}\ and\ \bibinfo {author} {\bibfnamefont {C.}~\bibnamefont
  {Muselli}},\ }\href {\doibase 10.1007/JHEP03(2016)122} {\bibfield  {journal}
  {\bibinfo  {journal} {JHEP}\ }\textbf {\bibinfo {volume} {03}},\ \bibinfo
  {pages} {122} (\bibinfo {year} {2016})},\ \Eprint
  {http://arxiv.org/abs/1511.05561} {arXiv:1511.05561 [hep-ph]} \BibitemShut
  {NoStop}%
\bibitem [{\citenamefont {Marzani}(2015)}]{Marzani:2015oyb}%
  \BibitemOpen
  \bibfield  {author} {\bibinfo {author} {\bibfnamefont {S.}~\bibnamefont
  {Marzani}},\ }\href@noop {} {\  (\bibinfo {year} {2015})},\ \Eprint
  {http://arxiv.org/abs/1511.06039} {arXiv:1511.06039 [hep-ph]} \BibitemShut
  {NoStop}%
\bibitem [{\citenamefont {Li}\ \emph {et~al.}(2011)\citenamefont {Li},
  \citenamefont {Mantry},\ and\ \citenamefont {Petriello}}]{Li:2011zp}%
  \BibitemOpen
  \bibfield  {author} {\bibinfo {author} {\bibfnamefont {Y.}~\bibnamefont
  {Li}}, \bibinfo {author} {\bibfnamefont {S.}~\bibnamefont {Mantry}}, \ and\
  \bibinfo {author} {\bibfnamefont {F.}~\bibnamefont {Petriello}},\ }\href
  {\doibase 10.1103/PhysRevD.84.094014} {\bibfield  {journal} {\bibinfo
  {journal} {Phys. Rev.}\ }\textbf {\bibinfo {volume} {D84}},\ \bibinfo {pages}
  {094014} (\bibinfo {year} {2011})},\ \Eprint {http://arxiv.org/abs/1105.5171}
  {arXiv:1105.5171 [hep-ph]} \BibitemShut {NoStop}%
\bibitem [{\citenamefont {Remiddi}\ and\ \citenamefont
  {Vermaseren}(2000)}]{Remiddi:1999ew}%
  \BibitemOpen
  \bibfield  {author} {\bibinfo {author} {\bibfnamefont {E.}~\bibnamefont
  {Remiddi}}\ and\ \bibinfo {author} {\bibfnamefont {J.~A.~M.}\ \bibnamefont
  {Vermaseren}},\ }\href {\doibase 10.1142/S0217751X00000367} {\bibfield
  {journal} {\bibinfo  {journal} {Int. J. Mod. Phys.}\ }\textbf {\bibinfo
  {volume} {A15}},\ \bibinfo {pages} {725} (\bibinfo {year} {2000})},\ \Eprint
  {http://arxiv.org/abs/hep-ph/9905237} {arXiv:hep-ph/9905237 [hep-ph]}
  \BibitemShut {NoStop}%
\bibitem [{\citenamefont {Bern}\ \emph {et~al.}(2005)\citenamefont {Bern},
  \citenamefont {Dixon},\ and\ \citenamefont {Smirnov}}]{Bern:2005iz}%
  \BibitemOpen
  \bibfield  {author} {\bibinfo {author} {\bibfnamefont {Z.}~\bibnamefont
  {Bern}}, \bibinfo {author} {\bibfnamefont {L.~J.}\ \bibnamefont {Dixon}}, \
  and\ \bibinfo {author} {\bibfnamefont {V.~A.}\ \bibnamefont {Smirnov}},\
  }\href {\doibase 10.1103/PhysRevD.72.085001} {\bibfield  {journal} {\bibinfo
  {journal} {Phys. Rev.}\ }\textbf {\bibinfo {volume} {D72}},\ \bibinfo {pages}
  {085001} (\bibinfo {year} {2005})},\ \Eprint
  {http://arxiv.org/abs/hep-th/0505205} {arXiv:hep-th/0505205 [hep-th]}
  \BibitemShut {NoStop}%
\bibitem [{\citenamefont {Dixon}\ \emph {et~al.}(2016)\citenamefont {Dixon},
  \citenamefont {von Hippel},\ and\ \citenamefont {McLeod}}]{Dixon:2015iva}%
  \BibitemOpen
  \bibfield  {author} {\bibinfo {author} {\bibfnamefont {L.~J.}\ \bibnamefont
  {Dixon}}, \bibinfo {author} {\bibfnamefont {M.}~\bibnamefont {von Hippel}}, \
  and\ \bibinfo {author} {\bibfnamefont {A.~J.}\ \bibnamefont {McLeod}},\
  }\href {\doibase 10.1007/JHEP01(2016)053} {\bibfield  {journal} {\bibinfo
  {journal} {JHEP}\ }\textbf {\bibinfo {volume} {01}},\ \bibinfo {pages} {053}
  (\bibinfo {year} {2016})},\ \Eprint {http://arxiv.org/abs/1509.08127}
  {arXiv:1509.08127 [hep-th]} \BibitemShut {NoStop}%
\bibitem [{\citenamefont {Goncharov}\ \emph {et~al.}(2010)\citenamefont
  {Goncharov}, \citenamefont {Spradlin}, \citenamefont {Vergu},\ and\
  \citenamefont {Volovich}}]{Goncharov:2010jf}%
  \BibitemOpen
  \bibfield  {author} {\bibinfo {author} {\bibfnamefont {A.~B.}\ \bibnamefont
  {Goncharov}}, \bibinfo {author} {\bibfnamefont {M.}~\bibnamefont {Spradlin}},
  \bibinfo {author} {\bibfnamefont {C.}~\bibnamefont {Vergu}}, \ and\ \bibinfo
  {author} {\bibfnamefont {A.}~\bibnamefont {Volovich}},\ }\href {\doibase
  10.1103/PhysRevLett.105.151605} {\bibfield  {journal} {\bibinfo  {journal}
  {Phys. Rev. Lett.}\ }\textbf {\bibinfo {volume} {105}},\ \bibinfo {pages}
  {151605} (\bibinfo {year} {2010})},\ \Eprint {http://arxiv.org/abs/1006.5703}
  {arXiv:1006.5703 [hep-th]} \BibitemShut {NoStop}%
\bibitem [{\citenamefont {Chetyrkin}\ and\ \citenamefont
  {Tkachov}(1981)}]{Chetyrkin:1981qh}%
  \BibitemOpen
  \bibfield  {author} {\bibinfo {author} {\bibfnamefont {K.~G.}\ \bibnamefont
  {Chetyrkin}}\ and\ \bibinfo {author} {\bibfnamefont {F.~V.}\ \bibnamefont
  {Tkachov}},\ }\href {\doibase 10.1016/0550-3213(81)90199-1} {\bibfield
  {journal} {\bibinfo  {journal} {Nucl. Phys.}\ }\textbf {\bibinfo {volume}
  {B192}},\ \bibinfo {pages} {159} (\bibinfo {year} {1981})}\BibitemShut
  {NoStop}%
\bibitem [{\citenamefont {Li}\ \emph {et~al.}(2015)\citenamefont {Li},
  \citenamefont {von Manteuffel}, \citenamefont {Schabinger},\ and\
  \citenamefont {Zhu}}]{Li:2014afw}%
  \BibitemOpen
  \bibfield  {author} {\bibinfo {author} {\bibfnamefont {Y.}~\bibnamefont
  {Li}}, \bibinfo {author} {\bibfnamefont {A.}~\bibnamefont {von Manteuffel}},
  \bibinfo {author} {\bibfnamefont {R.~M.}\ \bibnamefont {Schabinger}}, \ and\
  \bibinfo {author} {\bibfnamefont {H.~X.}\ \bibnamefont {Zhu}},\ }\href
  {\doibase 10.1103/PhysRevD.91.036008} {\bibfield  {journal} {\bibinfo
  {journal} {Phys. Rev.}\ }\textbf {\bibinfo {volume} {D91}},\ \bibinfo {pages}
  {036008} (\bibinfo {year} {2015})},\ \Eprint {http://arxiv.org/abs/1412.2771}
  {arXiv:1412.2771 [hep-ph]} \BibitemShut {NoStop}%
\bibitem [{\citenamefont {Bern}\ \emph {et~al.}(2002)\citenamefont {Bern},
  \citenamefont {De~Freitas}, \citenamefont {Dixon},\ and\ \citenamefont
  {Wong}}]{Bern:2002zk}%
  \BibitemOpen
  \bibfield  {author} {\bibinfo {author} {\bibfnamefont {Z.}~\bibnamefont
  {Bern}}, \bibinfo {author} {\bibfnamefont {A.}~\bibnamefont {De~Freitas}},
  \bibinfo {author} {\bibfnamefont {L.~J.}\ \bibnamefont {Dixon}}, \ and\
  \bibinfo {author} {\bibfnamefont {H.~L.}\ \bibnamefont {Wong}},\ }\href
  {\doibase 10.1103/PhysRevD.66.085002} {\bibfield  {journal} {\bibinfo
  {journal} {Phys. Rev.}\ }\textbf {\bibinfo {volume} {D66}},\ \bibinfo {pages}
  {085002} (\bibinfo {year} {2002})},\ \Eprint
  {http://arxiv.org/abs/hep-ph/0202271} {arXiv:hep-ph/0202271 [hep-ph]}
  \BibitemShut {NoStop}%
\bibitem [{\citenamefont {Kotikov}\ \emph {et~al.}(2003)\citenamefont
  {Kotikov}, \citenamefont {Lipatov},\ and\ \citenamefont
  {Velizhanin}}]{Kotikov:2003fb}%
  \BibitemOpen
  \bibfield  {author} {\bibinfo {author} {\bibfnamefont {A.~V.}\ \bibnamefont
  {Kotikov}}, \bibinfo {author} {\bibfnamefont {L.~N.}\ \bibnamefont
  {Lipatov}}, \ and\ \bibinfo {author} {\bibfnamefont {V.~N.}\ \bibnamefont
  {Velizhanin}},\ }\href {\doibase 10.1016/S0370-2693(03)00184-9} {\bibfield
  {journal} {\bibinfo  {journal} {Phys. Lett.}\ }\textbf {\bibinfo {volume}
  {B557}},\ \bibinfo {pages} {114} (\bibinfo {year} {2003})},\ \Eprint
  {http://arxiv.org/abs/hep-ph/0301021} {arXiv:hep-ph/0301021 [hep-ph]}
  \BibitemShut {NoStop}%
\bibitem [{\citenamefont {Becher}\ and\ \citenamefont
  {Bell}(2014)}]{Becher:2013iya}%
  \BibitemOpen
  \bibfield  {author} {\bibinfo {author} {\bibfnamefont {T.}~\bibnamefont
  {Becher}}\ and\ \bibinfo {author} {\bibfnamefont {G.}~\bibnamefont {Bell}},\
  }\href {\doibase 10.1103/PhysRevLett.112.182002} {\bibfield  {journal}
  {\bibinfo  {journal} {Phys. Rev. Lett.}\ }\textbf {\bibinfo {volume} {112}},\
  \bibinfo {pages} {182002} (\bibinfo {year} {2014})},\ \Eprint
  {http://arxiv.org/abs/1312.5327} {arXiv:1312.5327 [hep-ph]} \BibitemShut
  {NoStop}%
\bibitem [{\citenamefont {D'Alesio}\ \emph {et~al.}(2014)\citenamefont
  {D'Alesio}, \citenamefont {Echevarria}, \citenamefont {Melis},\ and\
  \citenamefont {Scimemi}}]{D'Alesio:2014vja}%
  \BibitemOpen
  \bibfield  {author} {\bibinfo {author} {\bibfnamefont {U.}~\bibnamefont
  {D'Alesio}}, \bibinfo {author} {\bibfnamefont {M.~G.}\ \bibnamefont
  {Echevarria}}, \bibinfo {author} {\bibfnamefont {S.}~\bibnamefont {Melis}}, \
  and\ \bibinfo {author} {\bibfnamefont {I.}~\bibnamefont {Scimemi}},\ }\href
  {\doibase 10.1007/JHEP11(2014)098} {\bibfield  {journal} {\bibinfo  {journal}
  {JHEP}\ }\textbf {\bibinfo {volume} {11}},\ \bibinfo {pages} {098} (\bibinfo
  {year} {2014})},\ \Eprint {http://arxiv.org/abs/1407.3311} {arXiv:1407.3311
  [hep-ph]} \BibitemShut {NoStop}%
\bibitem [{\citenamefont {Korchemsky}\ and\ \citenamefont
  {Sterman}(1995)}]{Korchemsky:1994is}%
  \BibitemOpen
  \bibfield  {author} {\bibinfo {author} {\bibfnamefont {G.~P.}\ \bibnamefont
  {Korchemsky}}\ and\ \bibinfo {author} {\bibfnamefont {G.~F.}\ \bibnamefont
  {Sterman}},\ }\href {\doibase 10.1016/0550-3213(94)00006-Z} {\bibfield
  {journal} {\bibinfo  {journal} {Nucl. Phys.}\ }\textbf {\bibinfo {volume}
  {B437}},\ \bibinfo {pages} {415} (\bibinfo {year} {1995})},\ \Eprint
  {http://arxiv.org/abs/hep-ph/9411211} {arXiv:hep-ph/9411211 [hep-ph]}
  \BibitemShut {NoStop}%
\bibitem [{\citenamefont {Beisert}\ \emph {et~al.}(2012)\citenamefont {Beisert}
  \emph {et~al.}}]{Beisert:2010jr}%
  \BibitemOpen
  \bibfield  {author} {\bibinfo {author} {\bibfnamefont {N.}~\bibnamefont
  {Beisert}} \emph {et~al.},\ }\href {\doibase 10.1007/s11005-011-0529-2}
  {\bibfield  {journal} {\bibinfo  {journal} {Lett. Math. Phys.}\ }\textbf
  {\bibinfo {volume} {99}},\ \bibinfo {pages} {3} (\bibinfo {year} {2012})},\
  \Eprint {http://arxiv.org/abs/1012.3982} {arXiv:1012.3982 [hep-th]}
  \BibitemShut {NoStop}%
\bibitem [{\citenamefont {Mueller}(1985)}]{Mueller:1984vh}%
  \BibitemOpen
  \bibfield  {author} {\bibinfo {author} {\bibfnamefont {A.~H.}\ \bibnamefont
  {Mueller}},\ }\href {\doibase 10.1016/0550-3213(85)90485-7} {\bibfield
  {journal} {\bibinfo  {journal} {Nucl. Phys.}\ }\textbf {\bibinfo {volume}
  {B250}},\ \bibinfo {pages} {327} (\bibinfo {year} {1985})}\BibitemShut
  {NoStop}%
\bibitem [{\citenamefont {Zakharov}(1992)}]{Zakharov:1992bx}%
  \BibitemOpen
  \bibfield  {author} {\bibinfo {author} {\bibfnamefont {V.~I.}\ \bibnamefont
  {Zakharov}},\ }\href {\doibase 10.1016/0550-3213(92)90054-F} {\bibfield
  {journal} {\bibinfo  {journal} {Nucl. Phys.}\ }\textbf {\bibinfo {volume}
  {B385}},\ \bibinfo {pages} {452} (\bibinfo {year} {1992})}\BibitemShut
  {NoStop}%
\bibitem [{\citenamefont {Beneke}\ and\ \citenamefont
  {Zakharov}(1993)}]{Beneke:1993yn}%
  \BibitemOpen
  \bibfield  {author} {\bibinfo {author} {\bibfnamefont {M.}~\bibnamefont
  {Beneke}}\ and\ \bibinfo {author} {\bibfnamefont {V.~I.}\ \bibnamefont
  {Zakharov}},\ }\href {\doibase 10.1016/0370-2693(93)91090-A} {\bibfield
  {journal} {\bibinfo  {journal} {Phys. Lett.}\ }\textbf {\bibinfo {volume}
  {B312}},\ \bibinfo {pages} {340} (\bibinfo {year} {1993})}\BibitemShut
  {NoStop}%
\bibitem [{\citenamefont {Bern}\ \emph {et~al.}(2007)\citenamefont {Bern},
  \citenamefont {Czakon}, \citenamefont {Dixon}, \citenamefont {Kosower},\ and\
  \citenamefont {Smirnov}}]{Bern:2006ew}%
  \BibitemOpen
  \bibfield  {author} {\bibinfo {author} {\bibfnamefont {Z.}~\bibnamefont
  {Bern}}, \bibinfo {author} {\bibfnamefont {M.}~\bibnamefont {Czakon}},
  \bibinfo {author} {\bibfnamefont {L.~J.}\ \bibnamefont {Dixon}}, \bibinfo
  {author} {\bibfnamefont {D.~A.}\ \bibnamefont {Kosower}}, \ and\ \bibinfo
  {author} {\bibfnamefont {V.~A.}\ \bibnamefont {Smirnov}},\ }\href {\doibase
  10.1103/PhysRevD.75.085010} {\bibfield  {journal} {\bibinfo  {journal} {Phys.
  Rev.}\ }\textbf {\bibinfo {volume} {D75}},\ \bibinfo {pages} {085010}
  (\bibinfo {year} {2007})},\ \Eprint {http://arxiv.org/abs/hep-th/0610248}
  {arXiv:hep-th/0610248 [hep-th]} \BibitemShut {NoStop}%
\bibitem [{\citenamefont {Moch}\ \emph {et~al.}(2004)\citenamefont {Moch},
  \citenamefont {Vermaseren},\ and\ \citenamefont {Vogt}}]{Moch:2004pa}%
  \BibitemOpen
  \bibfield  {author} {\bibinfo {author} {\bibfnamefont {S.}~\bibnamefont
  {Moch}}, \bibinfo {author} {\bibfnamefont {J.~A.~M.}\ \bibnamefont
  {Vermaseren}}, \ and\ \bibinfo {author} {\bibfnamefont {A.}~\bibnamefont
  {Vogt}},\ }\href {\doibase 10.1016/j.nuclphysb.2004.03.030} {\bibfield
  {journal} {\bibinfo  {journal} {Nucl. Phys.}\ }\textbf {\bibinfo {volume}
  {B688}},\ \bibinfo {pages} {101} (\bibinfo {year} {2004})},\ \Eprint
  {http://arxiv.org/abs/hep-ph/0403192} {arXiv:hep-ph/0403192 [hep-ph]}
  \BibitemShut {NoStop}%
\bibitem [{\citenamefont {Baikov}\ \emph {et~al.}(2009)\citenamefont {Baikov},
  \citenamefont {Chetyrkin}, \citenamefont {Smirnov}, \citenamefont {Smirnov},\
  and\ \citenamefont {Steinhauser}}]{Baikov:2009bg}%
  \BibitemOpen
  \bibfield  {author} {\bibinfo {author} {\bibfnamefont {P.~A.}\ \bibnamefont
  {Baikov}}, \bibinfo {author} {\bibfnamefont {K.~G.}\ \bibnamefont
  {Chetyrkin}}, \bibinfo {author} {\bibfnamefont {A.~V.}\ \bibnamefont
  {Smirnov}}, \bibinfo {author} {\bibfnamefont {V.~A.}\ \bibnamefont
  {Smirnov}}, \ and\ \bibinfo {author} {\bibfnamefont {M.}~\bibnamefont
  {Steinhauser}},\ }\href {\doibase 10.1103/PhysRevLett.102.212002} {\bibfield
  {journal} {\bibinfo  {journal} {Phys. Rev. Lett.}\ }\textbf {\bibinfo
  {volume} {102}},\ \bibinfo {pages} {212002} (\bibinfo {year} {2009})},\
  \Eprint {http://arxiv.org/abs/0902.3519} {arXiv:0902.3519 [hep-ph]}
  \BibitemShut {NoStop}%
\bibitem [{\citenamefont {Lee}\ \emph {et~al.}(2010)\citenamefont {Lee},
  \citenamefont {Smirnov},\ and\ \citenamefont {Smirnov}}]{Lee:2010cga}%
  \BibitemOpen
  \bibfield  {author} {\bibinfo {author} {\bibfnamefont {R.~N.}\ \bibnamefont
  {Lee}}, \bibinfo {author} {\bibfnamefont {A.~V.}\ \bibnamefont {Smirnov}}, \
  and\ \bibinfo {author} {\bibfnamefont {V.~A.}\ \bibnamefont {Smirnov}},\
  }\href {\doibase 10.1007/JHEP04(2010)020} {\bibfield  {journal} {\bibinfo
  {journal} {JHEP}\ }\textbf {\bibinfo {volume} {04}},\ \bibinfo {pages} {020}
  (\bibinfo {year} {2010})},\ \Eprint {http://arxiv.org/abs/1001.2887}
  {arXiv:1001.2887 [hep-ph]} \BibitemShut {NoStop}%
\bibitem [{\citenamefont {Gehrmann}\ \emph {et~al.}(2010)\citenamefont
  {Gehrmann}, \citenamefont {Glover}, \citenamefont {Huber}, \citenamefont
  {Ikizlerli},\ and\ \citenamefont {Studerus}}]{Gehrmann:2010ue}%
  \BibitemOpen
  \bibfield  {author} {\bibinfo {author} {\bibfnamefont {T.}~\bibnamefont
  {Gehrmann}}, \bibinfo {author} {\bibfnamefont {E.~W.~N.}\ \bibnamefont
  {Glover}}, \bibinfo {author} {\bibfnamefont {T.}~\bibnamefont {Huber}},
  \bibinfo {author} {\bibfnamefont {N.}~\bibnamefont {Ikizlerli}}, \ and\
  \bibinfo {author} {\bibfnamefont {C.}~\bibnamefont {Studerus}},\ }\href
  {\doibase 10.1007/JHEP06(2010)094} {\bibfield  {journal} {\bibinfo  {journal}
  {JHEP}\ }\textbf {\bibinfo {volume} {06}},\ \bibinfo {pages} {094} (\bibinfo
  {year} {2010})},\ \Eprint {http://arxiv.org/abs/1004.3653} {arXiv:1004.3653
  [hep-ph]} \BibitemShut {NoStop}%
\bibitem [{\citenamefont {Lee}(2012)}]{Lee:2012cn}%
  \BibitemOpen
  \bibfield  {author} {\bibinfo {author} {\bibfnamefont {R.~N.}\ \bibnamefont
  {Lee}},\ }\href@noop {} {\  (\bibinfo {year} {2012})},\ \Eprint
  {http://arxiv.org/abs/1212.2685} {arXiv:1212.2685 [hep-ph]} \BibitemShut
  {NoStop}%
\bibitem [{\citenamefont {Smirnov}(2008)}]{Smirnov:2008iw}%
  \BibitemOpen
  \bibfield  {author} {\bibinfo {author} {\bibfnamefont {A.~V.}\ \bibnamefont
  {Smirnov}},\ }\href {\doibase 10.1088/1126-6708/2008/10/107} {\bibfield
  {journal} {\bibinfo  {journal} {JHEP}\ }\textbf {\bibinfo {volume} {10}},\
  \bibinfo {pages} {107} (\bibinfo {year} {2008})},\ \Eprint
  {http://arxiv.org/abs/0807.3243} {arXiv:0807.3243 [hep-ph]} \BibitemShut
  {NoStop}%
\end{thebibliography}%
\end{document}